\documentclass{aa}
\usepackage{txfonts}
\usepackage{siunitx}
\usepackage{lipsum}
\usepackage{tabularx}

\usepackage{enumerate}
\usepackage{bm}

\usepackage{mleftright}
\let\left\mleft
\let\right\mright

\usepackage{amsmath}	

\usepackage{graphicx}   
\usepackage{subcaption}
\usepackage{siunitx}
\usepackage{xcolor}

\colorlet{lgray}{gray!50!white}
\colorlet{lred}{red!50!white}

\newcommand{\ud}{\,\mathrm{d}}

\newcommand{\bcdot}{\bm{\cdot}}

\DeclareMathOperator{\sinc}{sinc}

\usepackage{xpatch}
\usepackage{hyperref}
\hypersetup{
        colorlinks=true,
        breaklinks=true,
        citecolor=blue,
        allcolors=blue,
        frenchlinks=true
}

\makeatletter
\xpatchcmd\NAT@citex
{%
        \@citea\NAT@hyper@{%
                \NAT@nmfmt{\NAT@nm}%
                \hyper@natlinkbreak{\NAT@aysep\NAT@spacechar}{\@citeb\@extra@b@citeb}%
                \NAT@date
        }%
}
{%
        \@citea
        \NAT@nmfmt{\NAT@nm}%
        \NAT@aysep\NAT@spacechar
        \NAT@hyper@{\NAT@date}%
}
{}{}
\xpatchcmd\NAT@citex
{%
        \@citea\NAT@hyper@{%
                \NAT@nmfmt{\NAT@nm}%
                \hyper@natlinkbreak{\NAT@spacechar\NAT@@open\if*#1*\else#1\NAT@spacechar\fi}%
                {\@citeb\@extra@b@citeb}%
                \NAT@date
        }%
}
{
        \@citea
        \NAT@nmfmt{\NAT@nm}%
        \NAT@spacechar\NAT@@open\if*#1*\else#1\NAT@spacechar\fi
        \NAT@hyper@{\NAT@date}%
}
{}{}
\makeatother

\makeatletter
\renewcommand*\aa@pageof{, page \thepage{} of \pageref*{LastPage}}
\makeatother

\newcommand{\bigO}{\mathcal{O}}

\begin{document}
\title{Characterizing turbulence in galaxy clusters: Defining turbulent energies and assessing multi-scale versus fixed-scale filters}
 \titlerunning{Characterizing turbulence in galaxy clusters}

\author{Lorenzo Maria Perrone\inst{1}
                \and
                Thomas Berlok\inst{2,1}
                \and
                Ewald Puchwein\inst{1}
                \and
                Christoph Pfrommer\inst{1}
        }
        
        \institute{Leibniz-Institut f\"{u}r Astrophysik Potsdam (AIP), 
                An der Sternwarte 16, D-14482 Potsdam, Germany\\
                \email{lperrone@aip.de}
      \and
           Niels Bohr Institute, University of Copenhagen, Blegdamsvej 17, 2100 Copenhagen, Denmark
        }

\date{\today}

\abstract{
Disentangling turbulence and bulk motions in the intracluster medium (ICM) of galaxy clusters is inherently ambiguous, as the plasma is continuously stirred by different processes on disparate scales. This poses a serious problem in the interpretation of both observations and numerical simulations.
In this paper, we use filtering operators in real space to separate bulk motion from turbulence at different scales. We show how filters can be used to define consistent kinetic and magnetic energies for the bulk and turbulent component.
We apply our GPU-accelerated filtering pipeline to a simulation of a major galaxy cluster merger, which is part of the \textsc{PICO-Clusters} suite of zoom-in cosmological simulations of massive clusters using the moving mesh code \textsc{Arepo} and the \textsc{IllustrisTNG} galaxy formation model.
We find that during the merger the turbulent pressure fraction on physical scales of $\lesssim$50~kpc reaches a maximum of 5\%, before decreasing to 2\% after $\sim$1.3~Gyr from the core passage. These low values are consistent with recent observations of clusters with XRISM, and suggest that unless a cluster was recently perturbed by a major merger, turbulence levels are low. 
We then reexamine the popular multi-scale iterative filter method. In our tests, we find that its use can introduce artifacts, and that it does not reliably disentangle fluctuations living on widely separated length scales.
Rather, we believe it is more fruitful to use fixed-scale filters and turbulent energies to compare between simulations and observations. This work significantly improves our understanding of turbulence generation by major mergers in galaxy clusters, which can be probed by XRISM and next-generation X-ray telescopes, allowing us to connect high-resolution cosmological simulations to observations.
}


\maketitle



\section{Introduction}

The intracluster medium (ICM) that fills galaxy clusters is a hot and diffuse plasma, where motions are excited by different physical mechanisms on a wide range of temporal and spatial scales \citep{Simionescu2019a}. While constraining the Reynolds number of such a medium remains an area of open investigation \citep{Werner2016,Su2017,Wang2018,Zhuravleva2019,XRISMCollaboration2025c}, the ICM is understood to be turbulent, with density and velocity fluctuations probed down to scales of $\lesssim$10~kpc \citep{Walker2015}. The presence of turbulence has major implications for a number of physical processes, including, but not limited to, the amplification of magnetic fields through a small-scale dynamo, and its saturation at approximate equipartition levels with the turbulence \citep{Schekochihin2004b,Beresnyak2016,Dominguez-Fernandez2019,St-Onge2020a,Tevlin2025}; the reacceleration of relativistic electrons that are thought to produce the diffuse radio emission detected even at megaparsec-scale distances from the cluster center, as seen in radio observations \citep{Cuciti2022,Botteon2022,Beduzzi2023,Beduzzi2024,Rajpurohit2025} and modeled theoretically and in simulations \citep{Brunetti2007a,Pinzke2017a}; the turbulent heating of the ICM through dissipation of small-scale turbulence, which might contribute to offset radiative cooling in certain parts of the cluster \citep{Kunz2011b,Zhuravleva2014b,Yang2016a}; the nonthermal contribution to pressure support in the cluster, leading to hydrostatic mass bias \citep{Lau2009,Battaglia2012,Eckert2019b,Angelinelli2020a}; and, more generally, for our understanding of the effects of kinetic-scale plasma instabilities on macroscopic transport processes \citep{Zhuravleva2019,Berlok2021,Drake2021,Perrone2024,Perrone2024_bells,Yerger2025,Choudhury2025} and on the turbulent cascade \citep{Schekochihin2006a,Squire2019a,Squire2023,Arzamasskiy2023,Majeski2024b}.

Despite its importance for the formation and evolution of galaxy clusters, the properties of turbulence in the ICM are still not well understood. 
With the recent deployment of the XRISM X-ray microcalorimeter (and previously with Hitomi), it has finally become possible to directly probe small-scale gas velocities through measurements of turbulent line broadening \citep{HitomiCollaboration2016}. The first results indicate turbulent velocity amplitudes on the order of $(100-200) ~ \si{km.s^{-1}}$ in the central part of the cluster, with no obvious trend with respect to its dynamical state, i.e., whether it is a relaxed cluster or undergoing a major merger \citep{XRISMCollaboration2025c}. These values, corresponding to a turbulent pressure support of $2-4 \%$ \citep{XRISMCollaboration2025,XRISMCollaboration2025a}, are generally on the lower end of previous estimates obtained with X-ray spectrometers, although these measurements likely include contributions from different spatial scales, and in general the role of instrumental effects or data-reduction techniques is yet to be fully clarified \citep{Lau2017,ZuHone2018a,Groth2025,Vazza2025}.

The increased availability of state-of-the-art observations of several nearby clusters calls for a corresponding effort from the simulation side to 
develop a more comprehensive understanding of the drivers of turbulence and its evolution throughout cosmic time, in order to compare to observational data and constrain the physical models used to both simulate and interpret the observations.
A major stumbling block is the inherent ambiguity of defining what is turbulence in such a highly inhomogeneous environment as the ICM, which is radially stratified and where motions are stirred from megaparsec scales by cluster mergers \citep{Miniati2015a}, down to scales of $\sim$$10 ~ \si{kpc}$ by the central active galactic nucleus \citep[AGN;][]{Bourne2017,Ehlert2021} in combination with radiatively cooling gas \citep{Ehlert2023} or magnetohydrodynamic (MHD) instabilities driven by anisotropic heat conduction \citep{Perrone2022a}, and further below by kinetic plasma instabilities at observationally inaccessible scales comparable to the ion Larmor radius ($\sim$$10^{-9} ~ \si{pc}$; \citealt{Schekochihin2005a,Kunz2014a,Roberg-Clark2018,Komarov2018a,Ley2024}).

Different approaches have been employed to estimate the levels of ICM turbulence in numerical simulations and observations of galaxy clusters. Examples are spherical-shell averaging  or fitting \citep{Zhuravleva2013a,Zhuravleva2014b,Romero2024}, three-dimensional interpolation using shape functions \citep{Dolag2005,Valdarnini2011,Vazza2011}, sub-grid modeling \citep{Scannapieco2008,Iapichino2011}, power spectra estimators \citep{Gaspari2014a,Zhuravleva2014a,Shi2018a,Wang2024a}, and structure functions \citep{Miniati2014,Beresnyak2016,Li2020,Wang2021a}. In this work, we argue for an approach based on fixed-scale filtering and the computation of turbulent energies via statistical moments. We implement this method in a Python package that we parallelize for GPUs to handle large datasets efficiently. As a practical example, we apply it to a massive merging cluster from the \textsc{PICO-Clusters} simulation suite \citep{Berlok2026} run with the \textsc{Arepo} code \citep{Springel2010,Pakmor2016}, which includes magnetic fields \citep{Pakmor2011,Pakmor2013}, radiative cooling and heating, and the \textsc{IllustrisTNG} galaxy formation model \citep{Weinberger2017,Pillepich2018}. With this method we can extract detailed information on how much turbulence is generated during the merger and how it is partitioned on different scales. 
We also revisit the iterative multi-scale filtering technique \citep{Vazza2012}, which is a method that has gained traction in the community to separate bulk from turbulence and automatically identify the (spatially varying) integral and/or correlation scale of the turbulence in different parts of the cluster. We find that the detection of turbulence by iterative filters can often be unreliable even in simple test cases. For this reason, we believe they hinder the physical interpretation of the data, rather than clarify it, thereby making the fixed-scale filtering and the computation of the filtered energies our favored approach.

This paper is structured as follows. In Sect.~\ref{sec:averaging_operators} we discuss different filtering operators. In Sect.~\ref{sec:turbulent_energies} we introduce the notations for turbulent magnetic and kinetic energies through statistical moments, and examine their relation to the energy of small-scale fluctuations. We then show in Sect.~\ref{sec:turbulent_energies_application} the results of this decomposition to a massive merging cluster. Finally, in Sect.~\ref{sec:iterative_multiscale} we analyze the iterative multi-scale filters and highlight their potential pitfalls.
We draw our conclusions in Sect.~\ref{sec:conclusions}, and we outline the prospects for future work. 

The main text is complemented by the following appendices. In Appendix~\ref{app:filters_fourier_space} we compare two widely used filters in Fourier space and examine their convergence with iterative multi-scale filters. In Appendix~\ref{app:commutativity} we prove several useful mathematical results concerning the operations of volume averaging and filtering. In Appendix~\ref{app:properties-filt-en}
we show how the filtered energies change under Galilean transformations, and study the limit of subsonic turbulence. In Appendix~\ref{app:numerical_methods} we describe our numerical pipeline for filtering cosmological simulations. Finally, in Appendix~\ref{app:turbulent-energy-subvolumes}, we derive an expression for the turbulent energy in nonperiodic domains.

\section{Filtering operators}\label{sec:averaging_operators}

\subsection{Mathematical preliminaries}

In the study of turbulence, it is often useful to extract the statistical properties of chaotic fields by applying different types of mathematical operators, such as time or spatial averaging. Under the assumption of ergodicity averaging operators can be used in lieu of ensemble averaging, making them a practical tool to study chaotic flows \citep[see, e.g., discussion in][]{Frisch1995}.

A more general class of averaging operators are the filtering operators, which can be thought of as a convolution integral of the physical field with a kernel in time or space \citep[see, e.g.,][for a more in-depth discussion on smoothing filters and their applications to turbulent flows]{Pope2000}. In this work we focus on filtering operators in real space, defined by a kernel, $\mathcal{W}_\ell (\bm x', \bm x)$, parametrized by the smoothing length, $\ell$. Furthermore we assume that the filtering kernel is non-negative, normalized to unity, and centered around $\bm x$:
\begin{subequations}
\label{eq:kernel_properties}
\begin{align*}
        &\mathcal{W}_\ell (\bm x', \bm x) \geq 0, & \mathrm{(1a)} & &  & \int \mathcal{W}_\ell (\bm x', \bm x) \ud^3 x' = 1, \hspace{1em} \mathrm{(1b)}  \nonumber \\
    &\int \bm x' \mathcal{W}_\ell (\bm x', \bm x) \ud^3 x' = \bm x, & \mathrm{(1c)} 
\end{align*}
\end{subequations}
where, unless otherwise specified, the integration is carried out over the entire space. 
The application of such a filter in position space, denoted by $\langle ... \rangle_\ell$, can be written as
\begin{align}
        \langle f  \rangle_\ell (\bm x) \equiv \int f(\bm x') \mathcal{W}_\ell (\bm x', \bm x) \ud^3 x'. \label{eq:smoothing_filter}
\end{align}
Seen in spectral space, the effect of filtering is to suppress the power on certain scales but not others, depending on the shape of the kernel. In this paper we only deal with filters that leave the power on spatial scales larger than $\ell$ unchanged (low-pass filters); hence, the name ``smoothing filters.'' Thus, the filtering approach allows us to decompose a physical field into the sum of its smooth $\langle f \rangle_\ell (\bm x)$ and fluctuating $\delta f_{\ell} (\bm x)= f (\bm x) - \langle f \rangle_{\ell} (\bm x)$ components at each point in space.
The usual spatial average over a finite volume, $V$,
\begin{align}
    \langle f \rangle_V \equiv \frac{1}{V} \int_V f (\bm x') \ud^3 x' , \label{eq:volume_average}
\end{align}
can be understood as a special case of a filtering kernel, with a constant value when $\bm x' \in V$ and zero outside. The advantage of smoothing kernels over simple spatial averages is that they are local, i.e., the value of the filtered variable now depends on the position $\bm x$ and reflects the conditions in a specific part of the domain (we drop the explicit dependence of the filtered fields on $\bm x$ when there is no risk of ambiguity).

\subsection{Choice of smoothing kernel}

In this work we consider the following smoothing kernels (written here for $n=1,2,3$ dimensions): the spherical top-hat filter,
\begin{align}\label{eq:top-hat-kernel}
    \mathcal{W}_\ell (\bm x', \bm x) =
    \begin{cases}
            \Omega_n^{-1} \; \; &\text{if} \; |\bm x - \bm x'| \leq \ell ,\\
            0 \; \; &\text{if} \; |\bm x - \bm x'| > \ell ,         
    \end{cases}
\end{align}
where $\Omega_n = (\ell \sqrt{\pi})^{n} / \Gamma (n/2 + 1)$ is the volume of an $n$-dimensional sphere ($\Gamma (z)$ is the gamma function), such that $\Omega_1 = 2 \ell$, $\Omega_2 = \pi \ell^2$, and $\Omega_3 = (4/3) \pi \ell^3$ in one, two and three dimensions, respectively; and the Gaussian filter, 
\begin{align}\label{eq:Gaussian-kernel}
    \mathcal{W}_\ell (\bm x', \bm x) = \frac{1}{(2 \pi  \ell^2 )^{n/2}}
    \mathrm{e}^{-\frac{1}{2}|\bm x - \bm x'|^2 /\ell^2}.
\end{align}
The spherical top-hat and Gaussian kernels in Eqs.~\eqref{eq:top-hat-kernel}--\eqref{eq:Gaussian-kernel} are functions of the distance $|\bm x - \bm x'|$ only, which makes them isotropic around $\bm x$; nevertheless, we prefer to use the full notation $\mathcal{W}_\ell (\bm x', \bm x)$ to account for possible weighing functions, such as the density (see Favre averaging in Sect.~\ref{sec:LES}), or to mask certain regions of space (e.g., shocks; see Sect.~\ref{eq:shock_filtered_energies}).

These two kernels represent two very common choices in the literature. In particular, the top-hat filter is computationally cheap to implement but is characterized by oscillations in $k$ space. The Gaussian filter, on the other hand, is smoother and decays faster in $k$ space compared to the top-hat, but has infinite support in real space. In our implementation of the Gaussian kernel we truncate its support to $|\bm x - \bm x'| \leq 4 \ell$, but we also checked truncating at $6 \ell$ and noticed no significant differences (on the order of $0.2\%$). For their representation in Fourier space we refer the reader to Appendix~\ref{app:filters_fourier_space}.

Given that, for the same $\ell$, the spherical top-hat and the Gaussian kernels have very different shapes in real and wave number space, we introduce two additional measures of their effective spreads: the variance, $\sigma_{\ell}$, defined as
\begin{align}
    \sigma_\ell^2 = \int \bm x'^2 \mathcal{W}_{\ell} (\bm x',0) \ud^3 x',
\end{align}
and the ``cutoff'' filter length, $\lambda_{\mathrm{c}}$, which we take to be the half-width at half-maximum in wave number space (i.e., the wavelength at which the Fourier amplitude of the kernel is attenuated by a factor of $1/2$; see Appendix~\ref{app:kernels}). These measures are better suited than $\ell$ to quantify the performance of different kernels, allowing us to make a better comparison with other filtering implementations in the literature. In particular, the variance $\sigma_{\ell}$ is a good proxy for the beam size of the smoothing filter in real space, while the cutoff filter length $\lambda_{\mathrm{c}}$ is a useful measure of the filter pass-band in wave number space.
For our choice of smoothing filters, the variance and cutoff lengths are related to $\ell$ by $\sigma_\ell = \sqrt{3} \ell$, $\lambda_\mathrm{c} \simeq 5.34 \ell$ (Gaussian kernel) and $\sigma_\ell = \sqrt{3/5} \ell$, $\lambda_\mathrm{c} \simeq 2.51 \ell$ (spherical top-hat).

\begin{figure*}
        \centering
        \includegraphics[trim={0 2cm 0 0},clip,width=0.9\textwidth]{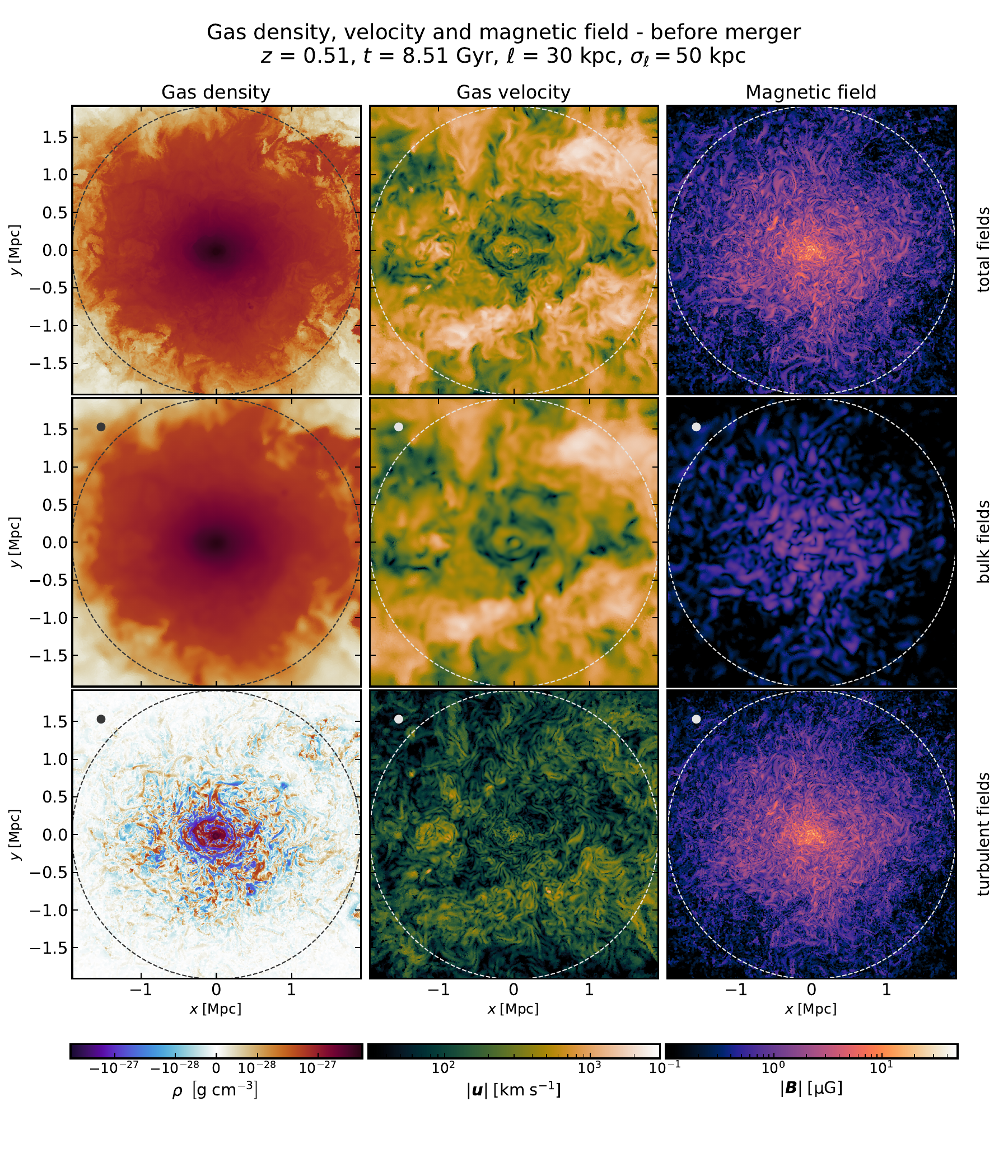}
        \caption{Decomposition of the gas density, gas velocity and magnetic field of \texttt{Halo3} before the merger ($z=0.51$) for a filter scale of $\ell=30 ~ \si{kpc}$ (represented by a beam of radius $\sigma_{\ell} = 50 ~ \si{kpc}$ in the top left corner). \textit{Top to bottom}: Slices of the total field, bulk and turbulent components across the center of the cluster. The coordinates are in physical units with respect to the center of mass of the cluster, while dashed lines denote the virial radius, $R_{200,\mathrm{c}}$. Each column has the same color bar. The cluster appears to be relatively relaxed, with a roughly spherical shape. 
    }
        \label{fig:before_all_fields}
\end{figure*}

\subsection{Connection with large-eddy simulations}\label{sec:LES}

Smoothing filters have long been used in turbulence theory to simultaneously resolve physical processes in both space and scale \citep{Leonard1974,Rogallo1984,Lesieur1996,Meneveau2000}, without placing restricting assumptions on scale separation in the flow. They are at the core of large-eddy simulation (LES) methods, where the Navier-Stokes equations \citep[or MHD, as in, e.g.,][]{Grete2015}  are filtered (``coarse-grained'') and simulated without the need to resolve the small Kolmogorov dissipation scale directly. As is well known, the downside is the appearance of an additional term in the momentum equation that represents stresses below the grid scale that needs to be modeled \citep[see discussion in][for applications of LES methods in astrophysics]{Maier2009,Schmidt2015}. 

In this work, we are less interested in the modeling of subgrid-scale stresses than in applying the same filtering machinery to study the contribution of different processes at the scales of choice. In particular, the filtering approach allows us to define notions of large-scale and small-scale energies, for which it is possible to write down rigorous equations for their time evolution.
One complication in compressible turbulence arises from the fact that filtered fields and energies are not not uniquely defined \citep[see, e.g.,][]{Aluie2013}. For instance, one may define the large-scale momentum as either $\langle \rho \bm u \rangle_{\ell}$ or  $\langle \rho \rangle_{\ell} \langle \bm u \rangle_{\ell}$. Certain choices are favored based on physical considerations; for example, the Galilean invariance of terms such as the subgrid-scale stress tensor (Appendix~\ref{app:galilean_invariance}). One popular option for compressible flows is to weigh all functions by the fluid density before averaging, also known as Favre averaging \citep{Favre1983}. For example, in this approach the Favre-filtered velocity is $\tilde{\bm u}_{\ell} \equiv \langle \rho \bm u \rangle_\ell / \langle \rho \rangle_\ell$, as opposed to simply $\langle \bm u \rangle_\ell$. 

In this work we follow Hollin's choice \citep{Hollins2022} which is based on the decomposition of the bulk kinetic energy density of compressible flows by \citet[][Sec.~6.4]{Monin1971}, replacing Reynolds averaging with filter averaging. This choice allows for simple and physically transparent interpretation of the bulk, turbulent and cross-transport terms of the kinetic energy, as outlined in Sec.~\ref{sec:kin_en_decom}.

\subsection{Filtering of a cosmological simulation}

To illustrate a practical application, we show the result of filtering in a cosmological simulation of a massive galaxy cluster, dubbed ``\texttt{Halo3}.'' The simulation is part of the \textsc{PICO-Clusters} suite of zoom-in galaxy cluster simulations \citep{Berlok2026} performed with the moving-mesh code \textsc{Arepo} (see Sect.~\ref{sec:numerics} for more details on the setup). Our filtering pipeline, a component of the \textsc{Turbocluster} library, is written in Python using the Numba-CUDA package to allow fast computations on NVIDIA graphical processing units (GPUs). For further numerical details on the implementation, see Appendix~\ref{app:numerical_methods}.

The \textsc{Arepo} code evolves the ideal MHD equations using comoving quantities in a uniformly expanding space-time defined by the Hubble function, $H(a) = \dot{a}/a$, where $a$ is the scale factor of the universe, and the dot represents the time derivative \citep{Pakmor2013,Weinberger2020a}; see, for example, Appendix B in \citet{Berlok2022} for a derivation of the comoving MHD equations. For instance, the Hubble flow 
due to the expansion of the Universe, $H(a) \bm r$, where $\bm r$ is the physical position vector, is subtracted from the total (physical) velocity, $\bm u$, and only the comoving ``peculiar'' velocity $\bm \varv$ is evolved:
\begin{align}
    \bm \varv \equiv a \dot{\bm x} = \bm u - H(a) \bm r,
\end{align}
where $\bm x = \bm r / a$ is the comoving position. However, here and in the rest of the paper it is understood that  all quantities are shown in physical (non-comoving) units. In particular, unless otherwise stated, we always compute the physical velocity, $\bm u$, with respect to the center of mass of \texttt{Halo3} and not the peculiar velocity. 

In Fig.~\ref{fig:before_all_fields}, we plot the total, bulk, and turbulent components of the density, velocity, and magnetic field, respectively, of \texttt{Halo3} at redshift $z=0.51$ before a major merger with a second massive cluster ($M_{200,\mathrm{c}} \simeq 5.38 \times 10^{14}~\mathrm{M}_{\odot}$ vs.\ $1.30 \times 10^{15}~ \mathrm{M}_{\odot}$ for \texttt{Halo3}), where $M_{200,\mathrm{c}}$ refers to the mass contained within the virial radius, $R_{200,\mathrm{c}}$, such that the mean density inside this sphere is 200 times the critical density. The smoothing kernel is Gaussian, with a physical smoothing length of $\ell = 30~\si{kpc}$ ($\sigma_{\ell} \simeq 50 ~ \si{kpc}$).
We define the smooth (bulk) and turbulent components of gas mass density, gas velocity, and magnetic field as
\begin{align}\label{eq:filtered_fields}
        \rho = \langle \rho \rangle_\ell + \delta \rho_\ell, \hspace{1em} \bm u = \langle \bm u \rangle_\ell + \delta \bm u_\ell, \hspace{1em} \bm B = \langle \bm B \rangle_\ell + \delta \bm B_\ell.
\end{align}
At the time of the snapshot in Fig.~\ref{fig:before_all_fields}, the cluster is still relatively relaxed, with a roughly spherical density distribution and low ($\lesssim$$600 ~ \si{km.s^{-1}}$) total velocities in the core region. There is a stark difference between the amplitudes of the total fields and the turbulent fluctuations. 
Filtering of the density field reveals inhomogeneous density fluctuations in the clusters: their absolute amplitude progressively decreases moving radially away from the center, and their morphology also changes from being spherically symmetric in the core region to being more isotropic further out.
Looking at the velocity fluctuations, we find that they amount to less than $200 ~ \si{km.s^{-1}}$ within the inner megaparsec region, which is about a third of the total fields. The opposite is true for the magnetic fields, as the amplitudes of the turbulent fluctuations is instead larger than that of the bulk field.
Finally, we see an increase in the amplitude of the velocity fluctuations with radius in agreement with previous studies \citep{Shi2018a}.

\section{Turbulent energies in the filtering approach}\label{sec:turbulent_energies}

\subsection{Definitions}\label{sec:definitions_turb_energies}

Having introduced the smoothing filters, we now introduce a way to define turbulent and bulk energies within the filtering approach that is both mathematically sound and physically meaningful. The solution comes from the generalized statistical moments of turbulent fields, which constitutes the cornerstone of LES of hydrodynamic turbulence \citep{Leonard1974,Rogallo1984}.
In particular, by introducing second- and third-order statistical moments, $\mu_2(f, g)$ and $\mu_3(f, g, h)$, defined as
\begin{align}
        \mu_2(f, g) &= \langle f g \rangle_\ell - \langle f \rangle_\ell \langle g \rangle_\ell, \label{eq:second_moment} \\
        \mu_3(f, g, h) &= \langle f g h \rangle_\ell - \langle f \rangle_\ell \mu_2(g,h) - \langle g \rangle_\ell \mu_2(h,f)  \nonumber \\
        & - \langle h \rangle_\ell \mu_2(f,g)- \langle f \rangle_\ell \langle g \rangle_\ell \langle h \rangle_\ell, \label{eq:third_moment}
\end{align}
where $f(\bm x)$, $g (\bm x)$, $h(\bm x)$ are generic turbulent fields, it is possible to define generalized turbulent energies, as well as their evolution equations \citep{Germano1992,Eyink2007}. Note that $f$, $g$, and $h$, can be vectors, in which case $\mu_2$, $\mu_3$ are higher-rank tensors. These statistical moments can be thought of as the correlation between the turbulent components of different fields, and vanish identically if any one of the arguments is a constant (in fact, this is the case for statistical moments of any order as can be shown, e.g., by induction). For a scalar field $\rho$, its second-order statistical moment with itself, $\mu_2 (\rho, \rho)$, can be be loosely interpreted as the variance of the field. A similar argument applies to a vector field $\bm u$ with the trace of the tensor $\mu_2(u_i, u_j) = \langle u_i u_j \rangle_\ell - \langle u_j \rangle_\ell \langle u_j \rangle_\ell $. 
It is worth noting that in the context of LES the full tensor $\mu_2(u_i, u_j)$ is known as the subgrid-scale Reynolds stress tensor, while $\mu_2 (B_i, B_j)$ is known as the Maxwell stress tensor (see Sect.~\ref{sec:LES} for further discussion on the connections with LES).

In this work, we closely follow the approach of \citet{Hollins2022} for the definitions of the kinetic and magnetic bulk and turbulent energies. 
The form of these relations depends on whether we are considering the magnetic or kinetic energy, since they involve second- or third-order correlations, respectively. We report them here below for convenience.
To avoid confusion, throughout the paper we employ the $\mathcal{E}$ and $\varepsilon$ to refer to the filtered energies and filtered energy densities, respectively, based on the statistical moments, while we use upper-case $E$ and lower-case $e$ whenever we refer to the commonplace definitions of energies and energy densities, respectively; for example, $e_{B} = \bm B^2 /8\pi$ and $e_{\mathrm{k}} = \rho \bm u^2 / 2$ for the magnetic and kinetic energy density, respectively.

\subsubsection{Magnetic energy}\label{sec:filtered_magnetic_energy}

We consider a decomposition of the magnetic field into a smooth and fluctuating component (Eq.~\ref{eq:filtered_fields}) with a smoothing filter of scale $\ell$.
Filtering the magnetic energy density $e_{B} = \bm B^2/8\pi$ leads to the following identity:
\begin{align}\label{eq:mag_energy_decomposition}
         \varepsilon_{B} = \varepsilon_{B,\mathrm{bulk}} + \varepsilon_{B,\mathrm{turb}},
\end{align}
where $\varepsilon_{B} = \langle e_{B} \rangle_\ell$ is the filtered energy density of the total field, $\varepsilon_{B,\mathrm{bulk}}$ is the energy density of the smooth magnetic field, and $\varepsilon_{B,\mathrm{turb}}$ is the filtered energy density of the fluctuations,
\begin{align}
        \varepsilon_{B} =  \frac{ \langle \bm B^2 \rangle_\ell}{8\pi} , \hspace{1em} \varepsilon_{B,\mathrm{bulk}} = \frac{\langle \bm B \rangle_\ell^2}{8\pi}, \hspace{1em} \varepsilon_{B,\mathrm{turb}} = \frac{\sum_{i} \mu_2 (B_i, B_i)}{8\pi}, \label{eq:filt_mag_energy_dens}
\end{align}
where $\mu_2$ is the second-order statistical moment of the magnetic field defined in Eq.~\eqref{eq:second_moment}, and the sum runs over the spatial indices $i=x,y,z$. 
For ease of reading, we dropped the subscript $\ell$ in the notation of the filtered energy densities (Eqs.~\ref{eq:mag_energy_decomposition}-\ref{eq:filt_mag_energy_dens}), but it is understood that they depend on the value of the filtering length. Integrating the filtered energy densities over the entire volume, one arrives at the following definitions of the filtered energies:
\begin{align}\label{eq:filt_mag_energy}
        &\mathcal{E}_{B} = \int_V \varepsilon_{B} \ud^3 x, \hspace{1em} \mathcal{E}_{B,\mathrm{bulk}} = \int_V \varepsilon_{B,\mathrm{bulk}} \ud^3 x, \nonumber \\
    &\mathcal{E}_{B,\mathrm{turb}} = \int_V \varepsilon_{B,\mathrm{turb}} \ud^3 x,
\end{align}
which satisfy
\begin{align}\label{eq:filt_mag_energy_volume}
        \mathcal{E}_{B} = \mathcal{E}_{B,\mathrm{bulk}} + \mathcal{E}_{B,\mathrm{turb}}.
\end{align}
It is worth highlighting that $\varepsilon_{B}$ (on the left-hand side of Eq.~\ref{eq:mag_energy_decomposition}) is not the energy density of the total magnetic field $e_{B} = \bm B^2 / 8\pi$, but only its filtered counterpart ($\varepsilon_{B} = \langle e_{B} \rangle_\ell \neq e_{B}$). Nevertheless, in an infinite volume or a periodic box the total volume-integrated filtered energy, $\mathcal{E}_{B} = \int_V \varepsilon_{B} \ud^3 x$, is equal to the energy of the total field, $E_B = \int_V e_{B} \ud^3 x$ (see Appendix~\ref{app:commutativity}). In the more general case of a nonperiodic box, we show in Appendix~\ref{app:commutativity} that the difference between the total filtered energy $\mathcal{E}_{B}$ and $E_B$ scales as $\mathcal{E}_{B} -  E_B \sim \bigO (\ell / L)$, where $L$ is the size of the box.

\subsubsection{Kinetic energy}\label{sec:kin_en_decom}

While in the bulk of the ICM turbulence is mostly subsonic, motions induced by mergers and sloshing can reach transonic Mach numbers, $\mathcal{M} = u / c_s \gtrsim 1$ \citep{Simionescu2019a}, where $u$ is the typical velocity amplitude and $c_s$ is the sound speed. In this case it becomes important to include contributions from the gas density fluctuations in the computation of the filtered kinetic energy, making it a third-order statistical moment. For compressible turbulence, by
decomposing the density and velocity field in terms of their smooth components and fluctuations (Eq.~\ref{eq:filtered_fields}),
it can be shown that the filtered kinetic energy density $\varepsilon_\mathrm{k}  = \langle e_{\mathrm{k}} \rangle_\ell$, where
\begin{align}
        e_{\mathrm{k}} = \frac{1}{2} \rho \bm u^2 \label{eq:kin_en_tot}
\end{align}
satisfies the following relationship:
\begin{align}
\varepsilon_\mathrm{k}  = \varepsilon_\mathrm{k,bulk} + \varepsilon_\mathrm{k,cross} + \varepsilon_\mathrm{k,turb}. \label{eq:kin_en_decomposition}
\end{align}
In Eq.~\eqref{eq:kin_en_decomposition}, $\varepsilon_\mathrm{k,bulk}$ is the energy density of the bulk flow,
\begin{align}
        \varepsilon_\mathrm{k,bulk} = \frac{1}{2} \langle \rho \rangle_\ell \sum_{i}   \langle u_i \rangle_\ell \langle u_i \rangle_\ell , \label{eq:kin_en_meanflow}
\end{align}
$\varepsilon_\mathrm{k,cross}$ is a cross term and represents the transport of the turbulent momentum by the bulk flow,
\begin{align}
        \varepsilon_\mathrm{k,cross} =  \sum_{i} \langle u_i \rangle_\ell \mu_2 (\rho, u_i), \label{eq:kin_en_transp}
\end{align}
and $\varepsilon_\mathrm{k,turb}$ is the filtered kinetic energy density of the fluctuations,
\begin{align}
        \varepsilon_\mathrm{k,turb} = \frac{1}{2} \langle \rho \rangle_\ell \sum_{i} \mu_2(u_i, u_i) + \frac{1}{2} \sum_{i} \mu_3(\rho, u_i, u_i), \label{eq:kin_en_turb}
\end{align}
where $\mu_2$ and $\mu_3$ are defined in Eqs.~\eqref{eq:second_moment}--\eqref{eq:third_moment} and the sum runs over the spatial indices $i=x,y,z$. Together $\varepsilon_\mathrm{k,cross}$ and $\varepsilon_\mathrm{k,turb}$ can be identified with the turbulent kinetic energy density, although in the rest of the paper we show them separately. In particular, $\varepsilon_\mathrm{k,turb}$ has the appealing property that it is Galilean invariant, while the filtered energy density $\varepsilon_\mathrm{k} $, the bulk energy density $\varepsilon_\mathrm{k,bulk}$, and the cross term $\varepsilon_\mathrm{k,cross}$ depend in general on the choice of frame of reference (see Appendix~\ref{app:galilean_invariance} for the derivation).

Similarly to the magnetic case, we also define the corresponding filtered energies by volume integrating Eq.~\eqref{eq:kin_en_decomposition}:
\begin{align}
        \mathcal{E}_{\mathrm{k}} = \mathcal{E}_{\mathrm{k,bulk}} + \mathcal{E}_{\mathrm{k,cross}} + \mathcal{E}_{\mathrm{k,turb}}. \label{eq:filt_kin_energy_volume}
\end{align}
As with the filtered magnetic energy, we stress that while $\varepsilon_\mathrm{k}$ is not the same as the total kinetic energy density $e_{\mathrm{k}} = (1/2) \rho \bm u^2$, when integrated over an infinite volume or a periodic box the two total energies coincide: $\mathcal{E}_{\mathrm{k}} = E_{\mathrm{k}}$, where $E_{\mathrm{k}} = \int_V e_{\mathrm{k}} \ud^3 x$. For a nonperiodic box of size $L$, their difference scales as $\mathcal{E}_{\mathrm{k}} -  E_\mathrm{k} \sim \bigO (\ell / L)$; see Appendix~\ref{app:commutativity}.

\begin{figure*}
        \centering
        \includegraphics[width=1.0\textwidth]{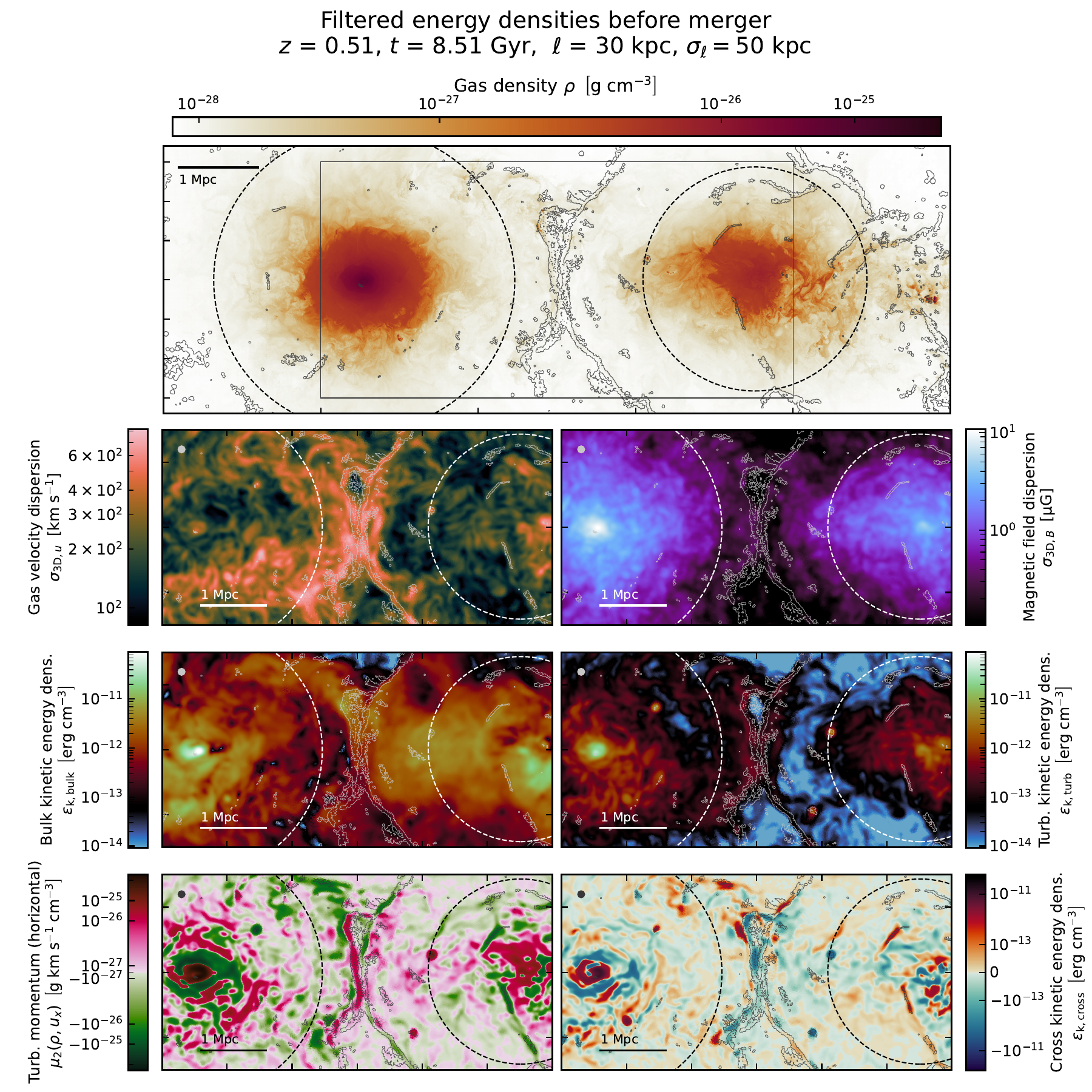}
        \caption{Filtered energy densities of the merging clusters for a filter scale of $\ell=30 ~ \si{kpc}$ (represented by a beam of radius $\sigma_\ell = 50 ~ \si{kpc}$ in the top left corner). \textit{Top to bottom}: Total density field; kinetic (left) and magnetic (right) turbulent dispersion; bulk (left) and turbulent (right) kinetic energy densities; turbulent momentum along merger axis (left) and cross kinetic energy density (right). Dashed circles are drawn at $R_{200,\mathrm{c}}$ of the two halos, while the contours represent the projected shock surfaces. Velocities are computed with respect to the center of mass of the two clusters. The velocity dispersion is larger in the periphery than in the core, while both the magnetic field dispersion and turbulent kinetic energy peak near the center and far from merger shocks.
        }
        \label{fig:filtered_energies_before_merger}
\end{figure*}

\subsection{Bulk and turbulent energy densities during a cluster merger}

In this section we show a practical application of the filtering method to compute the bulk and turbulent energy densities, shown in Fig.~\ref{fig:filtered_energies_before_merger}. We use the same snapshot as in Fig.~\ref{fig:before_all_fields} with the same filter length $\ell = 30~ \si{kpc}$ ($\sigma_\ell \simeq 50~ \si{kpc}$), 
this time displaying both clusters as they are rapidly approaching, with two strong shocks in between.
The physical velocities are now computed with respect to the center of mass of the system (only the bulk and cross kinetic energy densities are affected by the shift of frame of reference; see Appendix~\ref{app:galilean_invariance}). 
A great deal of information can be extracted from the plot: the local three-dimensional velocity, $\sigma_{\mathrm{3D},u}$, and magnetic field dispersion, $\sigma_{\mathrm{3D},B}$, below $50~ \si{kpc}$, which we computed through the second-order moment, $\mu_2$, as
\begin{align}\label{eq:3d-dispersion}
    &\sigma_{\mathrm{3D},u}^2  = \sum_{i=x,y,z} \mu_2 (u_i, u_i), &\text{and} && \sigma_{\mathrm{3D},B}^2 = \sum_{i=x,y,z} \mu_2 (B_i, B_i),
\end{align}
confirm the impression from Fig.~\ref{fig:before_all_fields} of a relatively quiescent core with higher levels of kinetic turbulence per unit mass at the virial radius, while the magnetic dispersion decreases with radius throughout. However this picture is partly modified when looking at the kinetic energy densities, which include the contribution of the smooth and turbulent density field: both bulk and turbulent kinetic energy densities, in fact, peak near the cores of the two clusters, where the gas density is much higher, with the turbulent component ($\varepsilon_\mathrm{k,turb}$) subdominant but better correlated with the magnetic dispersion, which is expected if the turbulent magnetic and kinetic energies are in equipartition. The cross kinetic energy term ($\varepsilon_\mathrm{k,cross}$) is dominated by the turbulent momentum along the merger axis, and is concentrated in the core regions and at large-scale features such as shocks.

\subsection{Physical interpretation of the filtered energies}\label{sec:physical_interpretation}

In the previous sections, we introduced the filtered energy densities for the magnetic and kinetic energies. From a mathematical point of view, their definitions follow directly from the second- and third-order statistical moments of the density, magnetic and velocity field. 
In this section, we briefly discuss some aspects that are meant to help build an intuitive physical interpretation. The reader interested in a practical application may want to jump ahead to Sect.~\ref{sec:turbulent_energies_application}, where we apply the filtering approach to simulations of a major cluster merger.
Here we address in particular four questions: (a) the different definitions of the turbulent energies with the filtering approach and with Reynolds averaging; (b) the connection between the filtered turbulent energy and the power spectrum; (c) the limit of the filtered kinetic energies in homogeneous subsonic turbulence; and (d) the influence of shocks on the turbulent energies.

\subsubsection{Filtered energies and Reynolds averaging}\label{sec:explicit_pseudo_energy}

In studies of turbulence it is common to use spatial averages to define fluctuating fields in a similar way to smoothing filters; for example, in the case of the magnetic field, $\delta \bm B = \bm B - \langle \bm B \rangle_V$. One advantage of spatial averages is that they satisfy the Reynolds rules of averaging,
\begin{align}\label{eq:reynolds-averaging}
        \langle f \langle g \rangle_V \rangle_V = \langle f \rangle_V \langle g \rangle_V,
\end{align}
where $f (\bm x)$ and $g (\bm x)$ are generic fields \citep[see, e.g.,][]{Monin1971}. This allows one to decompose the energy of a field (e.g., the magnetic field) into the sum of the energies of the constant-in-space volume average, $\langle \bm B \rangle_V$, and its spatially varying fluctuating component, $\delta \bm B$, respectively:
\begin{align}
        E_B &\equiv \int_V \frac{\bm B^2}{8\pi} \ud^3 x = \left.E_B\right|_{\mathrm{bulk}} + \left.E_B\right|_{\mathrm{turb}} \nonumber \\
    &= \int_V \frac{\langle \bm B \rangle_V^2}{8\pi} \ud^3 x + \int_V \frac{\delta \bm B^2}{8\pi} \ud^3 x. \label{eq:volume_averaging_magnetic_energy}
\end{align}
As we show below, this decomposition does not hold true for a general filtering operator that satisfies Eq.~\eqref{eq:kernel_properties}: if we were to define the energies of the bulk and of the fluctuating magnetic field in a analogous way to Eq.~\eqref{eq:volume_averaging_magnetic_energy}, i.e., ($\int_V \langle \bm B \rangle_\ell^2/8\pi  \ud^3 x $) and ($\int_V \delta \bm B_\ell^2/8\pi  \ud^3 x$) we would find that they do not sum up meaningfully.
This is the reason why a different formalism must be employed when computing the energies of the bulk component and of the fluctuations, as we described in Sects.~\ref{sec:filtered_magnetic_energy} and \ref{sec:kin_en_decom}.

It is instructive to compare the filtered turbulent energy $\mathcal{E}_{B,\mathrm{turb}}$ with the analogous expression from Eq.~\eqref{eq:volume_averaging_magnetic_energy} obtained via volume-averaging. We start by rewriting the second-order statistical moment $\mu_2$ using the definition in Eq.~\eqref{eq:second_moment} as
\begin{align}
    &\sum_{i} \mu_2 (B_i, B_i) = \langle B^2 \rangle_{\ell} - \langle \bm B \rangle_\ell^2 = \langle \left(\langle \bm B \rangle_\ell + \delta \bm B_\ell \right)^2 \rangle_{\ell} - \langle \bm B \rangle_\ell^2 \nonumber \\
    &= \left[ \langle \delta \bm B_\ell^2 \rangle_\ell + 2 \langle \delta \bm B_\ell \bcdot  \langle \bm B \rangle_\ell\rangle_\ell\right] + \left[ \langle \langle \bm B \rangle_\ell^2 \rangle_\ell - \langle \bm B \rangle_\ell^2 \right].
    \label{eq:mu_2B}
\end{align}
Substituting this expression into Eq.~\eqref{eq:filt_mag_energy_dens} and integrating over the volume leads to a substantial simplification, whereby
\begin{align}
    \mathcal{E}_{B,\mathrm{turb}} = \int_V \frac{\delta \bm B_\ell^2}{8\pi} \ud^3 x + \int_V \frac{2 \delta \bm B_\ell \bcdot \langle \bm B \rangle_\ell}{8\pi} \ud^3 x, \label{eq:magnetic_energy_explicit}
\end{align}
where we used the fact that the last square bracket of Eq.~\eqref{eq:mu_2B} vanishes when integrated over $V$ if the domain is periodic, or if the smoothing kernel $\mathcal{W}_\ell (\bm x', \bm x)$ is symmetric in its arguments and conserves the volume in $V$ (see Appendix~\ref{app:commutativity}).

Comparing Eq.~\eqref{eq:magnetic_energy_explicit} with the analogous expression for  volume averaging from Eq.~\eqref{eq:volume_averaging_magnetic_energy},
\begin{align}
        \left.E_B\right|_{\mathrm{turb}} &=  \int_V \frac{\delta \bm B^2}{8\pi} \ud^3 x,
\end{align}
shows that in addition to the contribution of the na\"{i}ve estimate for the turbulent energy, $\delta \bm B_\ell^2/8\pi$, in the filtering approach $\mathcal{E}_{B,\mathrm{turb}}$ contains a cross term proportional to the projection of the fluctuating magnetic field onto the bulk component, resulting in a higher turbulent energy if the two are preferentially aligned.

\subsubsection{Filtered turbulent energies and the power spectrum}\label{sec:turb-en-power-spectr}

In Eq.~\eqref{eq:mag_energy_decomposition}--\eqref{eq:filt_mag_energy_dens} we identified $\varepsilon_{B,\mathrm{turb}}$ and $\varepsilon_\mathrm{k,turb}$ with the turbulent energies of the magnetic and velocity field, respectively. In this section we make explicit their relationship with the power spectrum, which is often used as a tool to characterize turbulence.
For illustration, we focus again on the turbulent magnetic energy, $\mathcal{E}_{B,\mathrm{turb}}$, which involves a quadratic product, but we discuss the case of the turbulent kinetic energy toward the end of the section.
We consider here the case of a periodic domain, $V$. For a discussion on nonperiodic volumes (or sub-volumes of a larger periodic domain) see Appendix~\ref{app:turbulent-energy-subvolumes}.

We expand the magnetic field in Fourier series as 
\begin{align}
        \bm B (\bm x) = \sum_{\bm k}  \hat{ \bm B}_{\bm k} e^{i \bm k \bcdot \bm x }, \label{eq:synthetic_magnetic_field}
\end{align}
where the divergence-free condition for the magnetic field implies non-radial Fourier coefficients, $\hat{ \bm B}_{\bm k}$, i.e., $\bm k \bcdot \hat{ \bm B}_{\bm k} = 0, \forall \bm k$. Using the definition in Eq.~\eqref{eq:filt_mag_energy_dens}--\eqref{eq:filt_mag_energy} of the turbulent magnetic energy, we obtain (see Appendix~\ref{app:turb_periodic_volumes} for the derivation):
\begin{align}\label{eq:turbulent_energy_integrated}
        \mathcal{E}_{B,\mathrm{turb}} = \sum_{\bm k}  \frac{|\hat{\bm B}_{\bm k}|^2}{8 \pi}  \left( 1 -  |\tilde{\mathcal{W}}_{\ell, \bm k}|^2 \right) V,
\end{align}
where $\tilde{\mathcal{W}}_{\ell, \bm k}$ is the Fourier transform of the smoothing kernel multiplied by a factor of $V$, so that $\tilde{\mathcal{W}}_{\ell, \bm 0} = 1$.
For the ``ideal'' low-pass kernel (a step function that is 1 for $k<1/\ell$ and 0 for $k>1/\ell$), the turbulent magnetic energy is \begin{align}
\label{eq:Eps_B,turb}
    \mathcal{E}_{B,\mathrm{turb}} = \sum_{ k \ell > 1} \frac{|\hat{\bm B}_{\bm k}|^2}{8 \pi} V ,
\end{align}
which is the amount of energy present on scales smaller than the filter size, $\ell$. Conversely, the energy of the smooth magnetic field is the amount of energy present on scales greater than $\ell$ ($k \ell < 1$). The ideal low-pass kernel is a useful theoretical tool but is characterized by infinite support in real space, and thus it is not practical. For a generic low-pass kernel such as the spherical top-hat or a Gaussian filter, there will be in general deviations from the ideal step function: these deviations will introduce contamination from modes with $k \ell \approx 1$ and Eq.~\eqref{eq:Eps_B,turb} is only valid in an approximate sense.
Nevertheless, these results clarify the reason why we identify $\mathcal{E}_{B,\mathrm{turb}}$ with the ``turbulent'' magnetic energy, i.e., this is the energy contained in the scales smaller than the filter size.

\subsubsection{Turbulent kinetic energy in the subsonic limit}

For subsonic flows ($\mathcal{M} \ll 1$), it can be shown that when $\ell \ll H$, where $H$ is the typical scale-height of the background stratification, the ratio of the turbulent density fluctuations to the smooth density field is on the order of the Mach number $\delta \rho_\ell / \langle \rho \rangle_\ell \sim \mathcal{O} (\mathcal{M}) \ll 1$ (Appendix~\ref{app:subsonic-regime}). As a result, the transport term, $\varepsilon_\mathrm{k,cross}$, and the third-order moment in $\varepsilon_\mathrm{k,turb}$ are asymptotically small and the turbulent kinetic energy density reduces to  $\sum_i\langle \rho \rangle_\ell \mu_2(u_i, u_i)/2$ (see Appendix~\ref{app:subsonic-regime} for the derivation), which has the same form as the turbulent magnetic energy density in Eq.~\eqref{eq:filt_mag_energy_dens}. 

Under these assumptions, the turbulent kinetic energy in a volume $V$ is to leading order
\begin{align}\label{eq:turb_kin_power_spectr}
        \mathcal{E}_{\mathrm{k,turb}}  \approx \frac{1}{2} \rho_0 \sum_{ k \ell > 1} |\hat{\bm u}_{\bm k}|^2 V,
\end{align}
where $\rho_0$ is the mean density in the volume (assumed to be small compared to $H$), which is formally the same as Eq.~\eqref{eq:Eps_B,turb}. The same expression with the sum running over $ k \ell < 1$ gives instead the bulk kinetic energy, $\mathcal{E}_{\mathrm{k,bulk}}$. 

In the compressible or transonic regime, the correlations between density and velocity field cannot in general be neglected. As a result, the full expression of the turbulent kinetic energy is significantly more involved, as it now includes triadic interactions between the density and the velocity field, which make the turbulent (or the bulk) kinetic energy not straightforwardly identifiable with the power spectrum below (or above) $\ell$. Contributions from wave numbers that are all on large scales ($ k \ell \ll 1$) are still filtered out, but interactions between mixed wave numbers are in general unavoidable. This is not a problem of fixed-scale filters per se, but of compressible turbulence in general, where conventions exist to compute the power spectrum of turbulence (a quadratic quantity) by weighting the velocity with the density, its square root, or some other power thereof \citep[e.g.,][]{Kida1990,Grete2017}.

\subsubsection{The impact of shocks on the filtered energies}\label{eq:shock_filtered_energies}

When velocities exceed the sound speed, shocks and discontinuities can form in the flow, i.e., surfaces where one or more fluid variables are discontinuous. From the point of view of smoothing filters, these are thin surfaces (in numerical simulations their thickness is on the order of the cell size, while in reality it may shrink to the scale of several ion skin depths at collisionless shocks in galaxy clusters), so in general shocks and discontinuities would also contribute to the turbulent filtered energies. More specifically, from the definitions in Eq.~\eqref{eq:kin_en_transp}--\eqref{eq:kin_en_turb}, contact discontinuities where only the density is discontinuous (but with a uniform velocity field) do not contribute to the turbulent kinetic energy density, since the statistical moments vanish if any one of their arguments is spatially constant, but tangential discontinuities (with a nonzero velocity shear) and shocks do. 

From a physical perspective, it may appear odd that these spatially extended structures can directly contribute to the turbulent energy budget, while one might rather see them as potential sites for the excitation of turbulence (e.g., a shear layer going Kelvin--Helmholtz unstable). We believe that this is more of a conceptual wrinkle, instead of a weakness of the filtering approach, since supersonic turbulence is after all characterized by shocks, which are local in physical space but nonlocal in wave number space, and thus affect the entire power spectrum. Removing the contribution of shocks from the turbulent energies is thus equivalent to asking to remove the contribution of shocks from power spectra of supersonic turbulence.

Having said that, it might still be useful in certain cases to exclude shocks when filtering \citep[as in, e.g.,][]{Vazza2012}, for instance when one is interested in estimating the amount of subsonic turbulence in the bulk of the ICM, in which case filtering methods in real space to compute turbulent energies offer a decisive advantage over Fourier methods to compute power spectra. Our tests show that shocks typically contribute to the mean turbulent energies and turbulent pressure fractions within $R_{200,\mathrm{c}}$ only at the level of 2\% for the mean turbulent energy, and 7\% for the turbulent pressure (see Appendix~\ref{app:shocks_contribution}); we thus decided in the present work not to remove shocks from the turbulent energies. In the  proximity of shocks, however, their contribution might still be substantial.

\section{Turbulent energies during a cluster merger}\label{sec:turbulent_energies_application}

\subsection{PICO-Clusters simulations}\label{sec:numerics}

We now apply the methods outlined in the previous sections to a simulation of a massive galaxy cluster ($M_{200,\mathrm{c}} \simeq 2.72 \times 10^{15}~\mathrm{M}_{\odot}$ at $z=0$). This cluster, \texttt{Halo3}, is part of the PICO-Clusters suite \citep{Berlok2026}, which re-simulated 24 massive clusters (${\gtrsim} 10^{15}~\mathrm{M}_{\odot}$) from a $1.5 ~ \si{Gpc}$-wide cosmological volume using the zoom-in technique and the moving-mesh code \textsc{Arepo} \citep{Springel2010,Pakmor2016,Weinberger2020a}, using the updated version \textsc{Arepo-2} that evolved from the MilleniumTNG simulation \citep{Pakmor2023}. The \textsc{Arepo} code evolves the comoving form of the ideal MHD equations \citep{Pakmor2013} using Powell cleaning for divergence control \citep{Pakmor2026}, and PICO-Clusters adopts the galaxy formation model used in the \textsc{IllustrisTNG} simulations \citep{Weinberger2017,Pillepich2018}, which incorporates radiative gas cooling, star formation and associated feedback, and AGN feedback. The simulations shown in this paper have been run with a mass resolution for dark-matter particles of $5.9 \times 10^{7}~\mathrm{M}_{\odot}$ and with a target gas mass of $7 \times 10^{6}~\mathrm{M}_{\odot}$, which is comparable to that of the TNG300 cosmological volume simulation \citep{Springel2018}. The typical spatial resolution of the gas cells (assuming a spherical shape for simplicity) moderately increases with radius within $R_{200,\mathrm{c}}$: for example, at $z=0$ the median spatial resolution in a radial bin goes from about $4$ kpc at $R \simeq 100$ kpc, to approximately $11$ kpc at $R \simeq 1$ Mpc, further rising to ${\simeq} 27$ kpc at $R_{200,\mathrm{c}}$. Additional details on the PICO-Clusters suite can be found in \citet{Berlok2026}.

\begin{figure*}
        \centering
        \includegraphics[trim={0 2cm 0 0},clip,width=0.9\textwidth]{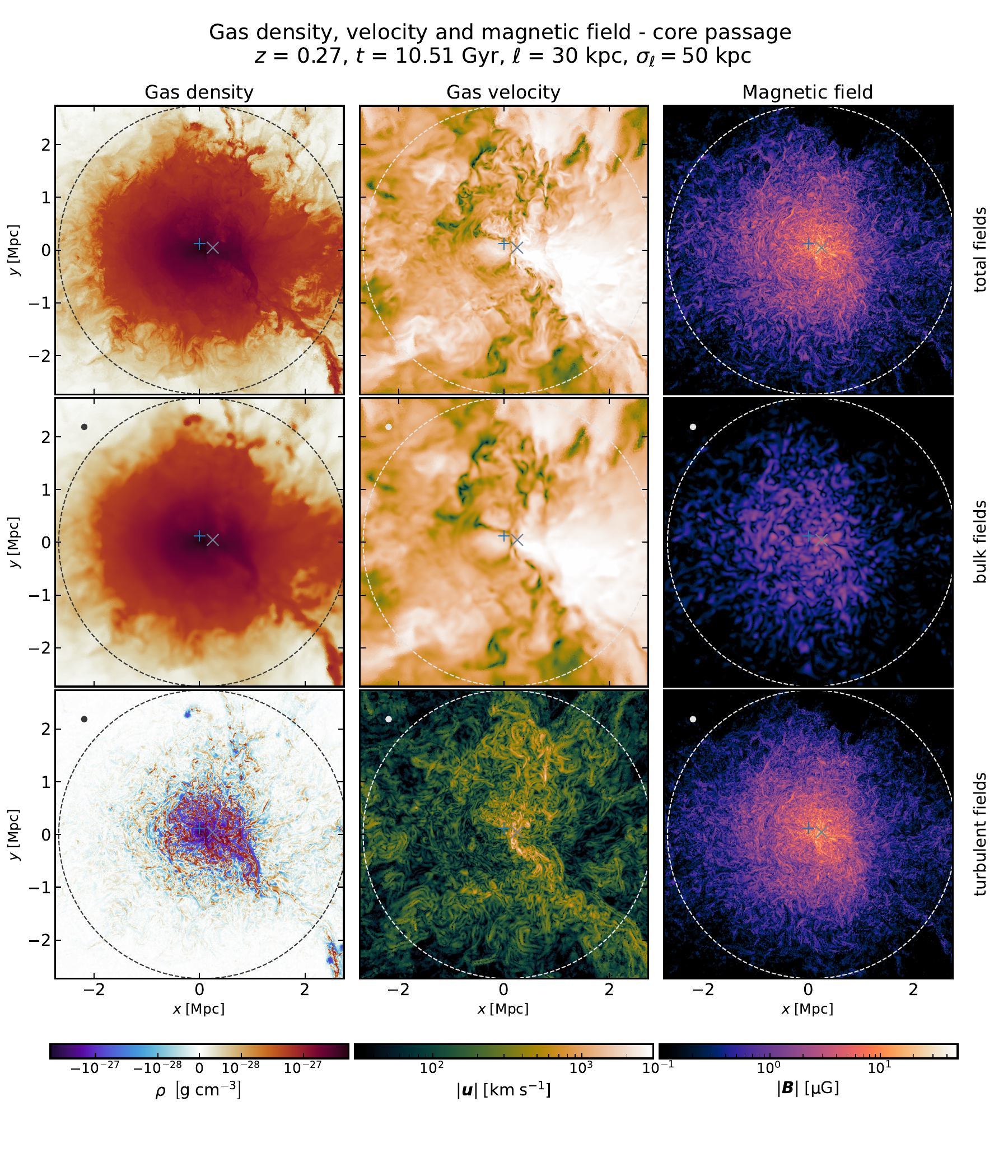}
        \caption{Decomposition of the density, velocity (in the rest system of the main cluster), and magnetic field of \texttt{Halo3} during the merger ($z=0.27$) for a filter scale of $30 ~ \si{kpc}$ (represented by a beam of radius $\sigma_\ell = 50 ~ \si{kpc}$ in the top left corner). The layout of the figure is the same as in Fig.~\ref{fig:before_all_fields}. The main cluster is highly disturbed by the infall of the less massive companion (along the horizontal axis, from the right side). The blue and gray markers indicate the positions of the centers of the main halo and its companion, respectively. The infall is accompanied by a large-scale inflow, without however significant small-scale fluctuations. The magnetic field remains more homogeneously distributed throughout the volume.
    }
        \label{fig:during_all_fields}
\end{figure*}

Of the 24 halos simulated in PICO-Clusters, we choose \texttt{Halo3} because it undergoes a major merger at low-redshift with a second massive halo ($M_{200,\mathrm{c}} \simeq 5.38 \times 10^{14}~\mathrm{M}_{\odot}$, mass ratio of $2.4$ before merger), which generates significant bulk motions and sloshing in the center. The snapshots available for \texttt{Halo3} have a high (but variable) cadence in time, with $\Delta t_{\mathrm{snap}}$ varying between (30--100) Myr for $t>6$ Gyr.
For these reasons, it represents a good candidate to test our decomposition of bulk and turbulent energies. Lastly, its time evolution was analyzed in detail in \citet{Tevlin2025}, allowing us to compare our results directly.

\subsection{Filtered fields during merger}

Similarly to Fig.~\ref{fig:before_all_fields}, we show now the total, bulk, and turbulent fields during the merger for the intermediate filter length $\ell = 30 ~ \si{kpc}$ (see Fig.~\ref{fig:during_all_fields}), 
when the core of the less massive cluster has rammed into the core of \texttt{Halo3}. The infalling less massive cluster is especially visible in the density and velocity field, where one can observe strong asymmetric features and the large radial inflow, respectively. The overall level of turbulent density fluctuations increases and high levels of turbulent velocities are excited along the contact discontinuity. On the contrary, except for a small region in the cluster center showing strong amplification (likely due to adiabatic compression), the turbulent magnetic field appears to be more homogeneously distributed in the entire volume.

\begin{figure*}
        \begin{subfigure}{0.5\textwidth}
                \centering
                \includegraphics[width=1.0\linewidth]{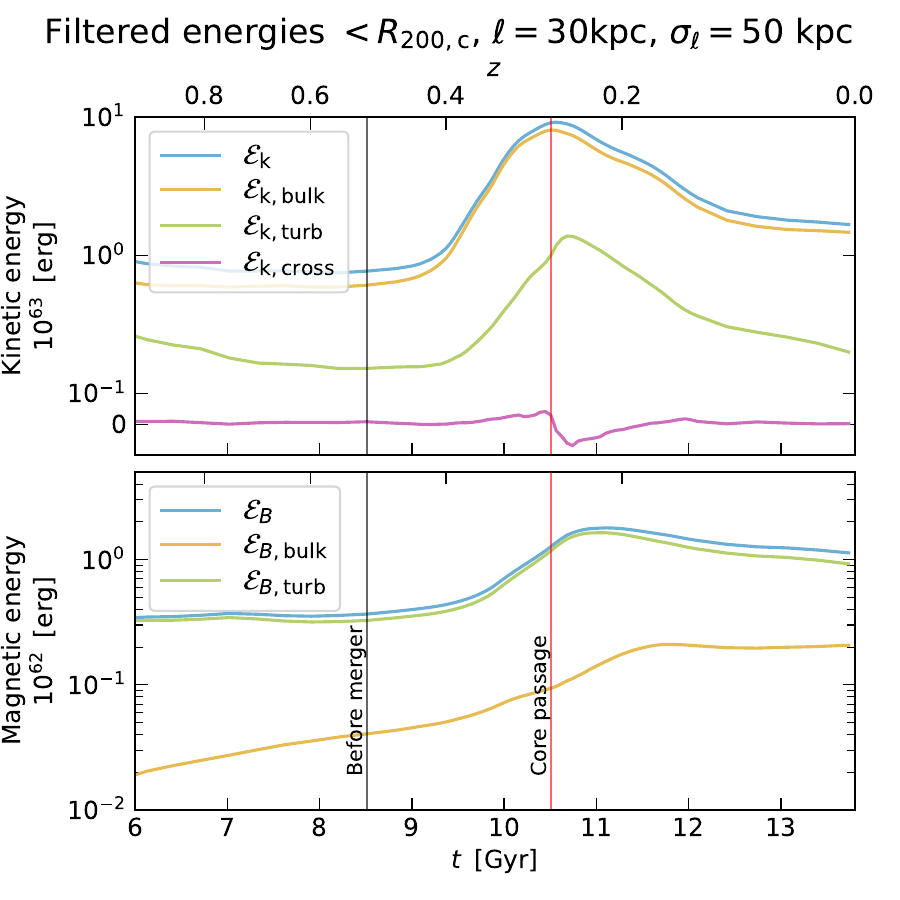}
                \caption{}
                \label{fig:kinetic_magnetic_energy_all_volume}
        \end{subfigure}
        \hfill
        \begin{subfigure}{0.5\textwidth}
                \centering
                \includegraphics[width=1.0\linewidth]{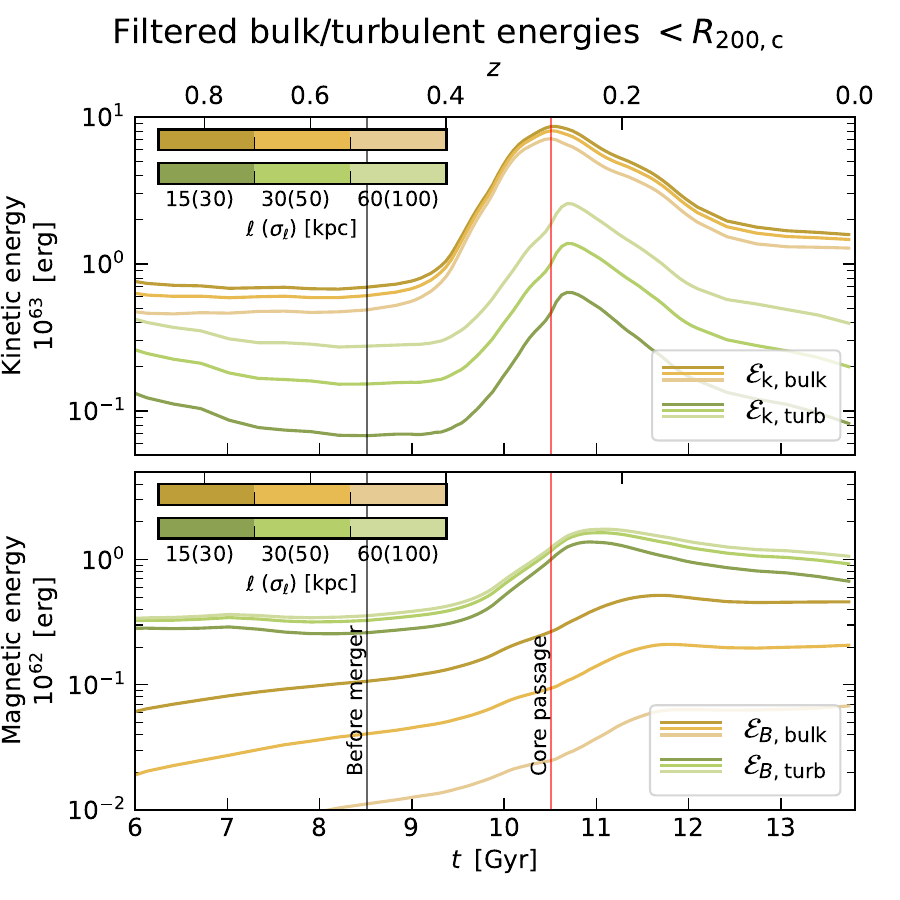}
                \caption{}
                \label{fig:kinetic_magnetic_energy_filter_lengths}
        \end{subfigure}
        \caption{Time evolution of the filtered kinetic and magnetic energies during the major merger. \textit{Left column}: Kinetic (top) and magnetic energy (bottom) and their decomposition according to Eqs.~\eqref{eq:filt_mag_energy_volume} and \eqref{eq:filt_kin_energy_volume} for a fiducial filter scale of $30 ~ \si{kpc}$ corresponding to a filter beam size of $\simeq$50 kpc. The vertical lines correspond to the time of the snapshots shown in Figs.~\ref{fig:before_all_fields} and \ref{fig:during_all_fields}. \textit{Right column}: Bulk and turbulent energies for different filter scales. No smoothing in time has been applied to produce this figure. For the top left panel we use a linear scale for the $y$ axis between $(-0.1,0.1)$ and logarithmic otherwise. For the scales considered, the turbulent kinetic energy is always subdominant compared to the bulk throughout the merger, while the reverse is true for the turbulent magnetic energy. }
        \label{fig:kinetic_magnetic_energy_filter}
\end{figure*}

\subsection{Turbulent and bulk energies}\label{sec:turb-bulk-energies}

We now apply the definitions of the bulk and turbulent kinetic and magnetic energies of Sect.~\ref{sec:turbulent_energies}, and follow their evolution in time throughout the merger starting from $t=6 ~ \si{Gyr}$ until the $z=0$. We use a Gaussian filter with  constant-in-physical-space filter length throughout the time series. 
Two other possible options for the filter length were considered: (1) to fix $\ell$ to be a constant fraction of $R_{200,\mathrm{c}}(z)$ for different redshifts; and (2) to choose a filter length that scales proportional to the scale-factor $a(t)$, i.e., constant in comoving units. We discarded option (1) because $R_{200,\mathrm{c}}$ is a meaningful quantity only for thermodynamically relaxed clusters, which is very much not the case during major mergers as in our simulations. This would have led to vastly different filter lengths across time, and dubious physical meaning of the resulting filtered energies. We also discarded option (2) because clusters are massive objects that are decoupled from the background cosmological expansion and they do not just expand isotropically with $a(t)$. This is definitely the case for $z<1$ (the time-interval considered in the time evolution plots), for which the inverse of the Hubble parameter $H(t) = \dot{a}/a$ (the typical timescale associated with the Universe expansion) is greater than $8$ Gyr, which is much longer than any dynamical timescale for the turbulence in galaxy clusters. 

While it is conceivable that the outermost scale of turbulence during mergers (which possibly reaches the cluster size) might increase with $a(t)$ because at low redshifts structures are bigger in physical units, the scales of the turbulence are set by the physical processes that take place in the ICM, which has largely decoupled from Hubble expansion. In particular, the viscous scale will be determined by microphysical properties, \citep[e.g., particle scattering, or plasma kinetic effects][]{Schekochihin2006a}, motivating our choice of a filter scale that is constant in physical space.

To better understand how the energies are partitioned between different scales, we selected three different lengths, i.e.,  
$\ell = \left\lbrace 15,~ 30,~ 60 \right\rbrace ~\si{kpc}$, which amount to $0.5\%$, $1\%$, and $2\%$ of $R_{200,\mathrm{c}}$ at $z=0$, respectively.
In terms of the effective smoothing beam size, $\sigma_\ell$, they approximately\footnote{The values quoted for the beam size $\sigma_\ell$ are rounded to the nearest tens.} correspond to $30$, $50$, and $100 ~ \si{kpc}$. 
The filtered energy densities are then integrated within $R_{200,\mathrm{c}}$ of \texttt{Halo3} to obtain the filtered energies. 

We show the results of the decomposition in Fig.~\ref{fig:kinetic_magnetic_energy_all_volume} for a fiducial filter scale of $\ell = 30 ~ \si{kpc}$ ($\sigma_\ell \simeq 50 ~ \si{kpc}$).
Several features can be observed. 
The kinetic and magnetic energies differ in that most of the kinetic energy resides in the bulk flow rather than the turbulent component for the interval shown in Fig.~\ref{fig:kinetic_magnetic_energy_all_volume}, while the reverse is true for the magnetic energy.  
During the merger, the total, bulk and turbulent kinetic energies increase substantially by roughly a factor of ten between the times of the snapshots in Fig.~\ref{fig:before_all_fields} and Fig.~\ref{fig:during_all_fields}, albeit at different rates. The total and the bulk kinetic energies (which also include the overall acceleration of the less massive cluster toward \texttt{Halo3}) begin to increase and reach their peak earlier than the turbulent energy, around the time when the cores collide. 
The transport term is subdominant throughout and reverses sign around the time of the core passage.
The magnetic energy is instead amplified by a factor of five, similar for both the bulk and turbulent components. Interestingly, the turbulent magnetic energy reaches its peak slightly delayed with respect to the turbulent kinetic energy ($\approx$$300 ~ \si{Myr}$), and decays more slowly afterward.

We repeated the same calculations for a lower and higher value of the filter scale, corresponding to a beam size of about $30 ~ \si{kpc}$ and $100 ~ \si{kpc}$. 
The bulk and turbulent components are shown in Fig.~\ref{fig:kinetic_magnetic_energy_filter_lengths}. As we increase the filter scale, the turbulent energies (both magnetic and kinetic) increase, while the bulk energies decrease. This is a consequence of the fact that, as we increase $\ell$, more and more modes are classified by the smoothing filter as ``turbulence,'' and consequently fewer of them as ``bulk.'' This trend is strictly monotonic for the magnetic energy, while this is not guaranteed in principle for the kinetic energy due to the presence of the cross term: while in our case it is subdominant, in strongly asymmetric configurations it could become highly negative \citep[this is the case of outflows from galactic disks;][]{Hollins2022}. Whereas the overall behavior of the bulk and turbulent energies for the different filter scale reflects that of Fig.~\ref{fig:kinetic_magnetic_energy_all_volume}, the change in the energy levels with respect to $\ell$ provides us with more information on how energy is partitioned across different scales. For instance, in the case of the kinetic energy doubling or halving the filter scale boosts the turbulent energy by roughly the same factor, whereas for the magnetic energy there is relatively little difference between the turbulent levels for the different filter scales. This suggests that, for this range of scales, the kinetic energy increases with $\ell$ following a power law, while the magnetic field is predominantly tangled on scales below $30 ~ \si{kpc}$. After the merger, both kinetic and magnetic turbulent energies decay faster the smaller the filter scale, which is to be expected for viscous and resistive numerical dissipation as we approach the grid scale.

\begin{figure}
        \centering
        \includegraphics[width=1.0\linewidth]{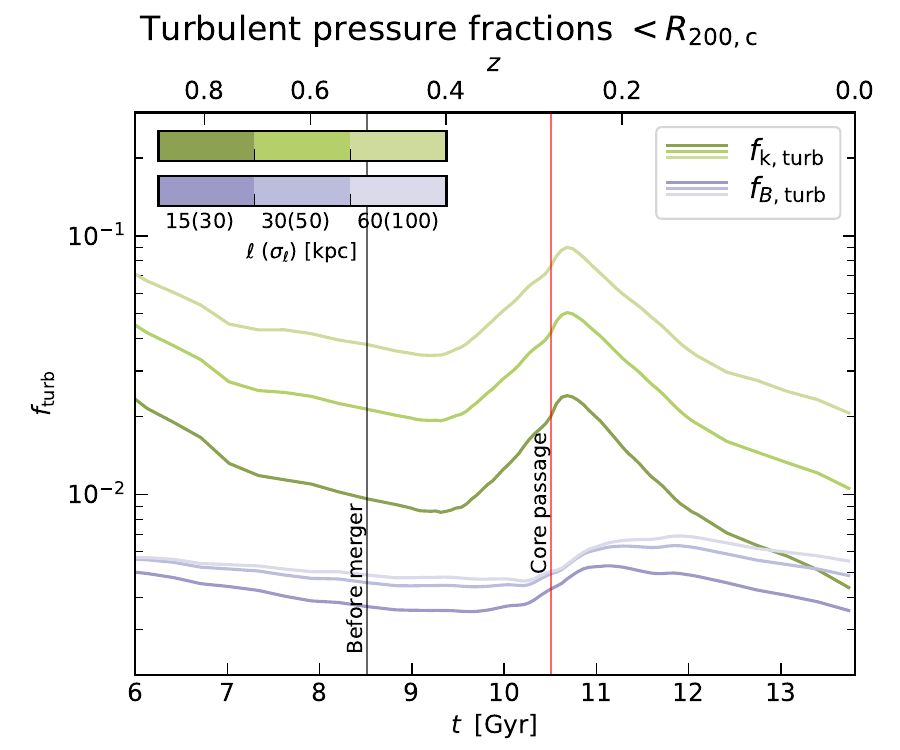}
        \caption{Time evolution of the kinetic and magnetic turbulent pressure fraction during a major merger for different filter lengths. The turbulent kinetic pressure decreases to pre-merger levels within $(1-1.5)$~Gyr from the core passage.}      \label{fig:nonthermal_ratio_filter_lengths}
\end{figure}

\subsection{Turbulent pressure fractions}

To compare the levels of turbulence detected with the filtering approach with other methods and with observations, we computed the turbulent kinetic, $f_{\mathrm{k,turb}}$, and magnetic pressure fraction, $f_{B,\mathrm{turb}}$, within $R_{200,\mathrm{c}}$ and for the different filter scales. In the filtering approach, we define them as
\begin{align}\label{eq:turbulent_pressure_fractions}
    &f_{\mathrm{k,turb}} \equiv \frac{\mathcal{E}_{\mathrm{k,turb}}}{\mathcal{E}_{\mathrm{k,turb}} + E_{\mathrm{th}}}, &f_{B,\mathrm{turb}} \equiv \frac{\mathcal{E}_{B,\mathrm{turb}}}{\mathcal{E}_{\mathrm{k,turb}} + \mathcal{E}_{B,\mathrm{turb}} + E_{\mathrm{th}}},
\end{align}
where the dependence on the filter scale, $\ell$, is implicit in the definition of the turbulent energies, $\mathcal{E}_{\mathrm{k,turb}}$ and $\mathcal{E}_{B,\mathrm{turb}}$. Note that the different normalization in the denominator was chosen to allow for a direct comparison of $f_{\mathrm{k,turb}}$ with the nonthermal pressure fraction, $f_{\mathrm{nth}}$, which is commonly reported in the context of XRISM and observational studies, and where contributions from magnetic fields and other nonthermal source of pressure are usually neglected\footnote{For typical ICM conditions the contributions of magnetic field \citep{Carilli2002,Tevlin2025} and cosmic ray pressure are each less than ${\approx}1\%$ \citep{Ackermann2010,Ackermann2014,Ruszkowski2023}.} \citep{Eckert2019b}.
It is useful to recast the kinetic pressure fraction in terms of the mean turbulent pressure, $p_{\mathrm{turb}}$, below a scale, $\ell$, and mean thermal pressure, $p_{\mathrm{th}}$, in a volume, $V$, where
\begin{align}\label{eq:turb_press_def}
        p_{\mathrm{turb}} \equiv \frac{2}{3} \frac{\mathcal{E}_{\mathrm{k,turb}}}{V}, 
\end{align}
and the thermal pressure is related to the thermal energy by the usual relation $p_{\mathrm{th}} =  (\gamma - 1) E_{\mathrm{th}}/V$. 
Our definition\footnote{A similar expression was already introduced in the context of smoothing filters \citep[][]{Wang2024}, although ours is more general since it does not place assumptions on subsonic or isotropic turbulence.} of the turbulent and nonthermal pressure (Eq.~\ref{eq:turb_press_def}) is consistent with, but not identical to that routinely employed in observational studies using the line-of-sight velocity dispersion $\sigma_{\mathrm{los}}$, \citep[$p_{\mathrm{turb}} \simeq \rho_0 \sigma_{\mathrm{los}}^2$][]{Eckert2019b}, where a homogeneous medium with gas density $\rho_0$ and isotropic turbulence is assumed.
Equation~\eqref{eq:turb_press_def}, instead, is constructed through the filtered turbulent kinetic energy, $\mathcal{E}_{\mathrm{k,turb}}$ (Eqs.~\ref{eq:kin_en_turb} and \ref{eq:filt_kin_energy_volume}), which depends on the filter scale, $\ell$. 
By taking the same limit of subsonic, homogeneous and isotropic turbulence, we show in Appendix~\ref{app:alternative-turb-press} that the turbulent pressure in Eq.~\eqref{eq:turb_press_def} reduces to $p_{\mathrm{turb}} \simeq \rho_0 \sigma_{\mathrm{1D},u}^2$, where $\sigma_{\mathrm{1D},u}^2 = \sigma_{\mathrm{3D},u}^2 / 3$ is the isotropic one-dimensional velocity dispersion (here it represents an rms value over the volume $V$). Replacing $\sigma_{\mathrm{1D},u}$ with $\sigma_{\mathrm{los}}$, we can see that the two formulations are indeed analogous. 

Using Eq.~\eqref{eq:turb_press_def} and assuming an adiabatic index of $\gamma = 5/3$, the turbulent kinetic pressure fraction can be expressed as
\begin{align}\label{eq:non-thermal-ratio}
    f_{\mathrm{k,turb}} = \frac{ p_{\mathrm{turb}}}{ p_{\mathrm{turb}} + p_{\mathrm{th}}},
\end{align}
which can be further simplified if we again assume subsonic, homogeneous turbulence (see Appendix~\ref{app:alternative-turb-press}): we first introduce the corresponding three-dimensional, $\mathcal{M}_{\mathrm{3D}}$, and one-dimensional turbulent Mach number, $\mathcal{M}_{\mathrm{1D}}$,
\begin{align}\label{eq:turb-mach-numbers}
    &\mathcal{M}_{\mathrm{3D}}^2 \equiv \frac{\sigma_{\mathrm{3D},u}^2}{c_\mathrm{s}^2}, & \mathrm{and} & & \mathcal{M}_{\mathrm{1D}}^2 \equiv  \frac{\sigma_{\mathrm{1D},u}^2}{c_\mathrm{s}^2},
\end{align}
and after substituting in Eq.~\eqref{eq:turb_press_def}, the turbulent kinetic pressure fraction becomes
\begin{align}
        f_{\mathrm{k,turb}} &\simeq \frac{\sigma_{\mathrm{3D},u}^2}{\sigma_{\mathrm{3D},u}^2 + \frac{3  k_{\mathrm{B}}T}{\mu m_\mathrm{u} }} = \frac{\mathcal{M}_{\mathrm{3D}}^2}{\mathcal{M}_{\mathrm{3D}}^2 + \frac{3}{\gamma}} \label{eq:f_turb_1}\\
    &= \frac{\mathcal{M}_{\mathrm{1D}}^2}{\mathcal{M}_{\mathrm{1D}}^2 + \frac{1}{\gamma}} \label{eq:f_turb_2} \begin{tabular}{p{12em}}
                (for isotropic turbulence), 
        \end{tabular}
\end{align}
where we used that $c_\mathrm{s}^2 = \gamma k_\mathrm{B} T/ (\mu m_\mathrm{u})$, with $T$ the mean temperature, $k_\mathrm{B}$ is the Boltzmann constant, $\mu$ the mean molecular weight, and $m_\mathrm{u}$ the atomic mass unit. Equations~\eqref{eq:f_turb_1} and \eqref{eq:f_turb_2}
are formally analogous to the nonthermal pressure fractions routinely used in observationally studies of galaxy clusters \citep[e.g., Eq.~10 of][]{Eckert2019b}, or by the XRISM Collaboration \citep[Eq.~1 of both][]{XRISMCollaboration2025c,XRISMCollaboration2025a}. 

After introducing the necessary quantities, we show in Fig.~\ref{fig:nonthermal_ratio_filter_lengths} the evolution of the turbulent kinetic and magnetic pressure fraction during the major merger for the three filter lengths considered. We notice a similar trend for the turbulent kinetic pressure with all filter lengths, with a sharp exponential-like increase during the merger that peaks near the core passage; the subsequent relaxation takes place in two phases: a first rapid decrease, followed by a slower decline. A similar dual trend was also observed in earlier work by \citet{Shi2018a}, where it was attributed to faster decay of turbulence in the inner regions of the cluster. For the intermediate filter length $\ell = 30 ~ \si{kpc}$ ($\sigma_{\ell} \simeq 50 ~ \si{kpc}$), 
we measure a maximum kinetic turbulent fraction of $f_{\mathrm{k,turb}} \simeq 5 \%$ near the core passage at $t\simeq10.7$~Gyr ($z\simeq0.26$), followed by a rapid decay to $2 \% $ by $t=12$~Gyr, and to $1 \%$  by the present time. For larger filter scales these numbers tend to increase, but even for our largest scale probed, $\ell \simeq 60$~kpc ($\sigma_{\ell} \simeq 100 ~ \si{kpc}$),
where $f_{\mathrm{k,turb}} \simeq 9 \%$ during the core passage, the kinetic turbulent fraction quickly decreases to $\simeq 3.5 \% $ by $t=12$~Gyr, and to only $2 \%$  by the present time. We stress that these estimates refer to the entire cluster within $R_{200,\mathrm{c}}$, and that the values may be different if we restrict ourselves to only the central region due to AGN feedback (we defer such an analysis with a more detailed comparison to XRISM data to future studies).

Similar to the results shown in Fig.~\ref{fig:kinetic_magnetic_energy_filter}, the turbulent magnetic pressure fraction is always lower than the corresponding turbulent kinetic pressure, although after the merger it decays much more slowly (e.g., for $\ell = \left\lbrace 30, 60 \right\rbrace$~kpc, corresponding to $\sigma_{\ell} \simeq \left\lbrace 50, 100\right\rbrace ~ \si{kpc}$, $f_{B,\mathrm{turb}}$ is still higher than pre-merger levels after $\gtrsim$3~Gyr from the core passage). 
Kinetic turbulent fractions of $\sim$$1-2\%$ after relaxation are roughly comparable to those found in \citet{Valdarnini2019}, but lower than those in \citet{Groth2025}, although multi-scale iterative filters were employed in both studies. A more direct comparison may be drawn  with the massive clusters in \citet{Vazza2011}, who also used fixed-scale filters of various sizes (but with a different kernel), and who found turbulent to thermal fractions below $300$~kpc of 20\% to 40\% within $R_{200,\mathrm{c}}$ for relaxed clusters and during major mergers, respectively, which are significantly larger than our estimates. This discrepancy could be due to the fact that for their kernel (triangular shape) a filter scale of $300$~kpc likely corresponds to a smoothing beam size much larger than our largest filtering scale considered 
($\ell = 60$ kpc, corresponding to $\sigma_\ell \simeq 100$ kpc).
By choosing a larger filtering scale, more of the energy is classified as ``turbulence,'' thus potentially explaining the difference. 

Our results suggest that, unless a cluster has undergone a major merger within the last $1-1.5$~Gyr, values of $2-3\%$ for the turbulent kinetic pressure fraction are not unusual, provided one carefully removes large-scale bulk motions. 
Our values of $f_{\mathrm{k,turb}}$ are in the same ballpark as the observational estimates from the XRISM cluster sample in the regions farther away from the core \citep{XRISMCollaboration2025f}, where we expect the effect of AGN feedback to be strongest. One caveat is that the estimates of the velocity dispersion with XRISM include all motions along the line-of-sight (LOS), which could in principle be a combination of both bulk motions (e.g., sloshing) and actual small-scale turbulence. Nevertheless, it is possible to associate these measurements of the velocity dispersion with an ``effective length,'' which measures the size of the region that contributes most to the X-ray flux \citep{XRISMCollaboration2025e}. This effective length provides an upper bound for the physical scales of the velocity fluctuations probed by XRISM, and may be compared to the smoothing length used in the filtering approach. We defer a more in-depth comparison with observations to a future publication.

\section{Pitfalls of multi-scale iterative filters}\label{sec:iterative_multiscale}

After showing a practical application of the filtering method at fixed scale, we discuss an alternative approach based on multi-scale iterative filters and highlight its potential pitfalls.

\subsection{Motivation}

Given the wide range of processes in the ICM that can excite turbulence, any one value of $\ell$ for the entire cluster will be inadequate to single out the different scales at which they operate. For this reason, only by repeating the operation of filtering and computing the filtered energies for different lengths (as we did in Sect.~\ref{sec:turbulent_energies_application}) can one gain a more complete picture.

To obviate this issue, a popular approach \citep[originally pioneered by][]{Vazza2012} is to employ an iterative scheme to locally determine the injection scale of the turbulence, and therefore distinguish the bulk from the turbulent component. In this iterative method, a kernel of increasingly larger size, $\ell_n$, is applied to a physical field (e.g., gas density), until a certain convergence criterion is met. Key to this approach is that the final $\ell_n$ of convergence (identified with the correlation scale of the turbulence) can differ from place to place.
For a scalar field $f (\bm x)$ with decomposition $f = \langle f \rangle_{\ell_{n}} + \delta f_{\ell_n}$ the convergence criterion chosen by \citet{Vazza2012} evaluated at the $n$-th iteration reads
\begin{align}
        \left| \frac{ \delta f_{\ell_n} - \delta f_{\ell_{n-1}} }{\delta f_{\ell_{n-1}}} \right| \leq \epsilon, \label{eq:vazza_convergence_criterion}
\end{align} 
where $\epsilon$ is a small tolerance parameter (typically $\epsilon = 0.02 - 0.1$). For a vector field the procedure is either repeated for the individual components \citep[as originally proposed by][although this could lead to different components being filtered on disparate scales]{Vazza2012}, or carried out using all components at once \citep{Vazza2017}. In this latter case, in Eq.~\eqref{eq:vazza_convergence_criterion} the absolute values are replaced by the norm of the difference between the two vectors $|\delta \bm u_{\ell_n} - \delta \bm u_{\ell_{n-1}}|$ (in the numerator), and $|\delta \bm u_{\ell_{n-1}}|$ (in the denominator). The iterative algorithm can optionally be complemented with a shock finder scheme in order to exclude from the averaging regions that are physically separated by a shock front, and that therefore do not communicate with each other \citep{Vazza2012,Vazza2017}. Other variants of the basic multi-scale filter have been adopted in the literature; for example, for use with smoothed-particle hydrodynamics codes \citep{Valdarnini2019}, with variable tolerance, $\epsilon$ \citep{Angelinelli2020a}, or with different convergence criteria \citep{Valles-Perez2021a,Valles-Perez2024a}. In particular, if the turbulence is Kolmogorov-like, the kinetic energy flux across scales is approximately conserved in the inertial range, and the convergence criterion in Eq.~\eqref{eq:vazza_convergence_criterion} may be modified to reflect the scaling of the velocity fluctuation, $|\delta \bm u_{\ell}| \propto \ell^{1/3}$ \citep[][]{Angelinelli2020a}. In this scenario, the iterative filter may be used to effectively constrain the kinetic energy flux \citep[][]{Vazza2017}.

Despite the attractiveness of a method that does not require the choice of an a priori filtering scale, an important limitation of multi-scale iterative filters is that they rely on the assumption that the typical scale of the bulk flow is much larger than the typical size of the turbulent eddies \citep[see, e.g., discussion in][]{Vazza2012}. When the scale separation between bulk flow and turbulence is limited, or when different physical processes drive motions on a continuous set of scales \citep[e.g., in the Perseus cluster where two independent drivers of turbulent motions on different scales were identified: the AGN feedback in the inner region, and mergers in the outer core,][]{XRISMCollaboration2025e}, such a distinction becomes blurry, and iterative filters may mix bulk motions and turbulence. As we show in the next section, we see evidence of this behavior even in idealized tests, which discourages us from applying multi-scale filters to simulations of galaxy clusters.

\begin{figure*}
        \centering
        \includegraphics[width=1.0\linewidth]{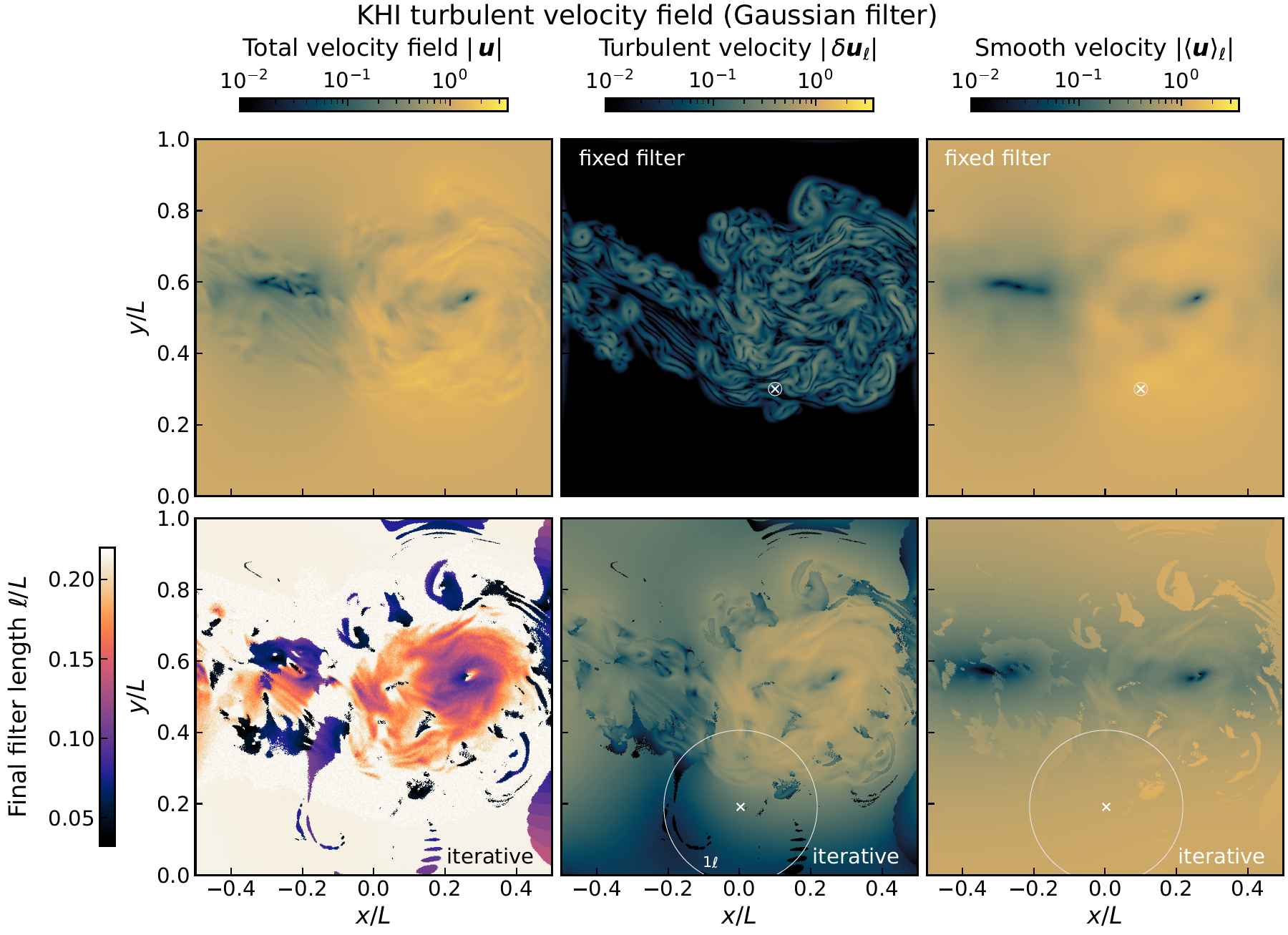}
        \caption{Comparison between iterative multi-scale filter (tolerance $\epsilon = 0.05$) and fixed-scale filter using a Gaussian kernel with a two-dimensional simulation of a Kelvin--Helmholtz instability run with \textsc{Arepo} (only the upper half of the domain is shown). In the second and third columns, the top (bottom) panel shows the result of fixed scale (iterative) filtering applied to the full 2D velocity field. Circles corresponding to $1\ell$ of the Gaussian kernel for two representative cells are drawn for comparison. The iterative filter fails to converge in regions where $|\delta \bm u_{\ell}|$ is small, where the smoothing length becomes very large, and also stops on very different lengths in nearby regions in the mixing layer. 
    }
    \label{fig:KHI_velocity_fixed_iterative_filter}
\end{figure*}

\subsection{Numerical tests}

We implemented the iterative multi-scale filter in our analysis code following \citet{Vazza2012} and performed a wide range of tests on idealized setups and simulated galaxy clusters. After extensive exploration, we concluded that multi-scale iterative filters suffer from certain issues that complicate the interpretation of the results and that render them unsuitable as a one-size-fits-all solution to determine the levels of turbulence in the ICM. These issues stem from the fact that, even in the ideal case when a significant scale separation exists between background flow and turbulence, the convergence of the iterative scheme is susceptible to noise and strongly depends on the initial conditions. In what follows, we discuss two representative test scenarios that highlight their pitfalls.

\subsubsection{Kelvin--Helmholtz instability}

In the first scenario we test the iterative filter against the Kelvin--Helmholtz instability, which is a ubiquitous hydrodynamical instability in the ICM caused, for example, by ram-pressure stripping in merging galaxies or near sloshing cores \citep{Walker2017b}. We set up a two-dimensional inviscid, hydrodynamic version of the simulation setup described in \citet{Lecoanet2016} and \citet{Berlok2019}, and with the same \textsc{Arepo} settings as in \citet{Berlok2020}. This is a subsonic simulation with a density contrast of two across the shear layer.

We focus on the velocity field after the instability is well into its nonlinear stage, and compare the iterative multi-scale filter to the fixed-scale filter where we choose a fiducial smoothing length, $\ell$. Both versions use a Gaussian kernel, which has better convergence properties than the spherical top-hat (see Appendix~\ref{app:filters_fourier_space} for a comparison). The results are shown in Fig.~\ref{fig:KHI_velocity_fixed_iterative_filter}. By looking at the top and bottom panels, we observe significant differences between the two filters: in the unperturbed region of the flow, the iterative filter did not correctly subtract the background shearing flow from the turbulent component, and thus returned a nonzero turbulent velocity field; conversely, near the mixing layer of the Kelvin--Helmholtz rolls, the smooth velocity field obtained with the iterative filter (bottom right panel) contains spurious small structures that are not present with the fixed-scale filter. 

We attribute this behavior to the fact that neighboring cells have iteratively converged to very different filter lengths, as can be seen from the lower left panel. In particular, the criterion in Eq.~\eqref{eq:vazza_convergence_criterion} is ill-conditioned when the fluctuating velocity field is close to zero, because $|\delta \bm u_{\ell}|$ appears in the denominator. For example, as we increase $\ell$, in the unperturbed part of the flow $\bm u$ is approximately constant and $\delta \bm u_{\ell} = \bm u -  \langle \bm u \rangle_\ell$ can be arbitrarily small (e.g., up to floating point precision errors due to the numerical integration). Consequently, even small changes between successive iterations may exceed the tolerance threshold $\epsilon$ and the iteration will not converge, leading to contamination of the local value of $\delta \bm u_{\ell}$ by regions very far away. This is visible, for example, in the upper and lower edges of Fig.~\ref{fig:KHI_velocity_fixed_iterative_filter}, where numerical artifacts (striations) are present in the turbulent velocity field. We expect a similar behavior to occur in simulations with enough resolution to resolve small-scale structures (e.g., in the tails of jellyfish galaxies, near discontinuities, and cold fronts). 

\begin{figure*}
    \centering
    \includegraphics[trim={0 0cm 0 0},clip,width=1.0\textwidth]{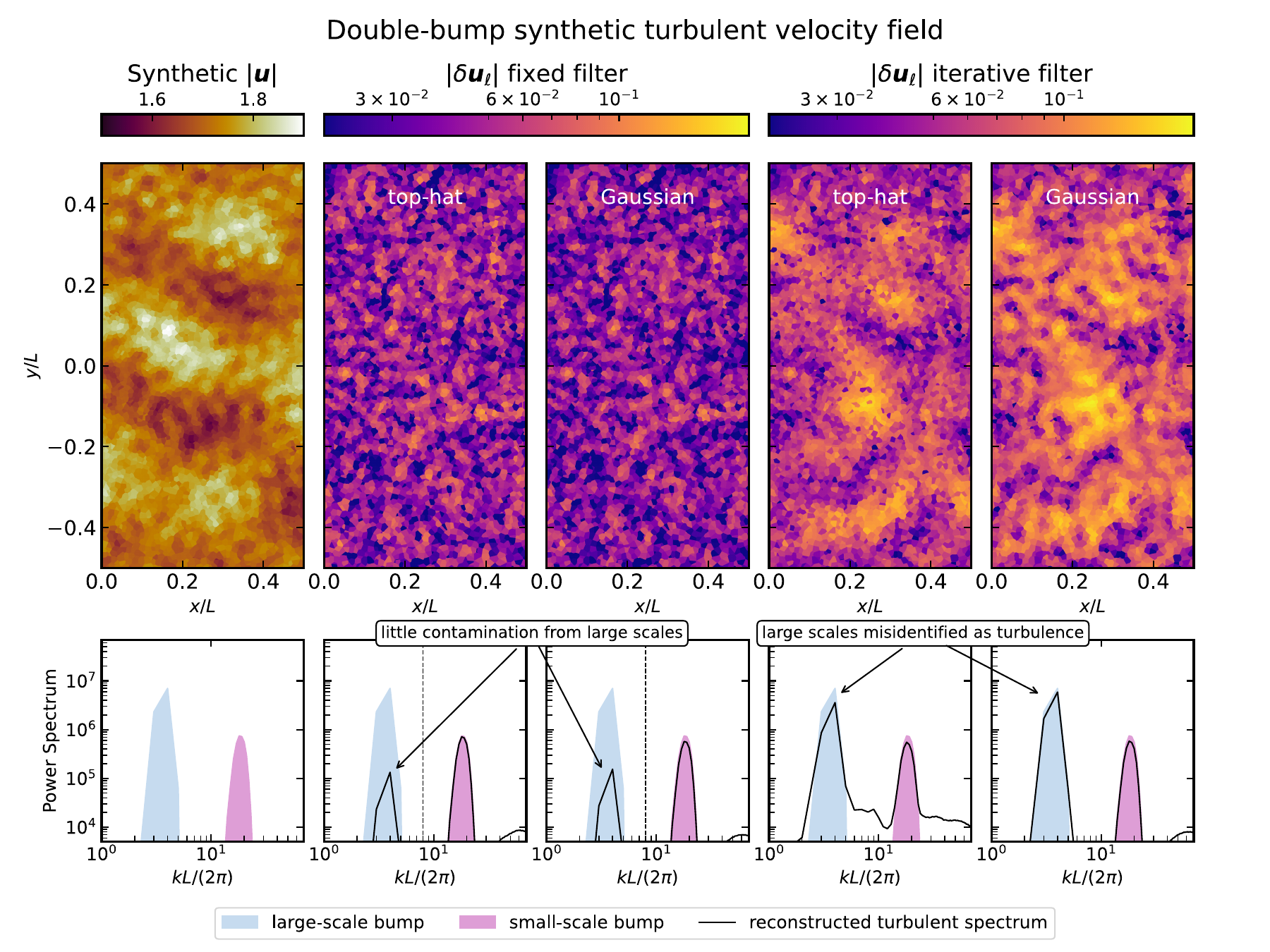}
        \caption{Comparison of fixed-scale and iterative filtering (tolerance $\epsilon = 0.05$) of a synthetic turbulent density field composed of two ``bumps'' in Fourier space.\textit{ Top row} (left to right): Total density field, reconstructed turbulent field with fixed and iterative filters (both with spherical top-hat and Gaussian kernels). The fixed-scale filters have been computed at the same $\lambda_{\mathrm{c}} = L/8$ (vertical dashed line in the lower row). \textit{Bottom row}: Input power spectra of the large- and small-scale bumps (light blue and pink, respectively) and power spectra of the resulting turbulent field after filtering (black line). The excess power at large $k$ in the reconstructed turbulent spectrum is due to numerical noise after interpolation on a uniform grid. The fixed-scale filters correctly separates the small-scale fluctuations. By contrast, in significant parts of the domain, the iterative filters converge on a scale outside of the spectral break between the two bumps, and thus erroneously include large-scale power into the turbulent field.}
        \label{fig:double_bump}
\end{figure*}

\subsubsection{Double-bump}

The second scenario examines the performance of the iterative filter in the presence of fluctuations that live on two sets of scales that are widely separated in wave number space. This test should be thought of as an idealized version of what may happen when turbulent fluctuations are superimposed on a large-scale variation due to, for example, the background density profile of the cluster; and is very similar to the numerical test originally proposed by \citet{Vazza2012} who used turbulent patches superimposed on a slowly varying background profile. The usefulness of these artificial problems is that we have exact knowledge of the initialized velocity field, which allows us to compare the performance of both filtering approaches. The expectation is that a successful iterative scheme would be able pick up the small-scale turbulent component and stop the iteration before reaching the smoothly varying large scales representing the background flow. 

We test this scenario in Fig.~\ref{fig:double_bump}, where we show a synthetic three-dimensional velocity field initialized on a three-dimensional Voronoi mesh made of a superposition of large-scale modes (representing a smoothly varying background flow) and small-scale modes, while exponentially suppressed elsewhere. 
The power is peaked around wave vectors $k L/ (2\pi ) \simeq 2-4$ and $15-20$, respectively ($L$ is the box size). 
For illustration purposes we compare the iterative filter with a fixed-scale filter whose smoothing length lies between the large- and small-scale bump (cutoff wave number\footnote{The cutoff wavelength $\lambda_\mathrm{c}$ and corresponding wave number $k_\mathrm{c}$ were introduced in Sect.~\ref{sec:averaging_operators}. They are are defined as the half-width at half-maximum of the smoothing kernel in wave number space, allowing us to compare different filtering kernels; see Appendix~\ref{app:filters_fourier_space} for further details. } of $k_\mathrm{c} L / (2 \pi) = 8$) and we repeat the procedure with the spherical top-hat (for which $k_\mathrm{c}$ corresponds to $\ell / L \simeq 0.023$) and Gaussian smoothing kernels ($\ell / L \simeq 0.050 $). We stress that in this test we are able to pick a suitable filter scale for the fixed-scale filters to separate the smooth background from the small-scale fluctuations only because we have perfect knowledge of the velocity field, which is obviously not the case in real simulations of galaxy clusters. Nevertheless it is useful to compare the performance of the iterative filter to the fixed-scale version to highlight their potential shortcomings.

\begin{figure}
        \centering
        \includegraphics[width=1.0\linewidth]{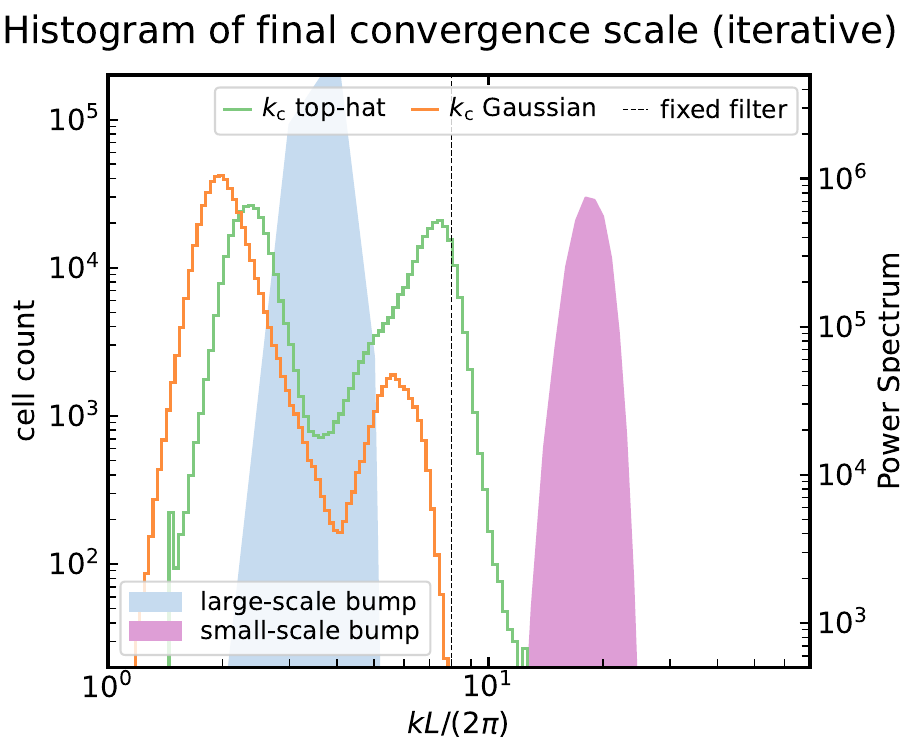}
        \caption{Distribution of the final convergence scale of the iterative filters (expressed in terms of the cutoff wave vector, $k_\mathrm{c}$) with spherical top-hat and Gaussian kernels. The vertical dashed black line corresponds to the cutoff wave number of the fixed-scale filter. For comparison we show on the right $y$ axis the power spectrum of the two bumps as in Fig.~\ref{fig:double_bump}. 
    A large fraction of the cells converges on scales much larger than the large-scale bump, misidentifying large-scale fluctuations as ``turbulence.''
        }
        \label{fig:double_bump_density_fixed_iterative_filter_histogram}
\end{figure}

Comparing the turbulent density field reconstructed using the two filtering schemes, we find that iterative filtering yields a small-scale field that still contains residual large-scale modes. We quantified the level of contamination by computing the power spectrum of the resulting density field and comparing it to the original synthetic field (bottom row of Fig.~\ref{fig:double_bump}). The top-hat kernel appears to perform better in this regard, although the reconstructed turbulent field is more noisy and visually looks less regular. If we look at the distribution of the final filter lengths of the iterative filter (Fig.~\ref{fig:double_bump_density_fixed_iterative_filter_histogram}) we find that with the top-hat kernel approximately $54 \%$ of the cells have converged to a scale above $k L/ (2\pi )  = 4$, which is where the large-scale bump starts and is well above the scales of the small-scale bump.
Similarly, with the Gaussian kernel ${\approx} 96 \%$ of the cells converged above $k L/ (2\pi )  = 4$, again skipping the small-scale bump. The contamination by large-scale modes in the iterative filters can also be quantified in terms of the energy: the reconstructed turbulent field with the top-hat kernel retains $48\%$ of the total energy contained in the large-scale modes ($1 \leq kL / 2 \pi \leq 4$), while this figure grows to $81\%$ with the Gaussian kernel. 
We conclude that the multi-scale iterative filter struggles to distinguish between fluctuations living on widely separated scales, even with homogeneous turbulence, and instead converges on disparate scales based on the initial conditions.

\subsection{Alternatives to iterative filters}

These considerations should warn against the use of an iterative filter in a realistic ICM environment where such a clean separation between scales is not at all warranted. Moreover, any further quantity obtained from the iteratively filtered fields and integrated over the volume (e.g., kinetic or magnetic turbulent energies, which we define in Sect.~\ref{sec:turbulent_energies}) would include contributions from different scales, making its physical interpretation unclear.

On a more conceptual level, as soon as we leave the idealized picture of statistically homogeneous, fully developed Kolmogorov turbulence -- with a single injection scale, an inertial range, and a dissipation scale -- an unambiguous and universal definition of what actually constitutes turbulence and what does not is in general not possible. For these reasons, we believe that the most transparent approach is to use a fixed scale filter with a physically motivated choice for the smoothing length that is appropriate for the problem and the phenomenon in question, and to repeat this procedure with several filter scales (as we did in Sect.~\ref{sec:turbulent_energies_application}). One tool to aid in the identification of suitable scales is to compute local power spectra. We plan to explore this in detail in future work.

\section{Conclusions}\label{sec:conclusions}

In this paper, we argue for a method based on smoothing filters to disentangle turbulence in cosmological simulations of galaxy clusters. With this approach, we can separate large-scale fields and small-scale fluctuations based on physical spatial scales, and systematically decompose magnetic and kinetic energies into ``bulk'' and ``turbulent'' components. 
By selecting different smoothing lengths, $\ell$, we can gain further insights into the properties of the fields in wave number space: in contrast to the power spectrum, it allows one to find the energy-containing scale in a certain region of the cluster by applying a fixed-scale filter on progressively larger scales and identifying where the small-scale energy stops increasing; moreover, smoothing filters make it possible to study the contribution of different physical processes and energy fluxes across scales. The two main advantages over the usual methods based on Fourier transforms are that (a) smoothing filters do not require periodic boundary conditions, they avoid interpolation errors, and can be applied to regions of irregular shape; and (b) as a result of the filtering, the turbulent (and smooth) fields are available to be used for further analysis, while power spectra or correlation function only yield volume-integrated measures.  

\subsection{Summary}

We demonstrate the application of smoothing filters and filtered energies to a cosmological simulation of a major cluster merger between two massive clusters (both $\gtrsim$$10^{15} ~\mathrm{M}_{\odot}$) from the PICO-Clusters suite of zoom-in simulations performed with the moving-mesh code \textsc{Arepo}. We follow the evolution of bulk and turbulent energies within $R_{200,\mathrm{c}}$ through the merger until $z=0$. We find that during the merger the turbulent kinetic pressure fraction on physical scales of
$\sigma_\ell \lesssim 50 ~ \si{kpc}$
reaches about $5 \%$ and then quickly decreases to $2\%$ after only $\sim$$1.3 ~ \si{Gyr}$ from the peak at the core passage (Fig.~\ref{fig:nonthermal_ratio_filter_lengths}). These low turbulent levels are consistent with observations of nearby massive clusters \citep{XRISMCollaboration2025f} and suggest that, unless the cluster has experienced a major merger in the last $(1-1.5) ~ \si{Gyr}$, it will appear as rather quiescent. A more in-depth analysis of the merger will be presented in a separate publication.

In the second part of the paper, we revisit the popular multi-scale iterative filtering method, which has been proposed in \citet{Vazza2012} as a technique to automatically distinguish turbulent from bulk motions. 
We perform a number of tests to evaluate its performance in simple but physically meaningful setups, using two widely used smoothing kernels: the Gaussian, and the spherical top-hat filter. 
We find that the convergence of iterative multi-scale filters does not always seem to reflect the morphology of the physical field either in real or wave number space.
For example, in simulations of the Kelvin--Helmholtz instability, the iterative filter often converges to very different scales for nearby regions, which produces visible artifacts in the bulk and turbulent fields (Fig.~\ref{fig:KHI_velocity_fixed_iterative_filter}).
On the other hand, in scenarios with fluctuations on two widely separated spatial scales (a factor of $\sim$10 in wavelengths), iterative filters struggle to differentiate between small and large-scales fluctuations in a significant fraction of the volume ($54\% - 96\%$ depending on the kernel), and the resulting filtered fields end up containing a mixture of both (Fig.~\ref{fig:double_bump_density_fixed_iterative_filter_histogram}). 
In actual simulations of galaxy clusters, we do not expect such wide scale separations; therefore, it is unclear how reliable iterative filters would be. Second, the mixing-up of different scales would not allow one to decompose the energies into turbulent and bulk, which only makes physical sense in the filter formalism with a fixed scale.

\subsection{Discussion: Whether smoothing filters can identify turbulence in galaxy clusters}

In this paper, we employed a method using fixed-scale smoothing filters to characterize turbulence in galaxy clusters. Leveraging previous work, we defined in Sec.~\ref{sec:definitions_turb_energies} filtered energies for the bulk and turbulent component of the magnetic and velocity field. In Sec.~\ref{sec:physical_interpretation} we drew a connection between the filtered energies and the energy contained in small-scale or large-scale Fourier modes in the power spectrum, which is the primary tool to study turbulent flows. 
Nevertheless, the question remains of whether the filtered turbulent energies can be read as a proxy for turbulence in galaxy clusters, and how to interpret their dependence on the filter size. The answer to this question inevitably rests on the concept of turbulence, which is often used loosely.

Methods such as spatial Fourier transforms and smoothing filters do not encode any physical pronouncement about the nature of turbulence, i.e., the presence of chaotic random fields, but only make a statement about the scales on which the fields vary. For this reason, the identification of ``small-scale fluctuations'' with turbulence is in general always fraught with uncertainty, and should be complemented with other diagnostics analyzing the time evolution of the flow or its chaotic behavior. For instance, an ordered small-scale fluctuation (smaller than the filter size) would contribute to the turbulence with both filtering methods and power spectra, but it is evidently not chaotic.

Finally, the ICM of galaxy clusters is a complex system where many physical processes combine to induce turbulent motions on a wide range of spatial and temporal scales. These processes will in general coexist, and the resulting fluctuations interact nonlinearly leading to a continuous distribution of energy across different scales, without a clear separation. This is a far cry from the simple picture of isotropic Kolmogorov turbulence with one injection scale, an inertial range and a dissipation scale. It is our opinion that in complex and heterogeneous systems as the ICM any definition of what constitutes turbulence will always be inherently ambiguous. For this reason, we recommend reframing this question as a scale-dependent statement, and ask what is the energy (or amplitude) of fluctuations below and above a certain physical scale. With this approach we can make meaningful comparisons across different simulations, and bridge the gap between theory and observations.

\begin{acknowledgements}

The authors thank the referee for providing constructive feedback, which led to a considerable improvement in the presentation of the paper. The data underlying this article will be shared on reasonable request to the corresponding author. LMP and CP gratefully acknowledge support by the European Research Council under ERC-AdG grant PICOGAL-101019746. 
Part of this work was done during a visit at the Niels Bohr Institute funded by the Rosenfeld Foundation.
TB gratefully acknowledges funding from the European Union’s Horizon Europe research and innovation programme under the Marie Skłodowska-Curie grant agreement No 101106080 and financial support by the Carlsberg Foundation via grant CF23-0417. The authors gratefully acknowledge the Gauss Centre for Supercomputing (GCS) for providing
computing time on the GCS Supercomputer SuperMUC-NG at the Leibniz Supercomputing Centre (LRZ) in Garching, Germany, under project pn68cu. The Tycho supercomputer hosted at the SCIENCE HPC center at the University of Copenhagen was used for supporting this work.

Our filtering code \textsc{Turbocluster} is written in Python, and depends on the following packages: \textsc{Paicos} for the handling of \textsc{Arepo} data (\url{https://github.com/tberlok/paicos}), \textsc{Numpy}, as well as \textsc{Numba} and \textsc{Cupy} for execution on GPUs. Plots were produced with \textsc{Matplotlib}, using the \textsc{Cmasher} package for colormaps.

\end{acknowledgements}




\bibliographystyle{aa}
\bibliography{AIP-GalClusTurb}



\newpage 

\appendix

\section{Comparison of spherical top-hat and Gaussian kernels for iterative filters}\label{app:filters_fourier_space}

In this appendix we discuss two of the most common choices of smoothing kernels, the spherical top-hat, and the Gaussian kernel. We implemented them in our filtering code and tested their convergence properties when used with the iterative multi-scale filters.

\subsection{Smoothing kernels in spectral space}
\label{app:kernels}

\begin{figure}
        \centering
        \includegraphics[width=1.0\linewidth]{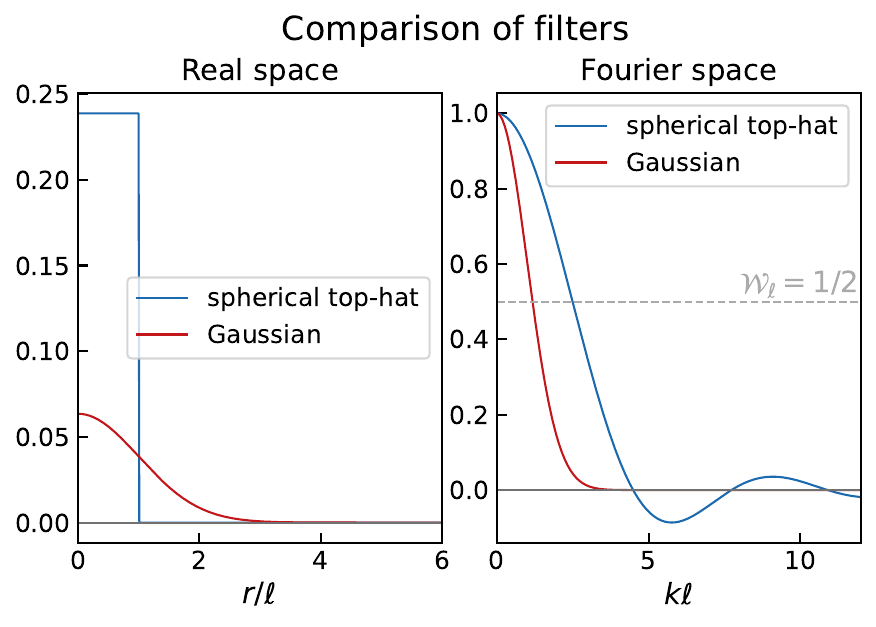}
        \caption{Comparison in real (\textit{left}) and Fourier space (\textit{right}) between the three-dimensional spherical top-hat and Gaussian filter. The points of intersection between the filter shapes and the $\mathcal{W}_\ell = 1/2$ line (dashed gray, right panel) represent the half-widths at half-maxima and define the cutoff wavelengths of both filters.}
        \label{fig:filters3D_comparison}
\end{figure}

We first compare the Fourier transform of the spherical top-hat and Gaussian kernel. We use the convention where the Fourier transform $\hat{f} (\bm k)$ of a function $f (\bm x)$ is defined as
\begin{align}\label{eq:fourier_transform}
        \hat{f} (\bm k) = \int_{\mathbb{R}^3} f(\bm x) e^{-i \bm k \bcdot \bm x} \ud^3 x.
\end{align}
In three dimensions the spherical top-hat filter is defined as
\begin{align}
        &\mathcal{W}_\ell (\bm x', \bm x) =
        \begin{cases}
                \left(\frac{4}{3}\pi \ell^3\right)^{-1} \; \; &\text{if} \; |\bm x - \bm x'| \leq \ell ,\\
                0 \; \; &\text{if} \; |\bm x - \bm x'| > \ell ,         
        \end{cases} 
\end{align}
while the Gaussian filter is given by
\begin{align}
        &\mathcal{W}_\ell (\bm x', \bm x) = \frac{1}{\left(2 \pi  \ell^2 \right)^{3/2}}
        e^{-\frac{1}{2}|\bm x - \bm x'|^2 /\ell^2}.
\end{align}
These kernels are functions of the single variable $ \bm r = \bm x' - \bm x$ only, and thus we sometimes use the shorthand notation $\mathcal{W}_\ell (\bm r)$ such that $\mathcal{W}_\ell (\bm r) = \mathcal{W}_\ell (\bm x - \bm x') = \mathcal{W}_\ell (\bm x', \bm x)$, and replace $\bm r$ with $\bm x$ when there is no risk of ambiguity. Their Fourier transforms can be easily computed, yielding:
\begin{align}
        &\hat{\mathcal{W}}_\ell (\bm k) = 
                \frac{3 j_1 (k \ell)}{k \ell}, & \text{(spherical top-hat)}, \\
        &\hat{\mathcal{W}}_\ell (\bm k) =
        e^{ -k^2 \ell^2 /2}, & \text{(Gaussian)},
\end{align}
where $j_1 (z) = \sin( z) / z^2 - \cos( z) / z$ is the first-order spherical Bessel function of the first kind, and $k = |\bm k |$.
We plot the smoothing kernels and their Fourier transform in Fig.~\ref{fig:filters3D_comparison} as a function of radius $r = |\bm x' - \bm x|$ and $k$, respectively. Both filters have qualitatively similar shapes in Fourier space: when convolved with a function, they suppress Fourier coefficients on scales smaller than $\ell$, while leaving those on larger scales unchanged, acting akin to a low-pass filter. Having said that, we note that despite having the same nominal $\ell$, the spherical top-hat has a broader half-width at half-maximum (HWHM) than the Gaussian kernel.

Therefore, to compare different filters we compute the HWHM, i.e., the frequency cutoff, $k_\mathrm{c}$, at which $\hat{\mathcal{W}}_\ell (k_\mathrm{c}) = 1/2$. We then convert it into a cutoff wavelength, $\lambda_\mathrm{c} = 2 \pi / k_\mathrm{c}$. This cutoff wavelength is related to the filter size, $\ell$, by a numerical coefficient which depends on the kernel used. For the spherical top-hat and the Gaussian kernels we have
\begin{align}
        \frac{\lambda_\mathrm{c}}{\ell} = \begin{cases}
                2.51503\ldots & \text{(spherical top-hat)}, \\
                2 \pi / \sqrt{2 \ln 2} = 5.33645\ldots & \text{(Gaussian)}.
        \end{cases}
\end{align}
Roughly speaking, we find that for the same $\ell$ the Gaussian kernel suppresses wavelengths twice as large as the spherical top-hat.

\subsection{Filtering of Fourier series}

In this section we show the action of a smoothing kernel on a generic function. We choose to expand the function in a Fourier series over discrete wave vectors, $\bm k$, since in all practical applications the domain of interest $V = L_x L_y L_z$ is finite. However, taking the limit $V \rightarrow \infty$ and rescaling the Fourier coefficients appropriately, the more general case of continuous $\bm k$ is recovered.

We consider a generic periodic function, $f(x)$, expanded in Fourier series:
\begin{align}\label{eq:fourier-coefficient}
        f (\bm x) = \sum_{\bm k} \hat{f}_{\bm k} e^{i \bm k \bcdot \bm x}, \hspace{2em} \hat{f}_{\bm k} = \frac{1}{V} \int_{V} f (\bm x) e^{-i \bm k \bcdot \bm x} \ud^3 x,
\end{align}
where $\bm k = 2 \pi \times (n_x/L_x, n_y/L_y, n_z/L_z)^T$, with $n_x, n_y, n_z = ..., -1, 0, 1, ...$, and where $\hat{f}_{\bm k}$ are the Fourier coefficients. Similarly for the smoothing kernel, $\mathcal{W}_\ell$:
\begin{align}\label{eq:fourier_smoothing_kernel}
        \mathcal{W} (\bm x ) = \sum_{\bm k} \hat{\mathcal{W}}_{\ell, \bm k} e^{i \bm k \bcdot \bm x}.
\end{align}
The application of a smoothing filter in real space can be represented by a product in Fourier space using the convolution theorem:
\begin{align}
                \langle f \rangle_{\ell} (\bm x) &\equiv \int_V f(\bm x' ) \mathcal{W}_\ell (\bm x - \bm x') \ud^3 x' \nonumber \\
                &= \sum_{\bm k} \hat{\mathcal{W}}_{\ell, \bm k} e^{i \bm k \bcdot \bm x} \int_V e^{-i \bm k \bcdot \bm x'} f(\bm x' ) \ud^3 x' \nonumber \\
                &= V \sum_{\bm k} \hat{f}_{\bm k} \hat{\mathcal{W}}_{\ell, \bm k} e^{i \bm k \bcdot \bm x}.
        \label{eq:convolution-theorem}
\end{align}
where we used the definitions of the Fourier coefficients and the inverse transform in Eq.~\eqref{eq:fourier-coefficient}--\eqref{eq:fourier_smoothing_kernel}.
Note that for a kernel with compact support in $V$, the Fourier coefficients, $\hat{\mathcal{W}}_{\ell, \bm k}$, are related to its Fourier transform (Eq.~\ref{eq:fourier_transform}) by $\hat{\mathcal{W}}_{\ell, \bm k} = V^{-1} \hat{\mathcal{W}}_{\ell} (\bm k)$.

\subsection{Convergence properties}

In this section we examine the convergence properties of the two kernels with the multi-scale iterative filter method proposed by \citet{Vazza2012}. The multi-scale iterative filter operates by successively filtering a scalar or vector field with kernels of increasingly larger size $\ell_n$ at every iteration, until a certain convergence criterion is met. The final $\ell_n$ when the iteration is stopped can differ from place to place, and might be identified with the local correlation scale of the turbulence.

In its original formulation \citep{Vazza2012}, for a scalar field, $f (\bm x)$, decomposed as $f = \langle f \rangle_{\ell_{n}} + \delta f_{\ell_n}$, the convergence criterion  evaluated at the $n$-th iteration is:
\begin{align}
        \left| \frac{ \delta f_{\ell_n} - \delta f_{\ell_{n-1}} }{\delta f_{\ell_{n-1}}} \right| \leq \epsilon, \label{eq:app_vazza_convergence_criterion}
\end{align} 
where $\epsilon$ is a small tolerance parameter (typically $\epsilon = 0.02 - 0.1$).

\begin{figure}
        \centering
        \includegraphics[width=1.0\linewidth]{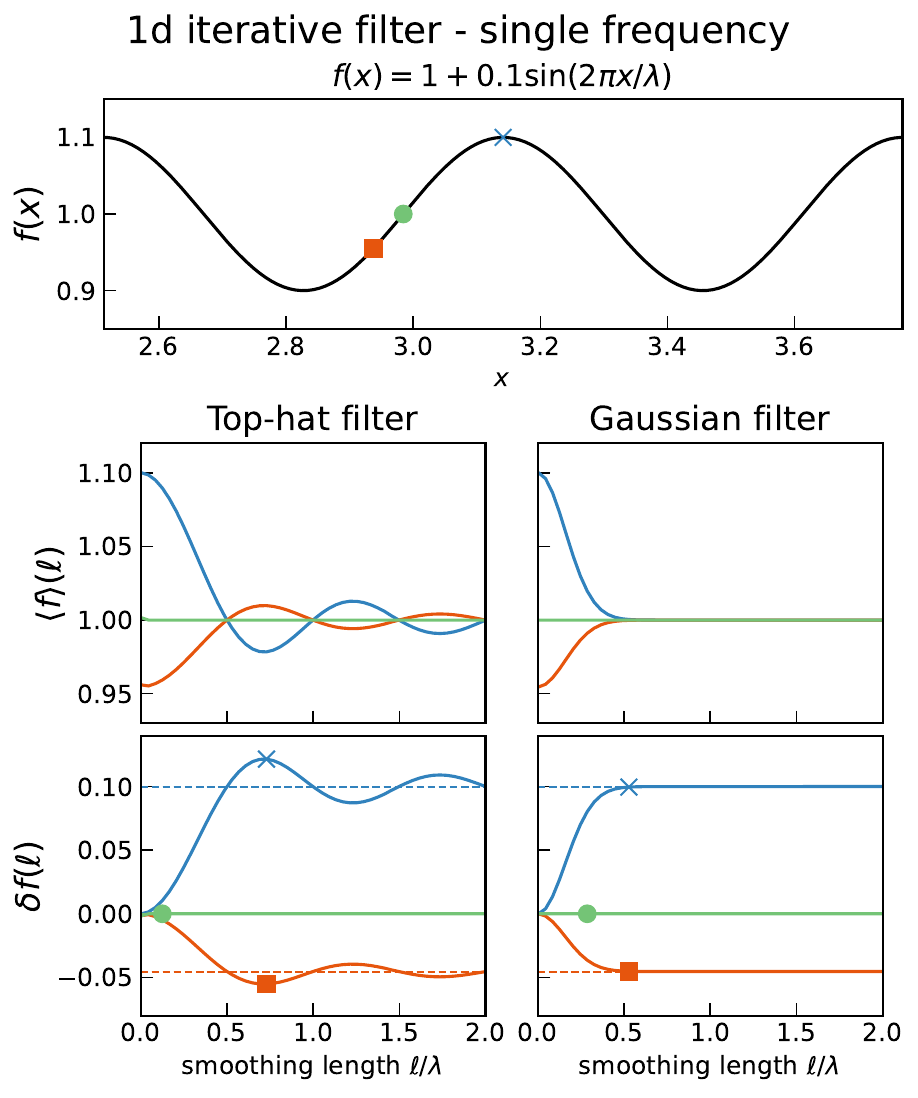}
        \caption{One-dimensional test of the convergence of multiscale iterative filters with a top-hat and Gaussian kernel on a uniform grid. The function consists of a constant background plus a sinusoidal perturbation of amplitude $a=0.1$ and wavelength $\lambda = 2 \pi / 10$. For the three selected points (colored markers in the top panel), we show the average (middle panel) and fluctuation (bottom panel) for different smoothing lengths $\ell$, using a spherical top-hat and Gaussian filter (left and right columns). Bottom panel: the markers indicate where the iterative algorithm stops (according to the criterion in Eq.~\eqref{eq:app_vazza_convergence_criterion} using $\epsilon = 0.01$), the dashed lines the ``true'' value of the fluctuation at the three selected points. The Gaussian filter has better convergence properties but is more expensive computationally.}
        \label{fig:mean_gaussian_filter_single_freq}
\end{figure}

While a spherical top-hat filter is used to compute the local bulk and turbulent component in the original work of \citet{Vazza2012}, we can show that the Gaussian kernel offers in general better convergence properties. To do so, we first note that if we consider $\delta f (\ell)$ as a function of the continuous variable $\ell$, for small increases in the smoothing length, $\Delta \ell$, the convergence criterion can be equivalently rewritten in terms of the logarithmic derivative of $\delta f (\ell)$,
\begin{align}
        \Delta \ell \left\lvert\frac{\mathrm{d} \ln \delta f}{\mathrm{d} \ell} \right\rvert \leq \epsilon, \hspace{2em} \text{for} \; \delta f \neq 0.
\end{align}
The criterion above suggests that for $\epsilon \ll 1$ the iterative algorithm tends to stop at the extrema of the function $\delta f (\ell)$. One issue is that, with a spherical top-hat filter, the reconstructed  $\delta f$ at these points is generally not a good approximation of the local turbulent field. This can be clearly seen in the simple case of a single-frequency fluctuation of wavelength $\lambda$ superimposed on a constant uniform background (Fig.~\ref{fig:mean_gaussian_filter_single_freq}). 
In this experiment, we sample three representative points (at a maximum of the sine, at a node and at an intermediate value), and numerically compute the values of the smooth field, $\langle f \rangle_\ell$, and fluctuation, $\delta f (\ell)$, as a function of $\ell$ for $\epsilon = 0.01$. From Fig.~\ref{fig:mean_gaussian_filter_single_freq}, we can see that when employing a spherical top-hat filter, the iterative algorithm stops near the second extremum of $\delta f (\ell)$ (for the two non-unity points, blue and red in Fig.~\ref{fig:mean_gaussian_filter_single_freq}). Regarding the point at the starting position $f(x)=1$ (green circle), we note that the expression in Eq.~\eqref{eq:app_vazza_convergence_criterion} becomes singular for $\delta f (\ell) \simeq 0$, which is the ``true'' value given that the sine is odd around that location. In a more realistic scenario, for example with a discretization error or on a nonuniform grid, the convergence of such points can be dominated by numerical noise. Finally, with the spherical top-hat filter the reconstructed $\delta f (\ell)$ has an oscillatory behavior (the analytical function has the form of a $\sinc(x) = \sin(x) / x$) and thus converges slowly to the ``true'' value  when $\ell \gg \lambda$.
With the Gaussian filter, we can get better convergence properties. As shown in the right panel of Fig.~\ref{fig:mean_gaussian_filter_single_freq}, the reconstructed fluctuating field $\delta f (\ell)$ does not oscillate and converges more quickly (at lower $\ell$) to the true value. This improved behavior, however, comes at the cost of more computations: in fact, for the same $\ell$, the support of the Gaussian filter extends much further than the spherical top-hat (we numerically truncate it beyond $4 \ell$ as a compromise between accuracy and speed), which greatly increases the number of operations.

\section{Commutativity between averaging and filtering operators}\label{app:commutativity}

In this Appendix we discuss under what conditions the volume average, $\langle ... \rangle_V$,
\begin{align}\label{eq:vol_avg_op}
        \langle f \rangle_V \equiv \frac{1}{V} \int_V f (\bm x) \ud^3 x,
\end{align}
commutes with the filtering operator, $\langle ... \rangle_\ell$,
\begin{align}
        \langle f \rangle_{\ell} (\bm x) \equiv \int f(\bm x' ) \mathcal{W}_\ell (\bm x', \bm x) \ud^3 x' .
\end{align}
As we go on to show, this property allows us to recover the total energies (kinetic and magnetic) when using the filtering approach; moreover, it allows us to rule out certain na\"{i}ve definitions of the turbulent energy density that identically integrate to zero.

\subsection{General considerations about commutativity}

In the following section we restrict ourselves to finite volumes $V = L_x L_y L_z$, for which the volume average is well defined. We first consider the case of a nonperiodic volume, $V$, where the only conditions on the smoothing kernel are that $\mathcal{W}_\ell$ is non-negative, symmetric in its arguments $(\bm x, \bm x')$ and normalized to unity in $V$:
\begin{align}
        \int_V \mathcal{W}_\ell (\bm x, \bm x') \ud^3 x' = \int_V \mathcal{W}_\ell (\bm x', \bm x) \ud^3 x' = 1, \hspace{2em} \forall \bm x \in V. \label{eq:symmetric_kernel}
\end{align}
Equation~\eqref{eq:symmetric_kernel} implies that, for points $\bm x$ near the boundaries of $V$, the integration of the kernel $\mathcal{W}_\ell$ is performed only over the cells belonging to the original volume of interest, without straying beyond it. 
In this case we have
\begin{align}
        \langle \langle f \rangle_{\ell} \rangle_V &= \frac{1}{V} \int_V \ud^3 x \left[ \int_V f(\bm x') \mathcal{W}_\ell (\bm x', \bm x) \ud^3 x' \right] \nonumber \\
        &= \frac{1}{V} \int_V \ud^3 x' f(\bm x') \left[ \int_V \mathcal{W}_\ell (\bm x', \bm x)  \ud^3 x \right] \nonumber \\
        &= \frac{1}{V} \int_V \ud^3 x' f(\bm x') = \langle f \rangle_V = \langle \langle f \rangle_V \rangle_{\ell}, \label{eq:commutativity_1}
\end{align}
where in the last equality we used the fact that applying the filtering operator to a constant leaves its value unchanged. Despite the apparent simplicity of this result, a smoothing kernel that satisfies Eq.~\eqref{eq:symmetric_kernel} is computationally hard to implement, as $\mathcal{W}_{\ell}$ needs to be defined so as to take into account all the edge cases of the points near the boundaries.

\begin{figure}
        \centering
        \includegraphics[width=0.6\linewidth]{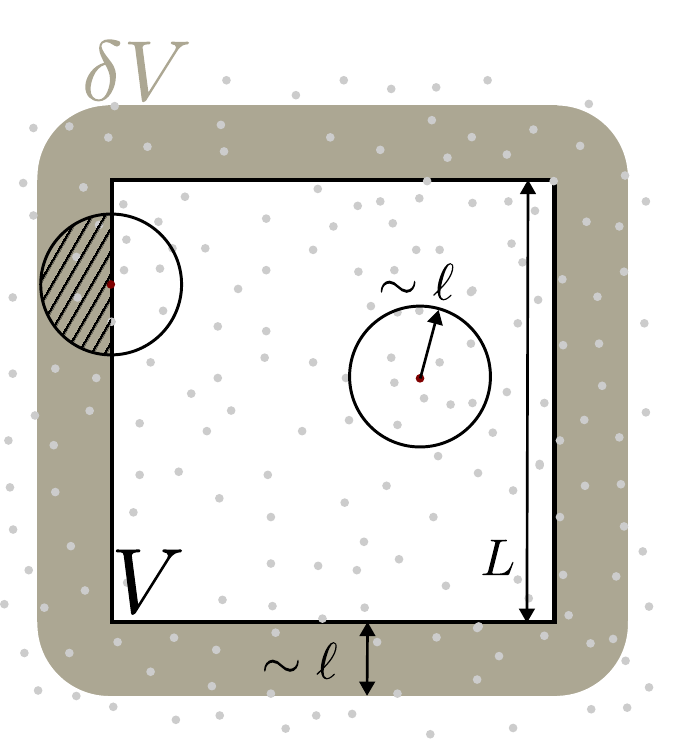}
        \caption{Cartoon of the smoothing filters within a finite domain $V$. For cells far from the boundaries, all of the support of the smoothing kernel (a ball of radius $\sim \ell$) is contained within $V$. For points near the edge and smoothing kernels of the form $\mathcal{W}_\ell (\bm x - \bm x')$, it is necessary to include a buffer region $\delta V$ of thickness $\sim \ell$.
        }
        \label{fig:sketch_filter_buffer}
\end{figure}

One common approach is to relax the conditions in Eq.~\eqref{eq:symmetric_kernel} and assume instead for simplicity that the smoothing kernel is a function of $(\bm x - \bm x')$ only. We then distinguish between the periodic and nonperiodic case. If the volume $V$ is periodic, it is straightforward to verify that the volume average and filtering operators also commute. To show it explicitly, we expand $f$ and $\mathcal{W}_{\ell}$ in Fourier series (Eq.~\ref{eq:fourier-coefficient}),
we convolve them using the convolution theorem (Eq.~\ref{eq:convolution-theorem}) and, after applying the volume-average operator (Eq.~\ref{eq:vol_avg_op}), we obtain
\begin{align}
        \langle \langle f \rangle_{\ell} \rangle_V &= \frac{1}{V} \int_V  \left[ V \sum_{\bm k} \hat{f}_{\bm k} \hat{\mathcal{W}}_{\ell, \bm k} e^{i \bm k \bcdot \bm x} \right] \ud^3 x \nonumber \\
        &=  V \hat{f}_{\bm 0} \hat{\mathcal{W}}_{\ell, \bm 0}.
\end{align}
%
%
From the definition of the Fourier series (Eq.~\ref{eq:fourier-coefficient}) it follows that  $\hat{\mathcal{W}}_{\ell, \bm 0} = 1/V$, since the smoothing kernel is normalized to unity, while $\hat{f}_{\bm 0} =  \langle f \rangle_V$. Hence, the volume average and filtering operators commute:
\begin{align}
        \langle \langle f \rangle_{\ell} \rangle_V = \langle f \rangle_V = \langle \langle f \rangle_V \rangle_{\ell}. \label{eq:commutativity}
\end{align}
When $V$ is not periodic, the convolution integral must be performed in real space, which means that near the boundaries of $V$ the smoothing kernel $\mathcal{W}_\ell$ overlaps with a buffer region $\delta V$ outside the volume of interest; see Fig.~\ref{fig:sketch_filter_buffer}. This region (of thickness $\sim \ell$) must now be included in the integration to correctly normalize the kernel $\mathcal{W}_\ell$:
\begin{align}
        \int_{V+\delta V} \mathcal{W}_\ell (\bm x - \bm x') \ud^3 x' = 1, \; \forall \bm x \in V, \label{eq:normalization_kernel}
\end{align}
and similarly to compute $\langle f \rangle_{\ell}$. In this case, which is computationally simpler and that we have implemented in our code, the volume average and the smoothing filter do not commute exactly anymore when $V$ is nonperiodic.
The difference between $\langle f  \rangle_V$ and $\langle \langle f \rangle_{\ell} \rangle_V$ can be given explicitly by first rewriting the volume integral of $f$ as:
\begin{align}
        &\int_V f (\bm x')\ud^3 x' = \int_V f (\bm x') \left[ \int_{V+\delta V} \mathcal{W}_\ell (\bm x - \bm x') \ud^3 x \right] \ud^3 x' , \label{eq:rewrite_volume}
\end{align}
where we used Eq.~\eqref{eq:normalization_kernel}. Similarly for the integral of $\langle f \rangle_{\ell}$:
\begin{align}
        &\int_V \langle f \rangle_{\ell} (\bm x') \ud^3 x' = \int_V  \left[ \int_{V+\delta V} f (\bm x) \mathcal{W}_\ell (\bm x' - \bm x) \ud^3 x \right] \ud^3 x' \label{eq:rewrite_filter}.
\end{align}
Splitting the inner integral in Eqs.~\eqref{eq:rewrite_volume}--\eqref{eq:rewrite_filter} between $V$ and the buffer volume, $\delta V$, allows one to cancel out the integration over $V$ (since $\bm x$ and $\bm x'$ are interchangeable within $V$), with only the cross integration remaining. Thus
\begin{align}
        \langle f  \rangle_V - \langle \langle f \rangle_{\ell} \rangle_V 
        = \frac{1}{V} \int_{\delta V} \ud^3 x \int_V \ud^3 x' \left[ f(\bm x') - f(\bm x) \right] \mathcal{W}_\ell (\bm x - \bm x'), \label{eq:difference_average_smoothing}
\end{align}
which involves a double integral that runs over all the pairs $\bm x \in V, \bm x' \in \delta V$ and weighs the difference between the function $f$ evaluated at these two points. If $f$ is bounded, it can be shown that Eq.~\eqref{eq:difference_average_smoothing} scales as $ \mathcal{O} (\delta V / V)$, or $\mathcal{O} (\ell / L)$ for a cubic volume of side $L$. In practice, when integrating over the cluster size this difference is usually very small (e.g., we calculated a relative difference of $0.04 \%$ for the total kinetic energy shown in Fig.~\ref{fig:kinetic_magnetic_energy_all_volume} at $z=0$).

\subsection{Recovering total energies}

As we discussed in Sect.~\ref{sec:turbulent_energies}, an apparent inconsistency of the filtering approach is that, when we define the filtered bulk and turbulent energy densities  their sum does not yield the total energy density. Let us take the example of the magnetic field, for simplicity: $\varepsilon_{B,\mathrm{bulk}} + \varepsilon_{B,\mathrm{turb}} = \varepsilon_B =  \langle e_B \rangle_\ell \neq e_B = \bm B^2 / 8\pi$;
mathematically speaking, $\varepsilon_B$ corresponds to the filtered version of $e_B$. The same applies to the kinetic energy. Using the results of the previous subsection, we can now show that with filtering it is in fact possible to recover the total energy within a volume, which is then repartitioned between bulk and turbulent energy in its entirety. 

We first assume that the volume average and filtering operator commute exactly (which they do if the domain is periodic or if the smoothing kernel, $\mathcal{W}_\ell (\bm x, \bm x')$, is symmetric in its arguments and the integration is restricted to $V$; see Eq.~\ref{eq:symmetric_kernel}). We use $e_{\star}$ to define either the kinetic or magnetic energy density, and $\varepsilon_{\star} = \langle e_{\star} \rangle_\ell$ its filtered counterpart. Similarly, their volume integrals are
\begin{align}
        E_{\star} = \int_V e_{\star} \ud^3 x, \hspace{2em} \mathcal{E}_{\star} = \int_V \varepsilon_{\star} \ud^3 x.
\end{align}
Using the volume average in Eq.~\eqref{eq:vol_avg_op}, we can write
\begin{align}
        \frac{1}{V}\mathcal{E}_{\star} &= \frac{1}{V} \int_V \varepsilon_{\star} \ud^3 x = \langle \varepsilon_{\star} \rangle_V = \langle \langle e_{\star} \rangle_\ell \rangle_V =  \langle \langle  e_{\star} \rangle_V \rangle_\ell \nonumber \\
        &= \frac{1}{V} E_\star \label{eq:proof_energies}
\end{align}
where we used Eq.~\eqref{eq:commutativity_1} to switch the order of the volume average and filtering operator. Thus $\mathcal{E}_{\star} = E_\star$. If the two operators do not commute exactly, the equality can be replaced by an approximate sign ($\mathcal{E}_{\star} \simeq E_\star$) with a relative error that scales as $\bigO (\ell / L)$; see Eq.~\eqref {eq:difference_average_smoothing}.

\subsection{A flawed definition of turbulent energies}

Another corollary of the commutativity between volume averaging and smoothing operators is that it allows us to rule out certain na\"{i}ve definitions of the turbulent energy densities. In fact, considering the kinetic or magnetic energy density $e_{\star}$ as a scalar field, one might be tempted to define the difference between $e_{\star}$ and its smooth counterpart as a measure of the turbulent energy density, similarly to how one would define the turbulent magnetic or velocity field: $e_{\star, \mathrm{turb}} \equiv e_{\star} - \langle e_{\star} \rangle_\ell$. However, by virtue of Eq.~\eqref{eq:proof_energies}, if we integrate this measure over the volume, $V$, we would obtain
\begin{align}
        \int_V e_{\star, \mathrm{turb}} \ud^3 x = E_{\star} - \mathcal{E}_\star,
\end{align} 
which is either zero (when either $V$ is periodic or $\mathcal{W}_\ell$ satisfies Eq.~\ref{eq:symmetric_kernel}) or asymptotically small.

\section{Properties of the filtered energies}\label{app:properties-filt-en}

\subsection{Galilean invariance of the turbulent kinetic energy}\label{app:galilean_invariance}

In this Appendix we show that the turbulent kinetic energy density ($\varepsilon_\mathrm{k,turb}$) as defined in Eq.~\eqref{eq:kin_en_turb} is Galilean invariant, i.e., it remains constant under a change of frame of reference, while the filtered total kinetic energy density ($\varepsilon_\mathrm{k} $), the energy density of the bulk flow ($\varepsilon_\mathrm{k,bulk}$), and the transport of turbulent momentum by the bulk flow ($\varepsilon_\mathrm{k,cross}$) do not. This result shows that $\varepsilon_\mathrm{k,turb}$ is a physically meaningful measure of the amount of turbulence on small scales.

We place ourselves in a frame of reference moving with uniform velocity, $\bm U_0$, with respect to the simulation box frame. This could be the center-of-mass velocity of a halo, of multiple pre-merger halos, or another physically motivated velocity. We define the velocity field in the new frame as $\bm \varv (\bm x) = \bm u (\bm x) - \bm U_0$ and replace it in Eqs.~\eqref{eq:kin_en_tot}--\eqref{eq:kin_en_turb}. Thus, the filtered total kinetic energy density becomes
\begin{align}
        \varepsilon_\mathrm{k}  (\bm \varv + \bm U_0) = \varepsilon_\mathrm{k}  (\bm \varv) + \frac{1}{2} \langle \rho \rangle_\ell \bm U_0^2 + \langle \rho \bm \varv \rangle_\ell \bcdot \bm U_0 
        , \label{eq:kin_en_tot_boost}
\end{align}
while the energy density of the bulk flow is
\begin{align}
        \varepsilon_\mathrm{k,bulk} (\bm \varv + \bm U_0) = \varepsilon_\mathrm{k,bulk} (\bm \varv) + \frac{1}{2} \langle \rho \rangle_\ell \bm U_0^2 + \langle \rho \rangle_\ell \langle \bm \varv \rangle_\ell \bcdot \bm U_0.
\end{align}
For the transport term $\varepsilon_\mathrm{k,cross}$ we need to calculate the second-order density-velocity moment in terms of the new velocity field, $\mu_2 (\rho, \bm \varv + \bm U_0)$. From the definition of the second-order statistical moment Eq.~\eqref{eq:second_moment}
it is straightforward to show that $\mu_2 (\rho, \bm \varv + \bm U_0) = \mu_2 (\rho, \bm \varv)$, i.e., it is invariant under a Galilean boost. This also applies to higher-order moments of any argument \citep[i.e., density, velocity, and magnetic field;][]{Germano1992}. The cross-transport term is then simply:
\begin{align}
        \varepsilon_\mathrm{k,cross} (\bm \varv + \bm U_0) = \varepsilon_\mathrm{k,cross} (\bm \varv) + \bm U_0 \bcdot \mu_2 (\rho, \bm \varv) . \label{eq:kin_en_transp_boost}
\end{align}
Finally, using Eq.~\eqref{eq:kin_en_decomposition} and Eqs.~\eqref{eq:kin_en_tot_boost}--\eqref{eq:kin_en_transp_boost} or by direct computation, we have
\begin{align}
        \varepsilon_\mathrm{k,turb} (\bm \varv + \bm U_0) =  \varepsilon_\mathrm{k,turb} (\bm \varv). 
\end{align}

\subsection{Turbulent kinetic energy in the subsonic regime}\label{app:subsonic-regime}

In this section we derive an expression for the turbulent kinetic energy in the subsonic regime. We assume a stably stratified atmosphere in hydrostatic equilibrium with no mean flow, and perturb the thermodynamic quantities around the background equilibrium:
\begin{align}
    &\rho (\bm x) = \rho_0 (\bm x) + \delta \rho(\bm x), &p(\bm x) = p_0 (\bm x) + \delta p (\bm x), \ldots \label{eq:subsonic-expansion}
\end{align}
Note that $\delta \rho$, $\delta p$, and so on are not the same as the fluctuations obtained through the application of the smoothing filter: $\delta \rho \neq \delta \rho_\ell$.
We then take the background thermodynamic quantities, $\rho_0 (\bm x)$, $p_0 (\bm x)$, $T_0 (\bm x)$, to vary on scales $H = | \mathrm{d} \ln \rho_0 / \mathrm{d} r|^{-1}$ much larger than those of the turbulent fluctuations (in principle, one ought to distinguish between the pressure, density and temperature scale-heights, but for the purpose of this argument we can take them to be all of the same order). If the flow is highly subsonic, $\mathcal{M} = u/c_\mathrm{s} \ll 1$, where $u$ is the amplitude of the typical velocities, it can be shown \citep[see, e.g.,][]{Zhuravleva2014a} that
\begin{align}
    \frac{\delta \rho}{\rho_0} \sim \mathcal{O} (\mathcal{M}) \ll 1. \label{eq:subsonic-density-fluct}
\end{align}
We now derive an analogous relation to Eq.~\eqref{eq:subsonic-density-fluct} for the smooth and turbulent density fluctuations obtained with smoothing filters, $\langle \rho \rangle_\ell$, and $\delta \rho_\ell = \rho - \langle \rho \rangle_\ell$, respectively. Our only assumption is that the filter scale, $\ell$, is much smaller than the typical scale-height, $H$. Applying the smoothing filter to the total density and expanding it as in Eq.~\eqref{eq:subsonic-expansion} we have
\begin{align}
    \langle \rho \rangle_\ell &(\bm x) = \int \rho_0 (\bm x') \mathcal{W}_\ell (\bm x', \bm x) \ud^3 x' + \int \delta \rho(\bm x') \mathcal{W}_\ell (\bm x', \bm x) \ud^3 x' \nonumber \\
    &= \int \left[ \rho_0 (\bm x) + (\bm x' - \bm x) \cdot \nabla \rho_0 (\bm x) + \ldots\right] \mathcal{W}_\ell (\bm x', \bm x) \ud^3 x'  + \nonumber \\ 
    &+ \int \delta \rho(\bm x')  \mathcal{W}_\ell (\bm x', \bm x) \ud^3 x' \nonumber \\
    &= \rho_0(\bm x) \left[1 + \mathcal{O}(\ell^2 / H^2) \right]+ \int \delta \rho(\bm x')  \mathcal{W}_\ell (\bm x', \bm x) \ud^3 x' ,
\end{align}
where in the second line we Taylor-expanded the background density, $\rho_0 (\bm x')$, around $\bm x$, and used the properties of the smoothing kernel (Eq.~\ref{eq:kernel_properties}c) to cancel the first term of the Taylor series; finally, we used the fact that $| \bm x' - \bm x| \lesssim \ell$ inside the smoothing integral. As a result, the density fluctuation through the smoothing filter is
\begin{align}
    \delta \rho_\ell (\bm x) &= \rho (\bm x) - \langle \rho \rangle_\ell (\bm x) \nonumber \\
    &= \delta \rho (\bm x) - \int \delta \rho(\bm x')  \mathcal{W}_\ell (\bm x', \bm x) \ud^3 x' + \mathcal{O}(\ell^2 / H^2),
\end{align}
and the ratio $\delta \rho_\ell / \langle \rho \rangle_\ell$ is
\begin{align}
    \frac{\delta \rho_\ell (\bm x)}{\langle \rho \rangle_\ell (\bm x)} &= \frac{\delta \rho (\bm x)}{\rho_0 (\bm x)} - \int \frac{\delta \rho(\bm x')}{\rho_0 (\bm x)}   \mathcal{W}_\ell (\bm x', \bm x) \ud^3 x' \nonumber \\ 
    &+ \mathcal{O}(\ell^2 / H^2) + \mathcal{O} (\mathcal{M}^2),
\end{align}
where the leading terms are both of order $\mathcal{O} (\mathcal{M})$.

We can now take the subsonic limit of turbulent kinetic energy, in which we also include the transport of turbulent momentum by the bulk flow (see Eq.~\ref{eq:kin_en_transp}--\ref{eq:kin_en_turb}):
\begin{align}
    \varepsilon_\mathrm{k,cross} + \varepsilon_\mathrm{k,turb} &= \sum_{i} \langle u_i \rangle_\ell \mu_2 (\rho, u_i) + \frac{1}{2} \langle \rho \rangle_\ell \sum_{i} \mu_2(u_i, u_i) \nonumber \\
    &+ \frac{1}{2} \sum_{i} \mu_3(\rho, u_i, u_i). \label{eq:app-turb-energy}
\end{align}
We use the fact that the second- and third-order statistical moments, $\mu_2$ and $\mu_3$, can be rewritten in integral form as \citep{Eyink2007,Hollins2022}:
\begin{align}
    \mu_2 (\rho, u_i) (\bm x) = \int &\left[ \rho (\bm x') - \langle \rho \rangle_\ell (\bm x) \right] \times \nonumber \\
    &\left[ u_i (\bm x') - \langle u_i \rangle_\ell (\bm x) \right] \mathcal{W}_\ell (\bm x', \bm x) \ud^3 x', \\
    \mu_3(\rho, u_i, u_i) (\bm x) = \int &\left[ \rho (\bm x') - \langle \rho \rangle_\ell (\bm x) \right] \times \nonumber \\
    &\left[ u_i (\bm x') - \langle u_i \rangle_\ell (\bm x) \right]^2 \mathcal{W}_\ell (\bm x', \bm x) \ud^3 x',
\end{align}
(note the different arguments $\bm x$ and $\bm x'$ in the square brackets). Replacing as before $\rho (\bm x') = \rho_0 (\bm x') + \delta \rho (\bm x')$ (Eq.~\ref{eq:subsonic-expansion}) and expanding $\rho_0 (\bm x')$ in Taylor series around $\bm x$, we obtain
\begin{align}
    &\mu_2 (\rho, u_i)(\bm x)  \simeq \int \delta \rho (\bm x') \left[ u_i (\bm x') - \langle u_i \rangle_\ell (\bm x) \right] \mathcal{W}_\ell (\bm x', \bm x) \ud^3 x' , \label{eq:turb-mom-subsonic} \\
    &\mu_3(\rho, u_i, u_i) (\bm x) \simeq \int \delta \rho (\bm x') \left[ u_i (\bm x') - \langle u_i \rangle_\ell (\bm x) \right]^2 \mathcal{W}_\ell (\bm x', \bm x) \ud^3 x' \nonumber \\
    &- \left( \int \delta \rho (\bm x') \mathcal{W}_\ell (\bm x', \bm x) \ud^3 x' \right) \times \mu_2 (u_i, u_i) \label{eq:mu3-subsonic},
\end{align}
where we neglected terms of order $\mathcal{O}(\ell / H)$. By simple inspection, we can see that the terms in Eq.~\eqref{eq:turb-mom-subsonic}--\eqref{eq:mu3-subsonic} are both order $\mathcal{O}(\mathcal{M})$ smaller than the leading term in Eq.~\eqref{eq:app-turb-energy}:
\begin{align}
\frac{1}{2} \langle \rho \rangle_\ell \sum_{i} \mu_2(u_i, u_i) \simeq \frac{1}{2} \rho_0 (\bm x) \sum_{i} \mu_2(u_i, u_i) ,
\end{align}
which we identify with the turbulent kinetic energy in the subsonic regime.

\subsection{Turbulent pressure in the homogeneous subsonic limit}\label{app:alternative-turb-press}

In the filtering approach we define the mean turbulent kinetic pressure, $p_{\mathrm{turb}}$, over a volume, $V$, as: 
\begin{align}
        p_{\mathrm{turb}} \equiv \frac{2}{3} \frac{\mathcal{E}_{\mathrm{k,turb}}}{V} = \frac{2}{3} \frac{1}{V} \int_V \varepsilon_{\mathrm{k,turb}} (\bm x') \ud^3 x'.
\end{align}
Assuming subsonic turbulence ($\mathcal{M} \ll 1$), the above expression can be approximated as
\begin{align}
    p_{\mathrm{turb}} \simeq \frac{1}{3} \frac{1}{V} \int_V \rho_0 (\bm x') \sigma_{\mathrm{3D},u}^2 (\bm x')  \ud^3 x'
\end{align}
where we used the subsonic limit of the turbulent energy density in Appendix~\ref{app:subsonic-regime}, and defined the three-dimensional local velocity dispersion, $\sigma_{\mathrm{3D},u}$, as $\sigma_{\mathrm{3D},u}^2 (\bm x) = \sum_{i} \mu_2 (u_i, u_i)$ (we drop the subscript ``$u$'' for readability). If the turbulence is homogeneous in the volume, $V$, the background quantities and the statistical properties of the fields do not depend on $\bm x$. This implies that $ \rho_0 \simeq \text{const}$, and the turbulent kinetic pressure reduces to
\begin{align}
        p_{\mathrm{turb}} \simeq \frac{1}{3} \rho_0 \sigma_{\mathrm{3D},\mathrm{rms}}^2 ,
\end{align}
where we defined the rms velocity dispersion, $\sigma_{\mathrm{3D},\mathrm{rms}}$, as 
\begin{align}
    \sigma_{\mathrm{3D},\mathrm{rms}}^2 \equiv \frac{1}{V} \int_V  \sigma_{\mathrm{3D}}^2 (\bm x')  \ud^3 x'.
\end{align}
If the turbulence is isotropic, then the volume averages of $\mu_2(u_x, u_x)$, $\mu_2(u_y, u_y)$ and $\mu_2(u_z, u_z)$ are the same, and we can define a one-dimensional rms velocity dispersion,
$\sigma_{\mathrm{1D},\mathrm{rms}}$, as
\begin{align}
        \sigma_{\mathrm{1D},\mathrm{rms}}^2 \equiv  \frac{1}{3} \sigma_{\mathrm{3D},\mathrm{rms}}^2,
\end{align}
whereby the turbulent pressure becomes $p_{\mathrm{turb}} \simeq \rho_0 \sigma_{\mathrm{1D},\mathrm{rms}}^2$.

\section{Contribution of shocks to the turbulent energies}\label{app:shocks_contribution}

  In this Appendix we estimate the contribution from shocks to the turbulent kinetic energy, and we compare the results to the baseline approach that we used in the Sect.~\ref{sec:turbulent_energies_application}. We employ two methods to remove shocks, both based on the on-the-fly shock finder included in \textsc{Arepo}  \citep{Schaal2015}:
\begin{enumerate}
    \item Method (1) consists in simply adding a weighing function in the convolution (a shock mask) that is 0 where shocks are detected, and 1 elsewhere. This means that all the shock cells\footnote{In simulations, shocks are numerically broadenend and exhibit several cells along the shock propagation direction with a negative velocity divergence. Hence, to be conservative in method (1), we mask all cells labeled ``shock zones'' in the terminolgy of \citet{Schaal2015}.} that are detected within the filter radius are zeroed out and ignored in the convolution. The shocked regions are then also neglected for the computation of turbulent energies. This method is the simplest but will still include cells ``behind'' the shock, which may have an increased velocity dispersion as pointed out, for example, in \citet{Vazza2012}.
    \item Method (2) removes the contribution of shocked cells as Method (1) but additionally zeroes out the entire convolution integral if at least one shocked cell is found within $1 \ell$ of the cell where we are performing the convolution (for this test we only use the Gaussian kernel, which extends up to $4 \ell$). In other words: we carve out a buffer region of size $\ell$ around all shocks, and we neglect this region entirely for the computation of quantities such as the turbulent energies.
\end{enumerate}

\begin{figure*}
    \centering
    \includegraphics[trim={0 0cm 0 0},clip,width=1.0\textwidth]{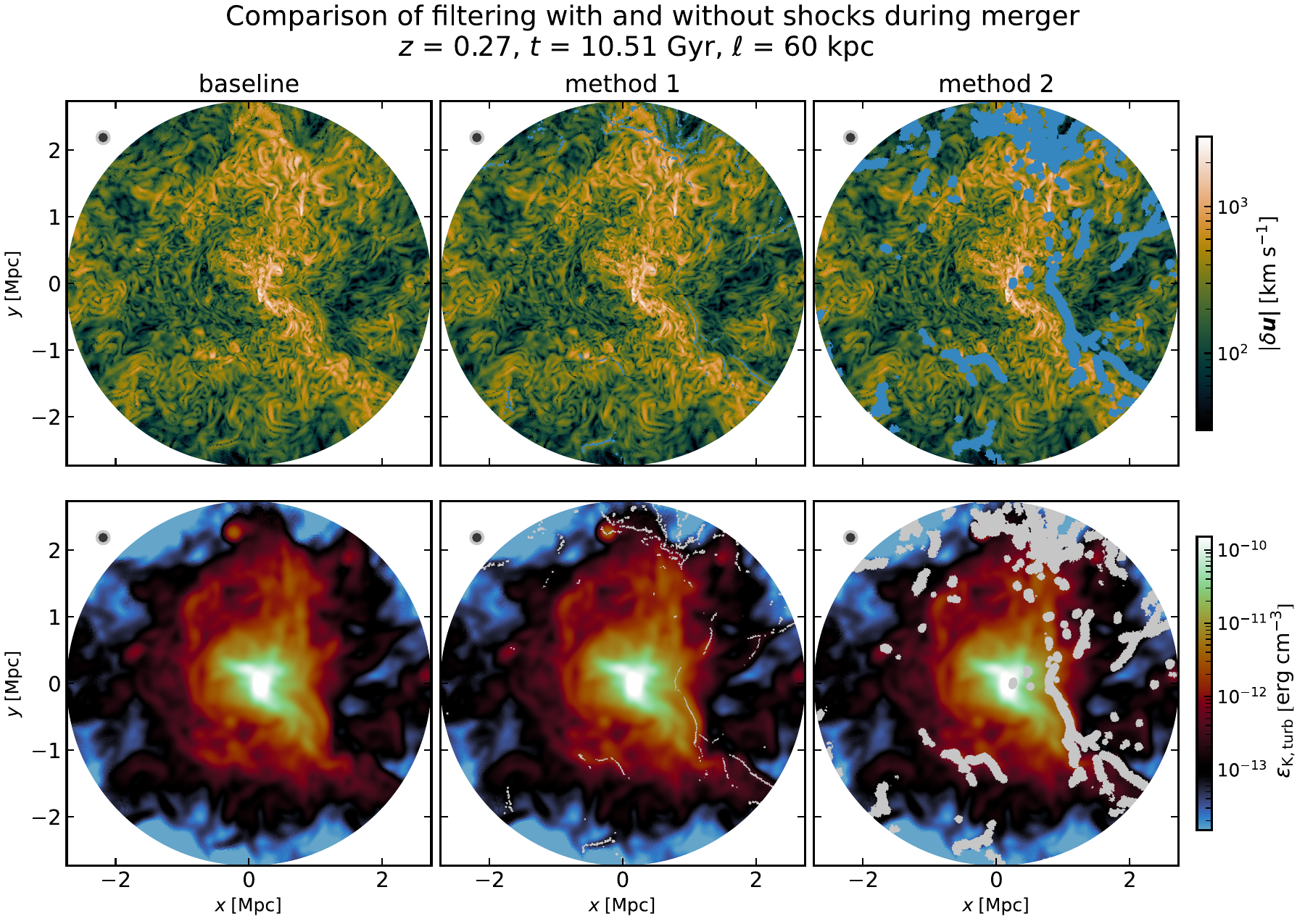}
        \caption{Comparison of turbulent velocities and turbulent kinetic energy densities with shocks (left, our baseline) and without shocks (middle and right columns, for Method 1 and Method 2, respectively). In cyan and gray we indicate the regions excluded from the computation of the filtered energies because of their proximity to shocks. We plot for comparison in the top left corner a black (gray) circle with radius $\ell = 60 ~ \si{kpc}$ ($\sigma_\ell = 100 ~ \si{kpc}$).}
        \label{fig:comparison_shock_noshock}
\end{figure*}

We show the turbulent velocity and the turbulent kinetic energy density, $\varepsilon_{\mathrm{k,turb}}$, obtained with the two methods in Fig.~\ref{fig:comparison_shock_noshock}, where we compare them with the baseline approach. We focus on the snapshot corresponding to the core passage (``during merger,'' Fig.~\ref{fig:during_all_fields}) because we expect more shocks within $R_{200,\mathrm{c}}$. In red/gray we show the regions corresponding to shocks (method 1), and in their proximity (method 2). We note already that the regions where $\varepsilon_{\mathrm{k,turb}}$ is largest are not necessarily near the merger shocks. Shocks are in general thin surfaces and thus only account for about $1\%$ of the volume within $R_{200,\mathrm{c}}$ at the core passage. Considering instead the buffer region of size $\ell$, about $15\%$ of the volume is discarded for the largest filter length examined (see Table~\ref{tab:1}). 

To quantify the contribution of shocks to the turbulent energies we integrate  $\varepsilon_{\mathrm{k,turb}}$ over the valid volume, $V_{\mathrm{valid}}$, (i.e., non-discarded cells) for each method, and we then divide by $V_{\mathrm{valid}}$ so as to allow for a fair comparison with the baseline approach. If we just compared at face-value the volume-integrated energies, $\mathcal{E}_{\mathrm{k,turb}}$, then we would have to somehow account for the missing volume (e.g., in the case of Method 2 with $\ell = 60$ kpc, the discarded volume is $15\%$ within $R_{200,\mathrm{c}}$), which would obviously lead to lower $\mathcal{E}_{\mathrm{k,turb}}$ when neglecting shocks. By dividing by the valid volume for each method, we can estimate instead how much more shocks contribute to the mean turbulent energy, $\mathcal{E}_{\mathrm{k,turb}} / V_{\mathrm{valid}}$. The results of this comparison are shown in Table~\ref{tab:2}, from which we see that, even with the largest filter length ($\ell = 60$ kpc), the mean turbulent energy per volume changes by only $\simeq 2 \%$ compared to the baseline. The results are similar with the other filter lengths. This means that regions in proximity of shocks contribute only marginally more to the turbulent energies compared to the average.

One last useful metric to compare is the turbulent pressure fraction within $R_{200,\mathrm{c}}$ as defined in Eq.~\eqref{eq:turbulent_pressure_fractions}. Since this quantity depends only on the ratio between turbulent and thermal pressures, it does not matter what volume to choose for the computation of these quantities as long as it is the same for both. We report the turbulent pressure fractions in Table~\ref{tab:3}, where we see that with the largest filter length ($\ell = 60$ kpc) and with shock removal with Method 2 the turbulent pressure fraction is decreased by about $7\%$ compared to the baseline. The decrease is less pronounced for smaller filter lengths. From these tests we conclude that shocks contribute only marginally to the turbulent kinetic energies and turbulent pressure fractions.

\begin{table}
\centering
\caption{Volumes discarded with shock removal because they are either shocks (method 1) or in their proximity (method 2).}
\begin{tabular}{p{1.8cm} |p{1.5cm} p{2cm} p{2cm}}
filter length $\ell$ [kpc] & baseline   & method (1)   & method (2)   \\
\hline
15 & - & 0.011 & 0.018 \\
30 & - & 0.011 & 0.055 \\
60 & - & 0.011 & 0.151
\end{tabular}
\tablefoot{The volumes quoted are relative to the volume of the sphere with radius $R_{200,\mathrm{c}}$ for the snapshot at $z=0.27$, $t= 10.5$ Gyr.}
\label{tab:1}
\end{table}

\begin{table}
\centering
\caption{Mean turbulent energy per volume ($10^{-13}$ erg cm$^{-3}$).}
\begin{tabular}{p{1.8cm} |p{1.5cm} p{2cm} p{2cm}}
filter length $\ell$ [kpc] & baseline   & method (1)   & method (2)   \\
\hline
15         & 1.84 & 1.84 & 1.79 \\
30         & 3.98 & 3.99 & 3.86 \\
60         & 7.44 & 7.48 & 7.27
\end{tabular}
\tablefoot{These values have been derived from the simulation snapshot at $z=0.27$, $t= 10.5$ Gyr.}
\label{tab:2}
\end{table}

\begin{table}
\centering
\caption{Turbulent kinetic pressure fraction $f_{\mathrm{k,turb}}$ with and without shock removal in units of per cent.}
\begin{tabular}{p{1.8cm} |p{1.5cm} p{2cm} p{2cm}}
filter length $\ell$ [kpc] & baseline   & method (1)   & method (2)   \\
\hline
15 & 2.00 & 1.98 & 1.94 \\
30 & 4.22 & 4.20 & 4.03 \\
60 & 7.62 & 7.60 & 7.05
\end{tabular}
\tablefoot{These values have been derived from the simulation snapshot at $z=0.27$, $t= 10.5$ Gyr.}
\label{tab:3}
\end{table}

\section{Filtered energies in periodic and nonperiodic domains}\label{app:turbulent-energy-subvolumes}

In this Appendix, we show the connection between the filtered energies and the power spectrum. We first consider the simple case of periodic boxes. We then extend the results to nonperiodic volumes. We focus on the magnetic field since its power spectrum involves a quadratic product, but the following arguments are formally the same for the velocity field in incompressible turbulence, whose power spectrum is also quadratic. 

\subsection{Periodic volumes}\label{app:turb_periodic_volumes}

We expand the magnetic field in a Fourier series in $V$:
\begin{align}
        \bm B (\bm x) = \sum_{\bm k}  \hat{ \bm B}_{\bm k} e^{i \bm k \bcdot \bm x }, \label{eq:app-fourier-B}
\end{align}
To obtain the smooth field, $\langle \bm B \rangle_\ell$, we convolve Eq.~\eqref{eq:app-fourier-B} with a smoothing kernel, $\mathcal{W}_\ell$, which yields
\begin{align}
        \langle \bm B \rangle_\ell = V \sum_{\bm k}  \hat{ \bm B}_{\bm k} e^{i \bm k \bcdot \bm x } \hat{\mathcal{W}}_{\ell, \bm k}, \label{eq:filtered_magnetic_field}
\end{align}
where $\hat{\mathcal{W}}_{\ell, \bm k}$ is the Fourier transform of the smoothing kernel and satisfies (see Appendix~\ref{app:filters_fourier_space})
\begin{align}
        \hat{\mathcal{W}}_{\ell, \bm 0} = 1 / V, \hspace{2em}  \lim_{k \ell \rightarrow \infty} \hat{\mathcal{W}}_{\ell, \bm k} = 0.
\end{align}
For convenience we absorb the factor of $V$ and define $\tilde{\mathcal{W}}_{\ell, \bm k} =  \hat{\mathcal{W}}_{\ell, \bm k} V$. To derive the turbulent energy density, $\varepsilon_{B,\mathrm{turb}}$, we use its definition in Eq.~\eqref{eq:mag_energy_decomposition}, for which we need to compute $\langle B^2 \rangle_{\ell}$ and $\langle \bm B \rangle_\ell^2$. Using the expansion in Fourier series (Eq.~\ref{eq:synthetic_magnetic_field}), the first term can be written as:
\begin{align}
    \langle B^2 \rangle_\ell &= \left\langle \sum_{\bm k} |\hat{ \bm B}_{\bm k}|^2 + \sum_{\bm k \neq \bm k'}   \hat{ \bm B}^*_{\bm k} \bcdot \hat{ \bm B}_{\bm k'} e^{i (\bm k' - \bm k) \bcdot\bm x }  \right\rangle_\ell \nonumber \\
    &= \sum_{\bm k} |\hat{ \bm B}_{\bm k}|^2 + \sum_{\bm k \neq \bm k'}  \hat{ \bm B}^*_{\bm k} \bcdot \hat{ \bm B}_{\bm k'} e^{i (\bm k' - \bm k) \bcdot\bm x }  \tilde{\mathcal{W}}_{\ell, \bm k' - \bm k} , \label{eq:filtered_B2}
\end{align}
where we applied again the convolution theorem (Eq.~\ref{eq:convolution-theorem}) to filter a sum of plane waves, in which each wave is filtered by multiplying it by $\tilde{\mathcal{W}}_{\ell, \bm k' - \bm k}$. Taking now the square of Eq.~\eqref{eq:filtered_magnetic_field} and subtracting it from Eq.~\eqref{eq:filtered_B2}, we obtain the turbulent energy density of the magnetic field at every point in space:
\begin{align}\label{eq:turbulent_magnetic_en_fourier}
        &8 \pi \varepsilon_{B,\mathrm{turb}} (\bm x) = \sum_{\bm k}  |\hat{ \bm B}_{\bm k}|^2 \left( 1 -  |\tilde{\mathcal{W}}_{\ell, \bm k}|^2 \right)  \\
        & + \sum_{\bm k \neq \bm k'}  \hat{ \bm B}^*_{\bm k} \bcdot \hat{ \bm B}_{\bm k'} e^{i (\bm k' - \bm k) \bcdot\bm x }  \left[\tilde{\mathcal{W}}_{\ell, \bm k' - \bm k} - \tilde{\mathcal{W}}_{\ell, \bm k} \tilde{\mathcal{W}}_{\ell, \bm k'} \right] . \nonumber
\end{align}
Equation~\eqref{eq:turbulent_magnetic_en_fourier} consists of a constant term (the first), and spatially varying term (the second). The constant term is a weighted sum of the energy in different Fourier modes, while the spatially varying cross term represents the interference between different wave vectors, $\bm k \neq \bm k'$.
Since the spatially varying term is a sum of complex exponentials, it vanishes exactly when integrated over the entire periodic volume, $V$. Thus, after integration we are left with 
\begin{align}\label{eq:turbulent_energy_integrated_appendix}
        \mathcal{E}_{B,\mathrm{turb}} = \sum_{\bm k}  \frac{|\hat{\bm B}_{\bm k}|^2}{8 \pi}  \left( 1 -  |\tilde{\mathcal{W}}_{\ell, \bm k}|^2 \right) V, 
\end{align}
which is a weighted sum of the energy in the Fourier modes, where low $k \ell \ll 1$ modes are suppressed, while $k \ell \gg 1$ modes are retained.

\begin{figure}
        \begin{subfigure}{1.0\linewidth}
                \centering
                \includegraphics[trim={0 1cm 0 0},clip,width=0.9\linewidth]{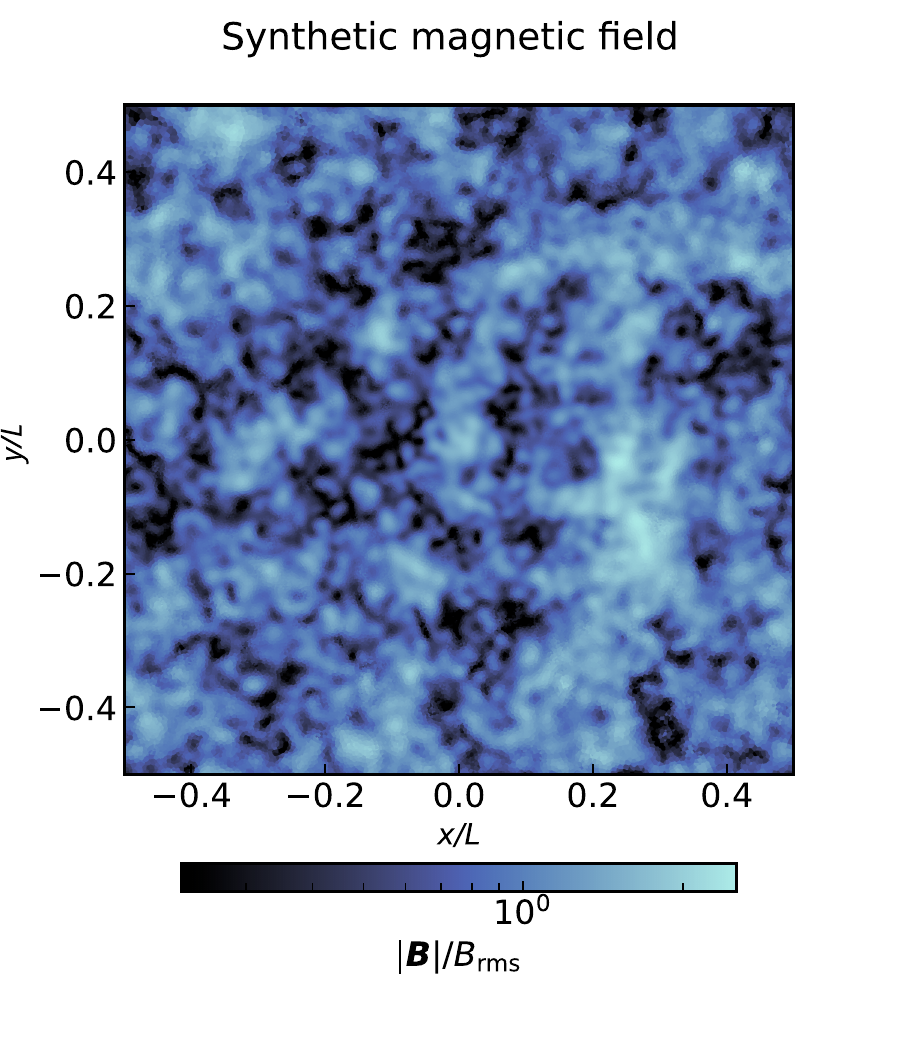}
                \caption{}
                \label{fig:turbulent_mag_field}
        \end{subfigure}
        \begin{subfigure}{1.0\linewidth}
                \centering
                \includegraphics[width=1.0\linewidth]{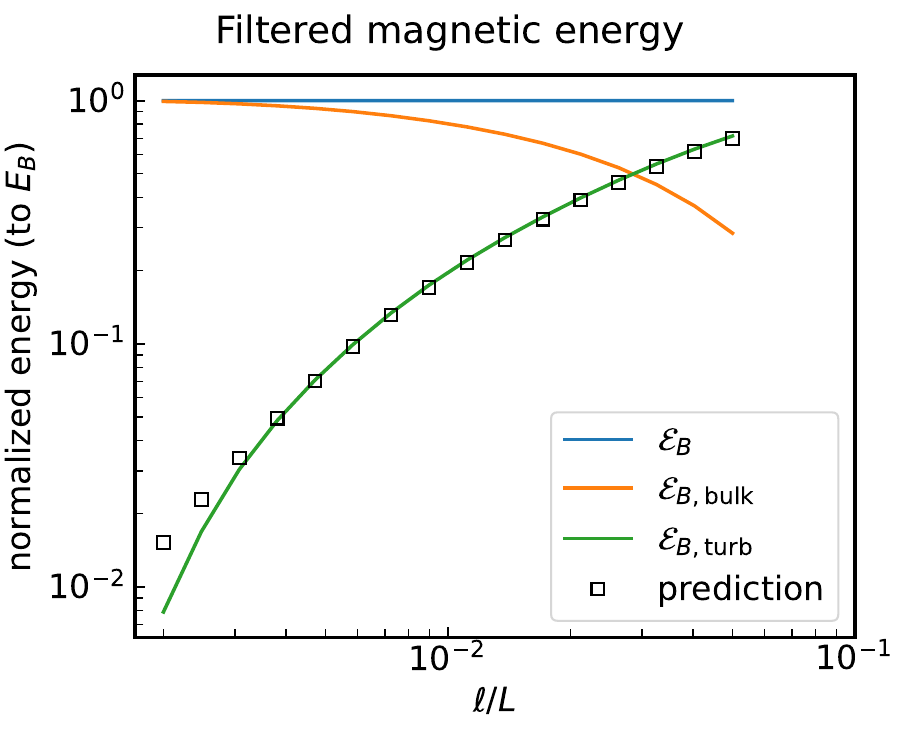}
                \caption{}
                \label{fig:turbulent_magnetic_energy}
        \end{subfigure}
        \caption{Decomposition of the magnetic energy according to Eq.~\eqref{eq:mag_energy_decomposition} as a function of the filtering length $\ell$ after integrating over the volume. \textit{Top}: Two-dimensional slice of a three-dimensional Kolmogorov synthetic magnetic field. \textit{Bottom}: Filtered total energy of the synthetic turbulent field (blue line), energy of the smooth component (orange), and energy of the turbulent field (green). The black squares show the analytical result of Eq.~\eqref{eq:turbulent_energy_integrated_appendix} obtained by applying the Gaussian filter with scale $\ell$ to the Fourier components of the synthetic field, and computing the energy of the modes. 
        }
        \label{fig:magnetic_energy_vs_filter_length}
\end{figure}

We use Eq.~\eqref{eq:turbulent_energy_integrated_appendix} to numerically verify our algorithm. We initialize a synthetic turbulent magnetic field that follows a Kolmogorov spectral energy density, $E(k) \propto k^{-5/3}$, on a nonuniform three-dimensional Voronoi mesh (Fig.~\ref{fig:turbulent_mag_field}). We then compute the total, bulk and turbulent filtered energy densities using a Gaussian kernel for different filter sizes, $\ell$, and integrate them over the periodic volume. The resulting filtered energies are shown in Fig.~\ref{fig:turbulent_magnetic_energy}, together with the analytical prediction from the power spectrum of the fluctuations (Eq.~\ref{eq:turbulent_energy_integrated_appendix}). The two measures agree very well, except for small $\ell$, where the finite resolution of the mesh plays a role. We note that, by construction, with a Gaussian kernel the turbulent energy (bulk energy) monotonically increases (decreases) with the filter size, $\ell$, as more of the Fourier modes get progressively classified as ``turbulence.'' This is not necessarily the case if the Fourier amplitude of the kernel is not monotonic (as with the spherical top-hat; see Appendix~\ref{app:filters_fourier_space}).
In the case of the kinetic energy, even with a monotonic kernel, it is not possible to predict a priori the behavior of the turbulent and bulk energies with $\ell$ because  the contribution from triadic interactions of small- and large-scale wave vectors can be either positive or negative. If the flow is highly subsonic, then the same considerations outlined above for the magnetic energy apply.

\subsection{Nonperiodic volumes}

For simplicity, we consider smaller Cartesian sub-volumes, $V^{(i)}$, of a larger periodic parent box, $V$, which is the more likely scenario when analyzing cosmological simulations. 

The steps leading up to Eq.~\eqref{eq:turbulent_magnetic_en_fourier} are the same as in the previous section. By contrast, we now integrate $\varepsilon_{B,\mathrm{turb}}$ over a sub-volume, $V^{(i)}$, of cubic shape contained in $V$, defined by its center, $\bm r^{(i)}$, and extension, $L_x^{(i)}$, $L_y^{(i)}$ and $L_z^{(i)}$ in the three dimensions, respectively. 
Integration of the turbulent energy density now yields two terms: one resulting from the first term (a constant)
\begin{align}
        &\left. \mathcal{E}_{B,\mathrm{turb}} (\bm r^{(i)}; V^{(i)})\right|_1 = \sum_{\bm k}  \frac{|\hat{\bm B}_{\bm k}|^2}{8 \pi}  \left( 1 -  |\tilde{\mathcal{W}}_{\ell, \bm k}|^2 \right) V^{(i)},
\end{align}
and a contribution from the second term which is now non-vanishing (in general), and depends on the location of the sub-volume:
\begin{align}
        &\left. \mathcal{E}_{B,\mathrm{turb}} (\bm r^{(i)}; V^{(i)}) \right|_2 = \sum_{\bm k \neq \bm k'}  \frac{\hat{ \bm B}^*_{\bm k} \bcdot \hat{ \bm B} _{\bm k'}}{8\pi} \left[\tilde{\mathcal{W}}_{\ell, \bm k' - \bm k}  - \tilde{\mathcal{W}}_{\ell, \bm k} \tilde{\mathcal{W}}_{\ell, \bm k'} \right] \times \nonumber \\
    &  \left[  \prod_{j=x,y,z} \sinc \left( \frac{(\bm k' - \bm k)_j L_j^{(i)}}{2} \right) \right] \cos \left((\bm k' - \bm k) \bcdot \bm r^{(i)}\right)  V^{(i)},  \label{eq:turb-mag-en-subvolume}
\end{align}
where we introduced the sinc function as $\sinc (z) = \sin(z)/z$. The second term represents interference between different modes and can be either constructive or destructive

In the interference term in Eq.~\eqref{eq:turb-mag-en-subvolume} both
large-scale modes (with respect to $\ell$) and small-scale modes can contribute. However, based on some general considerations, we can show that it is the small-scale modes that give the largest contribution. We focus our attention on the attenuation term due to filtering:
\begin{align}
    \tilde{\mathcal{W}}_{\ell, \bm k' - \bm k}  - \tilde{\mathcal{W}}_{\ell, \bm k} \tilde{\mathcal{W}}_{\ell, \bm k'}. \label{eq:attenuation}
\end{align}
We can distinguish three cases: (i) both wave vectors in the double summation are large-scale, $|\bm k| \sim |\bm k'| \ll 1/\ell$; (ii) both small-scale, $|\bm k| \sim |\bm k'| \gg 1/\ell$; (iii) one is large-scale and one small-scale ($| \bm k | \ll 1/\ell \ll | \bm k' |$ without loss of generality). We examine them below in turn:
\begin{enumerate}[ {(}i{)} ]
    \item when both wave vectors are large-scale, $\tilde{\mathcal{W}}_{\ell, \bm k} \approx \tilde{\mathcal{W}}_{\ell, \bm k'} \approx 1$, and their difference is also small: $| \bm k' - \bm k| \ll 1/\ell$; this implies $\tilde{\mathcal{W}}_{\ell, \bm k' - \bm k} \approx 1$ and the attenuation term (Eq.~\ref{eq:attenuation}) is approximately zero, and thus large-scale modes contribute little;
    \item when both wave vectors are small-scale, $\tilde{\mathcal{W}}_{\ell, \bm k} \approx \tilde{\mathcal{W}}_{\ell, \bm k'} \approx 0$, but $\tilde{\mathcal{W}}_{\ell, \bm k' - \bm k}$ can still be approximately 1 if $\bm k' \approx \bm k$ (although they cannot be exactly equal), and contribute to the interference term; 
    \item the final case where wave vectors are mixed also does not contribute significantly: the product $\tilde{\mathcal{W}}_{\ell, \bm k} \tilde{\mathcal{W}}_{\ell, \bm k'}$ is approximately zero (since $\tilde{\mathcal{W}}_{\ell, \bm k'} \approx 0$, and $| \bm k' - \bm k| \approx |\bm k'| \gg 1/\ell$, and thus $\tilde{\mathcal{W}}_{\ell, \bm k' - \bm k} \approx 0$. Therefore  mixed modes also do not contribute significantly.
\end{enumerate}
To summarize, we expect most of the contribution to the interference term in the turbulent energy to come from small-scale modes: this makes physical sense and agrees with the notion that, for a double product such as the magnetic energy, or the kinetic energy in the subsonic limit, only fluctuations on scales smaller than $\ell$ should enter in the turbulent energy (for compressible turbulence, contributions from mixed large- and small-scale modes are in general unavoidable).  In particular, we could imagine that modes with very similar wave vectors are more correlated than those with dissimilar wave vectors; for these correlated modes, there would be little cancellation in the double summation in Eq.~\eqref{eq:turb-mag-en-subvolume}, leading to a large contribution to the turbulent energy. If, on the other hand, the Fourier modes are uncorrelated, the interference term would be minimized.

\begin{figure}
        \centering
        \includegraphics[trim={0 1cm 0 0},clip,width=0.9\linewidth]{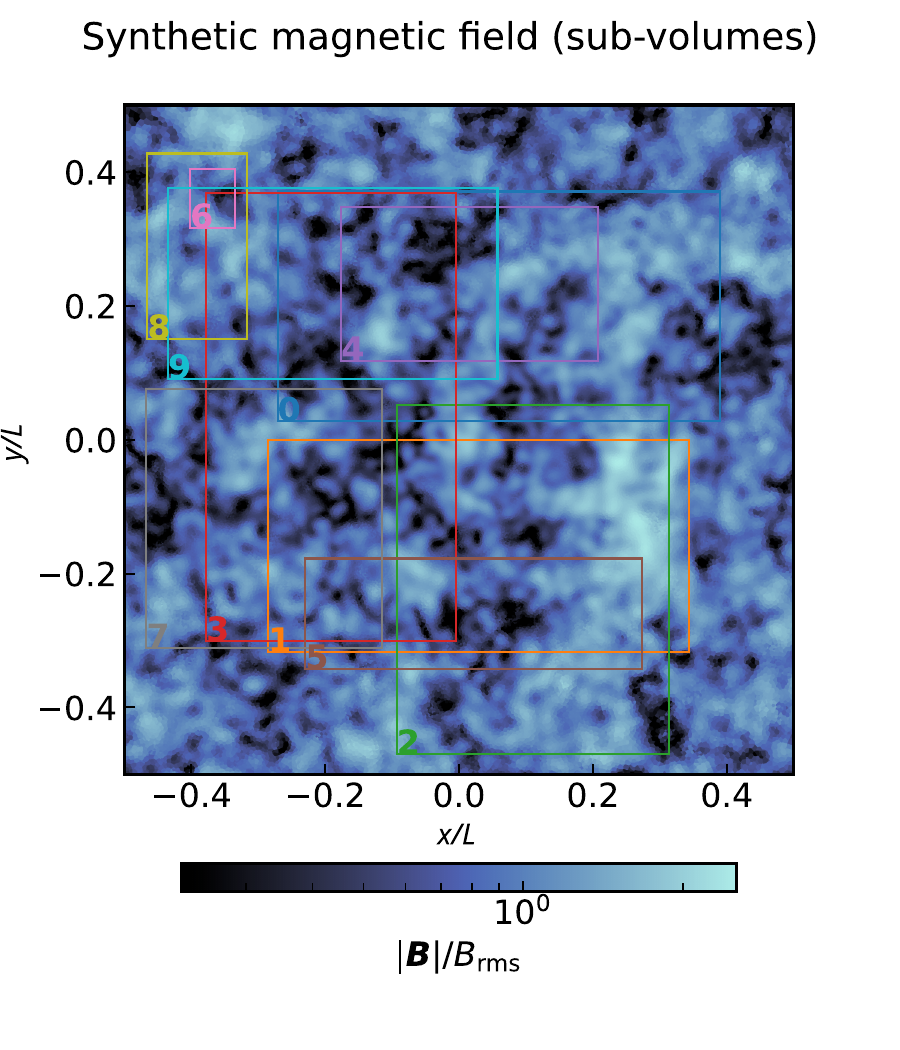}
        \caption{Two-dimensional slice of a three-dimensional turbulent synthetic magnetic field (same as in Fig.~\ref{fig:turbulent_mag_field}). Overplotted are the cross sections of the sub-volumes as seen in projection.
        }
        \label{fig:2a.magnetic_energy_fixed_filter_regions}
\end{figure}

\begin{figure}
        \centering
        \includegraphics[width=1.0\linewidth]{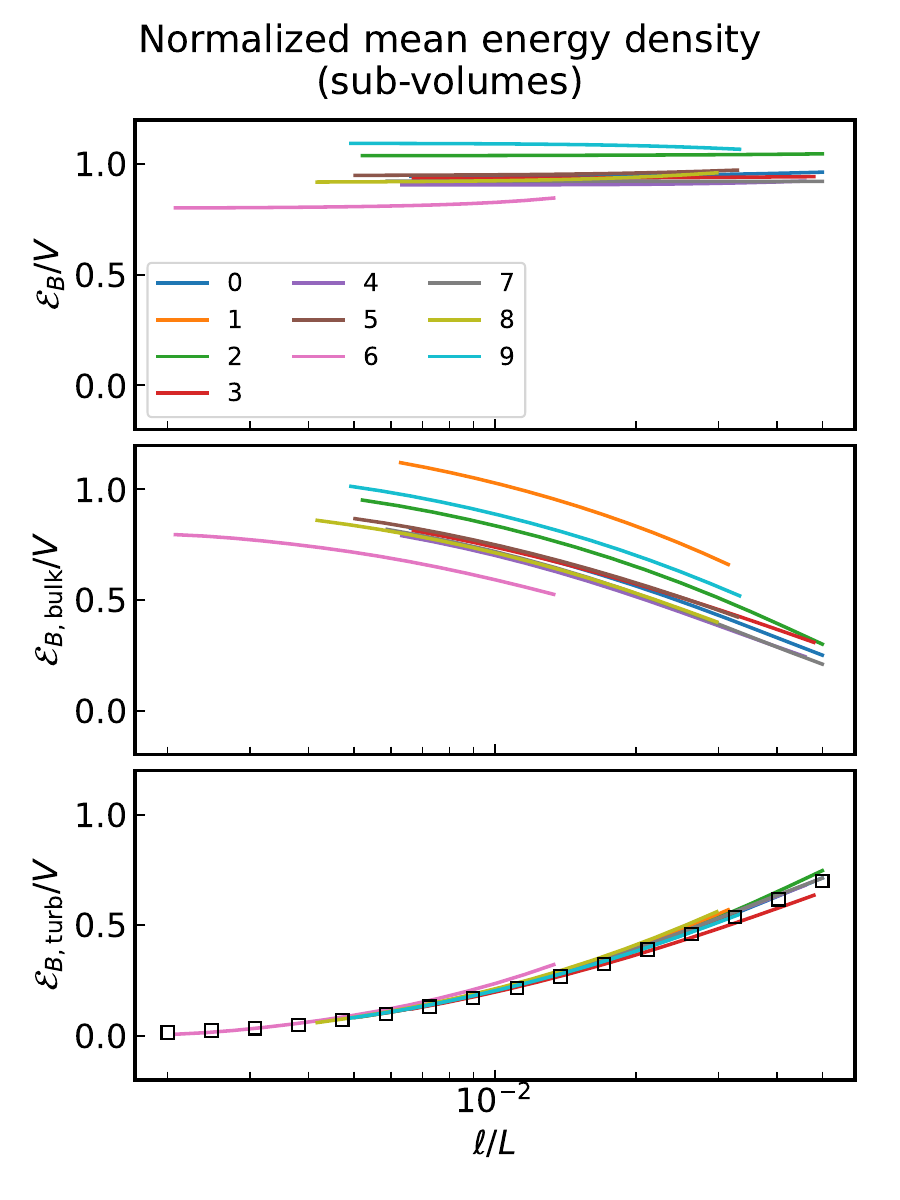}
        \caption{Filtered total (\textit{top}), bulk (\textit{middle}), and turbulent (\textit{bottom}) magnetic energies in small nonperiodic sub-volumes (see Fig.~\ref{fig:2a.magnetic_energy_fixed_filter_regions} for a slice of the turbulent magnetic field and a cross section of the sub-volumes). Since the Fourier modes are uncorrelated, the interference term due to nonperiodic boundaries is very small, and the mean turbulent energy density is approximately the same in all sub-volumes.  
        }
        \label{fig:2a.magnetic_energy_vs_filter_length_smaller_subregion_many}
\end{figure}

We can test our hypothesis by initializing a synthetic magnetic field as in Sect.~\ref{sec:turb-en-power-spectr}, whose power spectrum follows a known power-law (in this case Kolmogorov), and with uncorrelated Fourier amplitudes; see Fig.~\ref{fig:2a.magnetic_energy_fixed_filter_regions}. We then randomly select 10 nonperiodic Cartesian sub-volumes, $V^{(i)}$, with $i=0,\ldots,9$, and compute the total, $\mathcal{E}_{B}^{(i)}$, bulk, $\mathcal{E}_{B,\mathrm{bulk}}^{(i)}$ and turbulent filtered energies, $\mathcal{E}_{B,\mathrm{turb}}^{(i)}$, for different filter scales, $\ell$, which we plot in Fig.~\ref{fig:2a.magnetic_energy_vs_filter_length_smaller_subregion_many}. Since the sub-volumes are of various sizes, the total energy in each of them can in principle be different. However, because the magnetic field is statistically homogeneous, we expect the mean energy density in each sub-volume, $\mathcal{E}_{B}^{(i)}/ V^{(i)}$, to scatter around the mean value in $V$. On the other hand, if the Fourier modes are indeed uncorrelated, we predict the mean turbulent magnetic energy density in each sub-volume, $\mathcal{E}_{B,\mathrm{turb}}^{(i)}/ V^{(i)}$, to be the same for all sub-volumes for each $\ell$ according to Eq.~\eqref{eq:turb-mag-en-subvolume}, and equal to the energy contained in the power spectrum, with large-scale modes suppressed by the smoothing kernel:
\begin{align}
    \frac{\mathcal{E}_{B}^{(i)}}{V^{(i)}} \simeq \sum_{\bm k}  \frac{|\hat{\bm B}_{\bm k}|^2}{8 \pi}  \left( 1 -  |\tilde{\mathcal{W}}_{\ell, \bm k}|^2 \right). \label{eq:mean-turb-den-expect-subvol}
\end{align}
Indeed, by looking at Fig.~\ref{fig:2a.magnetic_energy_vs_filter_length_smaller_subregion_many}, we see that the turbulent mean energy densities closely follow the theoretical expectation based on Eq.~\eqref{eq:mean-turb-den-expect-subvol}, despite the larger scatter of the total and bulk energy densities.

\section{The \textsc{Turbocluster} library}\label{app:numerical_methods}

We implemented the spherical top-hat and Gaussian smoothing filters and the computation of the filtered energies in a new Python library called \textsc{Turbocluster}, which uses NVIDIA graphical processing units (GPUs) to speed up computations using the \textsc{Numba} library. Our code partially depends on \textsc{Paicos}, which is a Python software package for visualizing and analyzing \textsc{Arepo} simulations, but it is written in a way that it can easily be adapted to any simulation data that can be read in as a set of cell positions, volumes and associated physical quantities to be smoothed (e.g., density, velocity field, \ldots). At the moment \textsc{Turbocluster} runs on a single NVIDIA GPU, and is limited by its memory (40 GB for the NVIDIA A100 used in this work), since the data has to physically fit on the GPU before being filtered. If the volume to be filtered is too large, the domain can be split in smaller sub-regions to be processed one at a time, leading however to longer execution times. For a vector field in double precision, the maximum size of the domain that can be filtered in one pass with our A100 GPU is $\approx$$3\times 10^8$ cells (including the position coordinates and several scratch arrays that are necessary for the Cartesian tiling; see below for further details). We plan to release the \textsc{Turbocluster} library publicly in the near future.

\subsection{Neighbor search with Cartesian tiling}

An \textsc{Arepo} snapshot for a simulation with $N$ gas cells contains their positions,
but not their spatial connectivity. The main computational burden of applying a smoothing filter is thus to select the cells inside the spherical search radius, $r_{\mathrm{s}}$, of a cell ($r_{\mathrm{s}}$ is related to the spherical top-hat and Gaussian smoothing length, $\ell$, by $r_{\mathrm{s}} = \ell$ and $r_{\mathrm{s}} = 4\ell$, respectively). In computer science, this problem is known as a range search, which is related to the nearest neighbor problem.
It is well known that the brute-force approach for finding the nearest neighbor of a point in space requires $\mathcal{O} (N^2)$ distance comparisons, but spatial trees can reduce the algorithmic complexity to $\mathcal{O} (N \mathrm{\log} N)$. We have opted for a simpler approach, i.e., we select a cubic region with side length $L$ that encompasses a region of interest (e.g., the volume within $R_{200,\mathrm{c}}$) and divide it into $M^3$ cubic sub-regions with side lengths $d=L/M$ that we call ``tiles.'' Using this Cartesian tiling, we index and sort the cells based on the tile in which their mesh-generating points are located. This indexing has algorithmic complexity $\mathcal{O} (N)$ and the sorting, which uses an implementation of radix sort from the Thrust library (called via \textsc{Cupy}), is also extremely fast in practice. With this approach, the number of distance comparisons necessary for smoothing filters is reduced significantly (see below for further details).   

For a set of uniformly distributed points\footnote{In cosmological simulations with moving-mesh or smoothed particle-hydrodynamics codes, the gas cell positions are not uniformly distributed; rather, they follow the gas density distribution imposed by large-scale structure and galaxy formation.} with mean number density of cells, $n = N / L^3$, each tile contains on average $N_\mathrm{t} = N/M^3 = n d^3$ points. We then introduce a search efficiency (or hit ratio), $\eta$, measured as the number of cells found inside the search radius divided by the number of distance comparisons.
For example, with the brute-force approach, we need $\approx N^2$ evaluations, so the efficiency is
\begin{align}
        \eta = \frac{N (<r_{\mathrm{s}})}{N^2} \approx \frac{4.19}{N} \left(\frac{r_{\mathrm{s}}}{L}\right)^3,
\end{align}
where we used that, on average, $N (<r_{\mathrm{s}})  = 4\pi r_{\mathrm{s}}^3 n /3 $. 
With Cartesian tiling, for a given search radius $r_{\mathrm{s}}$, we only need to search for neighbors in the tiles that overlap (all or in part) with a sphere of radius, $r_{\mathrm{s}}$ centered on the point of interest, $\bm x_0$. For small search radii ($\leq d/2$), the worst case scenario is when $\bm x_0$ is located near the corner of a tile, in which case multiple tiles have to be checked. In this case, one needs to make at most $8 N_t$ evaluations to find the cells contained inside $r_{\mathrm{s}}$ and the worst case efficiency  is 
\begin{align}
        \eta = \frac{N (<r_{\mathrm{s}})}{8 N_\mathrm{t}} \approx 0.52 ~\left( \frac{r_{\mathrm{s}}}{d} \right)^3,
\end{align}
which for $r_{\mathrm{s}} = d / 2$ gives $\eta \simeq 0.065$.
However, as the ratio $r_{\mathrm{s}} / d$ increases, the position of $\bm x_0$ with respect to the tile edges and corners matters less, and the efficiency of Cartesian tiling increases, since smaller cubic tiles can better approximate a sphere of radius $r_{\mathrm{s}}$. For $r_{\mathrm{s}} \gg d$, $\eta \rightarrow 1$. 

\begin{figure}
        \centering
        \includegraphics[width=1.0\linewidth]{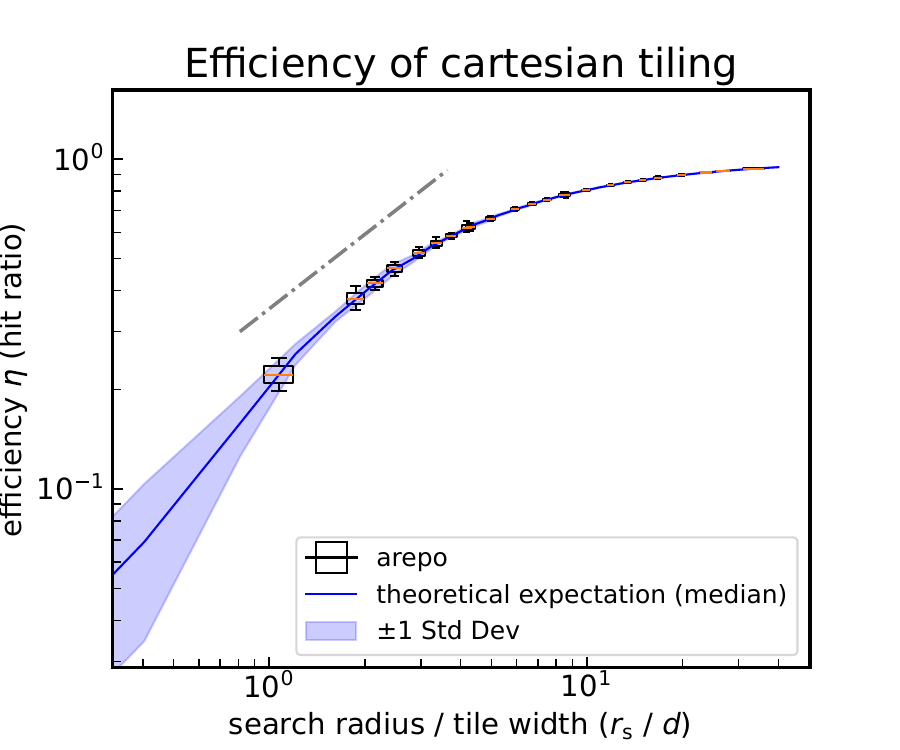}
        \caption{Cartesian tiling efficiency. The squares with error bars have been obtained by filtering a cubic region of size $L\simeq 1.49$~Mpc centered on the center of \texttt{Halo3}, and for different combinations of filter sizes $\ell$ and tiling parameter $M$. The solid blue line and shaded region represent the expected efficiency for a uniform random distribution of points (median value and one standard deviation, respectively). The dash-dotted gray line corresponds to a slope of $3/4$ and is plotted for reference.
        }
        \label{fig:tiling_efficiency}
\end{figure}

We measure the efficiency of our algorithm for \texttt{Halo3} at $z=0$, selecting a cubic region of size $L \simeq 1.49~\si{Mpc}$. We then apply the smoothing filter with spherical top-hat kernel and $\ell = \lbrace 30,  60,  90, 120, 150, 240 \rbrace ~\si{kpc}$, equal to the search radius, and for different tiling parameter, $M= \lbrace 64, 128, 256, 512 \rbrace$. For each pair ($\ell, M$), we compute the efficiency, $\eta$, and plot it against $r_{\mathrm{s}}/d$ in Fig.~\ref{fig:tiling_efficiency}. 
For $(r_s/d) \gtrsim 1$ the data points approximately follow a power-law with slope $3/4$ (shown for reference).
We also numerically determine the expected efficiency as a function of $r_{\mathrm{s}}/ d$ for a uniform distribution of randomly drawn points in a three-dimensional Cartesian grid; the median and $1 \sigma$ standard deviation are shown in Fig.~\ref{fig:tiling_efficiency} for comparison. As we can see, the theoretical expectation for a uniform distribution of points and the measured efficiency of our Cartesian tiling agree very well, which confirms the correctness of our implementation, and suggests that, within $L \simeq 1.49~\si{Mpc}$, the density of mesh-generating points is relatively homogeneous on the filter length scales employed here. 

\subsection{Timing benchmarks} 

\begin{figure}
        \centering
        \includegraphics[width=1.0\linewidth]{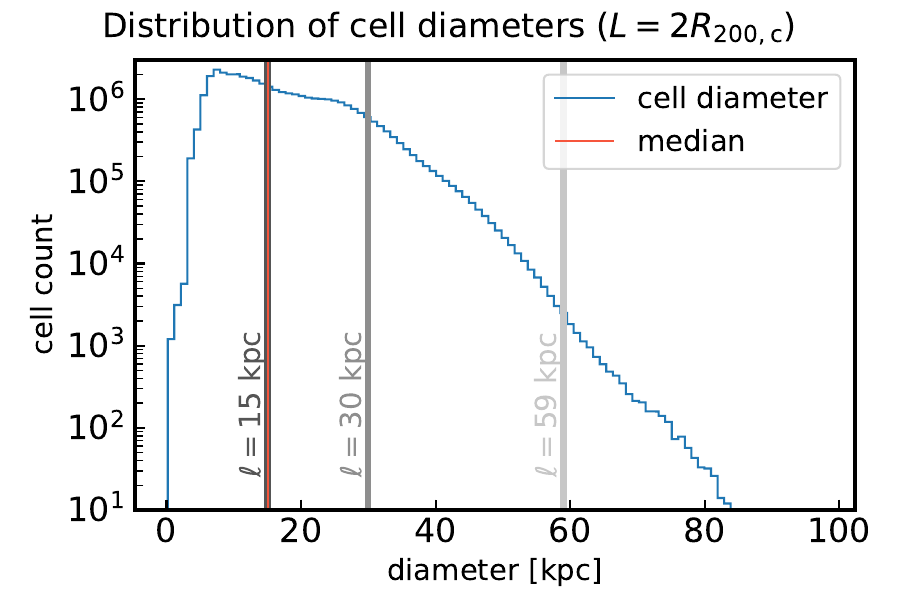}
        \caption{Histogram of the cell diameters inside a cubic region of $L=2R_{200,\mathrm{c}}$ centered on the cluster center. The vertical red line represents the median of the distribution, while the solid gray lines indicate the filter scales used with the Gaussian and spherical top-hat kernel.
        }
        \label{fig:diameter_distrib}
\end{figure}

We performed a number of timing tests on the $z=0$ snapshot of \texttt{Halo3}, using an NVIDIA A100  GPU. We kept the size of the cubic region fixed ($L = 2 R_{200,\mathrm{c}} \simeq 5.9~\si{Mpc}$), and filtered the data using three filter scales, $\ell = \left\lbrace 15,~ 30,~ 60 \right\rbrace ~\si{kpc}$, as in Sect.~\ref{sec:turb-bulk-energies}, both with spherical top-hat and with Gaussian kernels. This selection contains $N\simeq 39\times 10^6$ mesh-generating points of our computational Voronoi cells, with a median spatial resolution of ${\approx} 15$~kpc (assuming they are of spherical shape; see Fig.~\ref{fig:diameter_distrib}). This suggests that, for the spherical top-hat and for the lowest filter scale, in about $50\%$ of the cases the search radius around the cell is actually empty. From a physical perspective such a scenario should be avoided. For the Gaussian kernel, on the other hand, even for $\ell = 15$~kpc, the search radius is $60$~kpc, which is inadequate only for a low number of cells in the cluster outskirts.  

Using our default parameters ($M=256$, for which $d\simeq 23$~kpc), it takes 20.4 (0.4) seconds to apply a Gaussian (spherical top-hat) smoothing filter for the intermediate scale $\ell = 30~$kpc. These timings increase substantially for the largest scale $\ell = 60~$kpc, particularly for the Gaussian kernel (137.4 s), which by construction has a search radius four times as large as $\ell$. We repeated this procedure using different values of $M= \lbrace 64, 128, 256, 512 \rbrace$ for the Cartesian tiling, in order to assess whether there is an optimal choice that minimizes the execution time. As it turns out, $M=256$ indeed represents the value for which the execution time is lowest, while choosing $M<256$ and $M=512$ leads to longer execution. This makes intuitive sense, since for $M=512$ the total number of Cartesian tiles becomes higher than the number of particles in our selection: as a result many tiles end up being empty, and computational resources are wasted looping over empty tiles. On the contrary, if $M$ is too small, the ratio $r_{\mathrm{s}}/d$ decreases, and neighbor searching using Cartesian tiling is less efficient (see Fig.~\ref{fig:tiling_efficiency}).

\begin{table}  
        \centering 
        \caption{Benchmark of smoothing filter with an  NVIDIA A100 GPU.
        }
        \label{tab:parameters} 
        \begin{tabular}{l|ccc}
        $M$ & $\ell=15$~kpc & $\ell=30$~kpc & $\ell=60$~kpc \\ \hline
        64 & 13.3 (2.5) & 46.1 (3.1) & 218.9 (8.4) \\
        128 & 5.6 (0.4) & 26.5 (0.9) & 158.1 (3.0) \\
        256 & 3.5 (0.1) & 20.4 (0.4) & 137.4 (1.7) \\
        512 & 3.9 (0.1) & 27.3 (0.4) & 194.5 (2.1) \\
        \hline
        \end{tabular}
        \tablefoot{Timings in seconds of the smoothing filter applied to \texttt{Halo3} at $z=0$, for different grid sizes of the Cartesian tiling ($M$) and for different smoothing lengths ($\ell$). The values quoted in the Table refer to the application of a Gaussian filter, with the spherical top-hat filter in brackets. The region filtered is a cube of size $L = 2 R_{200,\mathrm{c}} \simeq 5.9~\si{Mpc}$.
        }
\end{table}

\end{document}